\documentclass[10pt]{article}
\usepackage{amsmath,amssymb,amsthm,amscd,mathtools, graphicx, mathtext,color,wrapfig, verbatim, slashed}
\usepackage{tikz-cd}
\usepackage{bbm}
\usepackage{slashed}

\usepackage{mathrsfs}
\usepackage{pifont}
\usepackage{graphicx}
\usepackage[font={small,sl}]{caption}
\usepackage[all]{xy}

\usepackage{tikz}
\usetikzlibrary{matrix,arrows}

\usepackage[numbers,sort&compress]{natbib}
\makeatletter

\@addtoreset{equation}{section}
\makeatother
\newcommand{\slzero}{\mathord{\mathpalette\slzeroaux\relax}}
\newcommand{\slzeroaux}[2]{%
  \ooalign{%
    \hfil\raisebox{.12ex}{\scalebox{.75}{$#1/$}}\hfil\crcr
    $#1 0$\crcr
  }%
}

\newcommand{\beq}{\begin{equation}}
\newcommand{\eeq}{\end{equation}}

\newcommand{\rd}{/\!\!/\!\!/\!\!/}

\newcommand{\Vertex}{\mathsf{Vertex}}

\newcommand{\nc}{\newcommand}
\nc{\mc}{\mathcal}
\newcommand{\scQ}{\ensuremath{\mathcal{Q}}}

\newcommand{\fg}{\mathfrak{g}}

\newcommand{\fq}{\mathfrak{q}}

\newcommand{\fgh}{\widehat{\mathfrak{g}}}

\newcommand{\MDX}{{\mathscr{D}_{\cal X}}}
\newcommand{\MDY}{{\mathscr{D}_{\cal Y}}}
\newcommand{\MDx}{{\mathscr{D}_{X}}}
\newcommand{\MDy}{{\mathscr{D}_{Y}}}
\newtheorem{theorem}{Theorem}[section]

\theoremstyle{definition}

\begin{document}
\setcounter{tocdepth}{2}

\begin{titlepage}


\renewcommand{\thefootnote}{\fnsymbol{footnote}}

\vskip 1.0cm

\begin{center}
{\LARGE \bf
Knot Categorification from
\vskip 0.5 cm
Mirror Symmetry
\vskip 0.5cm}
{\it \large
Part III: String Theory Origin
\vskip 0.5cm }

{\large
Mina Aganagic}

\medskip

\vskip 0.5cm

{\it
Leinweber Institute for Theoretical Physics, University of California, Berkeley\\
Department of Mathematics, University of California, Berkeley\\
}

\end{center}

\vskip 0.5cm

\centerline{{\bf Abstract}}
\medskip
In Parts I and II of this series \cite{A1,A2}, I provided two geometric approaches to the categorification of Chern-Simons link invariants associated to an arbitrary simply laced Lie algebra $\fg$: one based on B-type branes, or coherent sheaves, the other on A-branes, or Lagrangians, related by homological mirror symmetry \cite{ADZ}. The resulting theory is explicitly solvable \cite{ALR}. 

In this paper, I describe the string-theory origin of the theory. It arises from self-dual strings of the six-dimensional $(2,0)$ theory in the abelianized phase. A link $K$ comes from self-dual string defects supported on $K\times{\mathbb R}$, and its homology is the space of BPS states in their presence. The two approaches originate from two descriptions of the self-dual string world-sheet theory, related by T-duality. The six-dimensional origin makes it manifest that the homology groups are invariants of links. 

The theory generalizes to non-simply laced Lie algebras and to links in $\Sigma \times \mathbb R$, where $\Sigma$ is an arbitrary open surface. The category of A-branes is equivalent to the derived category of modules of the KLRW algebra of $\Sigma$ \cite{ALSZ}, an equivalence which categorifies a key step of the quantum geometric Langlands correspondence.

\noindent\end{titlepage}
\setcounter{page}{1} 

\setcounter{section}{0}
\tableofcontents 
\newpage
\section{Introduction}

This is the third in a sequence of papers on categorification of quantum link invariants \cite{A1,A2}. In the previous two papers \cite{A1,A2}, I gave two geometric solutions to the link categorification problem, related by homological mirror symmetry. In this paper, I describe the string-theory origin of the theory and the generalizations it leads to.

\subsection{Chern-Simons theory and quantum groups}

Witten showed in 1989 that Chern-Simons theory with gauge group $G$, Lie algebra ${\fg}$, and level $k$ leads to invariants of knots and links in three dimensions \cite{Jones}. He also showed how to solve Chern-Simons theory, reducing the problem of computing its partition functions to matrix multiplication. The relevant matrices describe the action of braiding on the space of conformal blocks of the affine Lie algebra ${\fgh}_k$.
Reshetikhin and Turaev \cite{RT1,RT2} obtained the same link and three-manifold invariants from R-matrices of the quantum group $U_{\fq}(\fg)$.
Drinfeld and Kohno proved that the braiding matrices of ${\fgh}_k$ and the R-matrices of $U_{\fq}(\fg)$ coincide when $k$ and $\fq$ are related by ${\fq}= \exp(2\pi i/(k+h^{\vee}))$ \cite{Drinfeld, Koh}. For link invariants alone, ${\fq}$ may be a generic complex number, and $k$ need not be an integer. When $k$ is not an integer, as we will assume throughout, Chern-Simons theory is analytically continued \cite{WittenACS}.

\subsection{The problem}
For any link $K$ in $\mathbb R^3$, a Lie algebra $\fg$, and representations coloring the components of $K$, the Chern-Simons invariant is a Laurent polynomial\footnote{Up to a prefactor ${\fq}^{\#}$.} in ${\fq}$
\beq\label{PKf}
P_K(\fq)= \sum_{j\in\mathbb Z}a_j(K)\,\fq^{j},
\eeq
with integer coefficients $a_j(K)\in\mathbb Z$. This integrality suggests that $P_K$ is the Euler characteristic of a more refined homological invariant.

\subsubsection{}
In 1998, Khovanov \cite{Kh} showed that when $P_K$ is the Jones polynomial, so that $\fg=\mathfrak{su}_2$ and the links are colored by its defining representation, there is a bigraded chain complex
\beq
C^{*,*}(K)
\eeq
associated to $K$, with differential
\beq
\partial:C^{i,j}(K)\longrightarrow C^{i+1,j}(K),
\eeq
whose homology groups
\[
{\cal H}^{i,j}(K)= {\rm ker}\,\partial^i / {\rm im}\,\partial^{i-1}
\]
are link invariants and whose graded Euler characteristic 
\beq\label{EJ}
\chi(K)(\fq)= \sum_{i, j\in\mathbb Z} (-1)^i {\fq}^j \,{\rm dim}\, {\cal H}^{i, j}(K)
\eeq
is the Jones polynomial.

\subsubsection{}
From the physics perspective, the generators of the complex are analogues of perturbative supersymmetric ground states, the differential should be generated by instantons, and its cohomology gives the exact supersymmetric ground states \cite{WittenM}. The Morse theory approach to supersymmetric quantum mechanics is the only known way to produce the space of supersymmetric ground states.
\subsubsection{}
Khovanov's construction is simple and explicit. Its limitation is that it has no evident geometric or physical origin, and it does not readily generalize to arbitrary Lie algebras and representations. Webster gave a general framework for categorifying $U_{\fq}(\fg)$ link invariants for arbitrary simple Lie algebras $\fg$ in 2013 \cite{webster}. His theory is based on KLRW algebras, which generalize the algebras used by Khovanov and Lauda \cite{KhL} and by Rouquier \cite{Rh}, for categorifying the quantum group itself. Webster's construction is purely formal. It does not by itself provide a method for computing the resulting homology groups, not even for the unknot.
\subsubsection{}
The problem addressed in Parts I and II was to find the theory categorifying $U_{\fq}(\fg)$ link invariants for general $\fg$, which is geometric and explicitly solvable. There is in fact a pair of geometric theories, one based on the category of B-branes, or coherent sheaves \cite{A1}, and the other on a category of A-branes, or Lagrangians \cite{A2}. They extend the link homology from links in $\mathbb R^3$ to links in $\mathbb R^2\times S^1$. 

The two theories are equivalent, related by homological mirror symmetry \cite{ADZ}. For links in ${\mathbb R}^3$, the homology theory that results from \cite{A1,A2} is equivalent to Webster's, based on KLRW algebras \cite{ADZ, W1, W2}. Thanks to homological mirror symmetry, which makes complementary aspects of the theory easy to understand, the resulting homology theory is explicitly computable \cite{ALR}.

The categories of A-branes in \cite{A2} and in this paper are generalizations of Heegaard Floer theory to arbitrary simple Lie algebras $\fg$. Heegaard Floer theory categorifies the Alexander polynomial \cite{Osz, Rasmussen}, corresponds to $\fg=\mathfrak{gl}_{1|1}$ and is explicitly solvable.
\subsubsection{}

The aim of this paper is to derive the theories of Parts I and II from string theory. Long before \cite{A1,A2}, string theory had suggested that knot homology should have a physical origin in the six-dimensional $(2,0)$ conformal field theory \cite{OV,GSV,WF}. Now we learn precisely how this comes to bear.
\subsection{The solution}\label{solution}

The theory discovered in \cite{A1,A2,ALR,ADZ}, and generalized here, produces homological invariants of links, colored by minuscule representations of a simple Lie algebra $\fg$, in
\beq
M_{3,{\rm CS}}=\mathbb R\times\Sigma,
\eeq
where $\Sigma$ is a non-compact surface. In the original \cite{A1, A2}, ${\fg}$ was a simply laced Lie algebra and $\Sigma={\mathbb R}\times S^1$.

\subsubsection{}
The theory has a description based on a category of A-branes on a manifold ${\cal Y}={\cal Y}(\Sigma)$, equipped with a top holomorphic form $\Omega$ and a multivalued superpotential ${\cal W}$. The multivaluedness of ${\cal W}$ gives rise to the equivariant gradings, which include a grading associated to ${\fq}$. The category is the derived Fukaya-Seidel category of ${\cal Y}(\Sigma)$ with potential ${\cal W}$. We denote the category by
\beq
{\mathscr  D}_{\cal Y}=D Fuk({\cal Y}(\Sigma),{\cal W}).
\eeq
Throughout, we will identify ${\mathscr  D}_{\cal Y}$ with its subcategory split-generated by a collection of branes described in sections \ref{HMS} and \ref{s-exact}. The subcategory is expected to be the whole category, an assumption which does not affect the link homology theory, or its string origin.
The target ${\cal Y}(\Sigma)$ projects to
\beq\label{Ysim}
Y(\Sigma)=\prod_{a=1}^{{\rm rk}\fg}{\rm Sym}^{d_a}(\Sigma),
\eeq
the configuration space of points on $\Sigma$ colored by the simple roots of $\fg$. Above, $d_a$ is the number of points colored by the simple root $e_a$ of $\fg$.

${\cal Y}(\Sigma)$ is known as the $G'$ open Zastava space of $\Sigma$, where $G'$ is a Lie group of adjoint type, whose Lie algebra $\fg'$ is the Langlands dual of $\fg$. It has played a central role in the geometric Langlands program \cite{Gaitsgory08, Raskin}.

\subsubsection{}
Take a link $K\subset M_{3, CS}=\Sigma\times\mathbb R$, with components colored by minuscule representations of $\fg$, and present it as the plat closure of a braid; this is a closure of a braid of $2m$ strands by $m$ consecutive caps at the top, and $m$ cups at the bottom. This data determines a collection of $2m$ marked points on $\Sigma$, the multiplicities $d_a$, and the superpotential ${\cal W}$. It also determines two A-branes in ${\mathscr  D}_{\cal Y}$ -- a brane ${\cal U}$ associated to the collection of cups, and the brane ${\mathscr B}{\cal U}$ which corresponds to the image of the collection of caps under the braid. A virtue of ${\cal Y}(\Sigma)$ is that these A-branes have an explicit description in terms of products of curves on $\Sigma$. The homological invariant of $K$ is the space ${\rm Hom}^{*,*}_{{\mathscr  D}_{\cal Y}}
({\mathscr B}{\cal U},{\cal U})$ of morphisms between the two branes in ${\mathscr  D}_{\cal Y}$:
\beq\label{KA}
{\cal H}^{*,*}(K)
=
{\rm Hom}^{*,*}_{{\mathscr D}_{\cal Y}}
({\mathscr B}{\cal U},{\cal U}).
\eeq
In Heegaard Floer theory the target is just ${\rm Sym}^{d}(\Sigma)$, but the construction is otherwise analogous.
\subsubsection{}
In a category of A-branes on ${\cal Y}$ with superpotential ${\cal W}$, morphisms between two branes are the supersymmetric ground states of the theory on a strip, with those branes as boundary conditions. The space of morphisms $Hom^{*,*}_{\MDY}({\mathscr B}{\cal U},{\cal U})$ is defined by Floer theory. This corresponds to viewing the theory on the strip as an effective supersymmetric quantum mechanics whose target is the space of maps from the interval to ${\cal Y}$ \cite{GMW2}. This gives 
\beq\label{ww}
Hom^{*,*}_{\MDY}({\mathscr B}{\cal U},{\cal U}) ={\rm ker}\,\partial / {\rm im}\,\partial,
\eeq
as the cohomology of a Floer complex
\beq\label{FCI0}
C^{*,*}({\mathscr B}{\cal U},{\cal U})=\bigoplus_{i,j} C^{i,j}({\mathscr B}{\cal U},{\cal U}),
\eeq
spanned by the intersection points of the two Lagrangians ${\mathscr B}{\cal U}$ and ${\cal U}$. The complex is graded by the fermion number $i$ and additional equivariant gradings, which include a grading associated to ${\fq}$. The differential
\beq\label{FCI1}
\partial: C^{*,j}({\mathscr B}{\cal U},{\cal U})\longrightarrow C^{*+1,j}({\mathscr B}{\cal U},{\cal U})
\eeq
is the Floer differential, obtained by counting holomorphic maps from the strip to ${\cal Y}$. Thus, as in Khovanov's construction, the invariant is defined at the level of a complex, not only at the level of its homology groups. Here the complex and its differential arise geometrically from Floer theory.

\subsubsection{}
By closing the strip to an annulus, the theory computes the Euler characteristic of the link homology
\beq\label{BM}
\chi({\mathscr B}{\cal U},{\cal U})
=
\sum_{{\cal P}\in {\mathscr B}{\cal U}\cap {\cal U}}
(-1)^{M({\cal P})}\,{\fq}^{J({\cal P})}.
\eeq
The sum is over all intersection points ${\cal P}$ of the two branes, with $M({\cal P})$ their Maslov, or homological, grading and $J({\cal P})$ their equivariant grading. Thus the decategorified invariant is a signed, ${\fq}$-graded count of intersections, giving the corresponding Chern-Simons, or $U_{\fq}(\fg)$, link invariant.

\subsubsection{}
When $\Sigma=\mathbb R\times S^1$ as in \cite{A1, A2} and $\fg$ is a simply laced Lie algebra, the theory has another description in terms of a category of B-type branes on ${\cal X}$, with an equivariant action ${\rm T}$. ${\cal X}$ is the moduli space of $G'$ monopoles on $\mathbb R^3=\mathbb R\times\mathbb C$. The torus action ${\rm T}$ contains a $\mathbb C^\times_{\fq}$ action that scales $\mathbb C$ with weight $\fq$.

The category of B-type branes is the derived category of ${\rm T}$-equivariant coherent sheaves on ${\cal X},$ which we will denote by
\beq
{\mathscr  D}_{\cal X}=D{\rm Coh}_{\rm T}({\cal X}).
\eeq
The sigma models on $({\cal X}, {\rm T})$ and $({\cal Y}, {\cal W})$ are mirror. Mirror symmetry identifies the A-side invariant \eqref{KA} with
\beq
{\cal H}^{*,*}(K)={\rm Hom}^{*,*}_{{\mathscr D}_{\cal X}}({\mathscr B}{\cal U},{\cal U}).
\eeq
In ${\cal X}$, the B-brane ${\cal U}$, mirror to the A-brane of the same name, is the structure sheaf of a holomorphic Lagrangian supported on a vanishing cycle. The vanishing cycle is a product of the minuscule Grassmannians associated to the colors in the matching, one for each cup. For $\fg=\mathfrak{su}_2$, these are vanishing $\mathbb P^1$'s. 

We will explain in this paper how this description generalizes to an arbitrary simple Lie algebra ${\fg}$; the generalization is more subtle than on the A-side.

\subsubsection{}
The theory is exactly solvable \cite{ALR}. This rests on four facts: First, on both sides of mirror symmetry, suitable collections of branes generate the relevant categories. Second, the algebra of open strings between those branes is explicitly computable \cite{W1, W2, ADZ}. It turns out to be a cylindrical generalization of the KLRW algebra from \cite{webster}. Thus, the theory of \cite{A1, A2} recovers, from geometry, the algebraic categorification of $U_{\fq}(\fg)$ link invariants developed by Webster. Third, in this particular family of theories, one can describe the ${\mathscr B}{\cal U}$ brane as a complex, a bound state, of the generating branes. Finally, the complex which resolves ${\mathscr B} {\cal U}$ in terms of the generators suffices to obtain $Hom_{{\mathscr D}_*}^{*,*}({\mathscr B}{\cal U}, {\cal U})$ from \eqref{ww}, as the cohomology of the complex \eqref{FCI0} with a differential \eqref{FCI1}, giving homologies of arbitrary links in $M_{3,CS}$. 
\subsubsection{}
The fact that for ${\fg}=\mathfrak{su}_2$ the theory in \cite{A2, ALR} reproduces Khovanov homology was proven recently in \cite{LeS}. It follows from \cite{MackaayWebster} that the theories of \cite{A1,A2} should reproduce the previously known type A link homologies, constructed by Khovanov and Rozansky \cite{KhR} and generalized in \cite{Wu, Yonezawa} to all anti-symmetric representations.

As an elementary example of the theory, we will compute the homology of the unknot colored by a minuscule representation for an arbitrary Lie algebra $\fg$. This generalizes \cite{A1} where ${\fg}$ was a simply laced Lie algebra. Outside of type $A$, even this simple result is new.\footnote{Recently, a different $\fg =B_n$ homology of links colored by the spinor representation was constructed in \cite{Elias}.}

\subsection{String theory origin}

This paper explains the superstring origin of the theories from \cite{A1, A2}. String theory played a crucial role in discovering \cite{A1,A2}, and it also leads to their generalizations, two of which we will describe here: to non-simply laced Lie algebras and to invariants of knots and links in $M_{3, CS}= \Sigma\times{\mathbb R}$, where $\Sigma$ is any non-compact Riemann surface.

The starting point is type IIB string theory on an ADE surface singularity. In a limit, the theory becomes the six-dimensional conformal field theory with $(2,0)$ supersymmetry, associated to a simply laced Lie algebra $\fg$. 

\subsubsection{}
One key result of this paper will be a description of the $(2,0)$ conformal field theory on a six-manifold
$$M_6=\Sigma\times{\mathbb C}\times{\mathbb R}^2,$$
in the abelianized phase, in terms of the theory on its self-dual strings. 

The self-dual strings of the 6d $(2,0)$ theory originate from D3-branes of IIB string, wrapping two-cycles of the ADE surface. 
We will use this to derive the theories on self-dual strings of $(2,0)$ CFT. 

\subsubsection{}\label{Tdual}
The worldsheet theory on strings probing $\Sigma\times{\mathbb C}\subset M_6$ is a sigma model to ${\cal Y}(\Sigma)$ with superpotential ${\cal W}$ given in section \ref{ss-limit}. This description is obtained by T-duality on an $S^1$ isometry of ${\mathbb C}.$

When $\Sigma={\mathbb R}\times S^1$, the world-sheet theory has a second description, obtained by T-duality with respect to the $S^1$ in $\Sigma$ instead. The result is a sigma model to ${\cal X}$, with equivariant ${\rm T}$ action, also derived in section \ref{ss-limit}. 

Both ${\cal Y}$ and ${\cal X}$ are monopole moduli spaces. The light self-dual strings, which are finite tension excitations of $(2,0)$ theory, are associated to smooth, 't Hooft-Polyakov monopoles. The heavy strings, which are defects of $(2,0)$ theory with infinite tension, correspond to singular, Dirac monopoles. 

\subsubsection{}
The world-sheet theory has a special class of domain walls and boundary conditions that let us describe heavy self-dual strings supported on
$$K\times{\mathbb R}\subset M_{3, CS} \times {\mathbb R} =\Sigma\times{\mathbb R}^2,$$
where $K$ is a framed link in $M_{3, CS}= \Sigma\times{\mathbb R}$, in terms of the theory on light self-dual strings on a strip
$$[-1,1]\times{\mathbb R}\subset{\mathbb R}^2,$$
with pairs of branes as boundary conditions. These branes are what we called ${\cal U}$ earlier, or ${\mathscr B}{\cal U}$ after applying a braid-group action. The construction of the ${\cal U}$ and ${\mathscr B}{\cal U}$-branes relies on the description of ${\cal X}$ and ${\cal Y}$ as monopole moduli spaces, specifically on monopole bubbling. While for most of this paper we are focused on invariants of framed links, the theory has an extension to ribbon graphs.

\subsubsection{}
In the theory based on ${\cal Y}$, the ${\cal U}$-branes preserve A-type supersymmetry; in the description based on ${\cal X}$, they preserve B-type supersymmetry. The T-dualities of section \ref{Tdual} are needed to avoid the mixed A/B type boundary conditions. 
\subsubsection{}
The theory thus assigns to a link $K$ the space of supersymmetric ground states of the theory on the strip with the corresponding pair of branes as boundary conditions. These are, by definition, the spaces of morphisms
$${\cal H}^{*,*}(K)=Hom_{{\mathscr D}_*}^{*,*}({\mathscr B}{\cal U},{\cal U})$$
between the branes in the corresponding category. The space of supersymmetric ground states of the theory of self-dual strings is also the space of BPS states that the 6d $(2,0)$ theory assigns to the link.
\subsubsection{}
As we will review in sections \ref{sec2} and \ref{sec3}, the fact that the Euler characteristic of $Hom_{{\mathscr D}_*}^{*,*}({\mathscr B}{\cal U},{\cal U})$ is the Chern-Simons link invariant holds by construction. It also holds by construction that $Hom_{{\mathscr D}_*}^{*,*}({\mathscr B}{\cal U},{\cal U})$ is a braid invariant.

It is not a priori manifest, from the descriptions based on $\MDY$ and $\MDX$, that it is a link invariant. The fact that ${\cal H}^{*,*}(K)= Hom_{{\mathscr D}_*}^{*,*}({\mathscr B}{\cal U},{\cal U})$ is an isotopy invariant of a link $K$ in $M_{3,CS}$ can be proven explicitly, as in \cite{ALR}. 

The six-dimensional origin makes isotopy invariance of ${\cal H}^{*,*}(K)$ manifest \cite{WF}. The 6d $(2,0)$ theory on $M_6$ is topologically twisted along $\Sigma\times{\mathbb R}^2$ (with holomorphic twist along ${\mathbb C}$). Adding self-dual strings supported on $K\times{\mathbb R}$ preserves topological invariance, and therefore guarantees invariance under isotopy of $K$.

\subsubsection{}
 It is shown in \cite{ALSZ} that the category ${\mathscr D}_{{\cal Y}(\Sigma)}$ of A-branes on ${\cal Y}(\Sigma)$ is exactly solvable, for any $\Sigma$, as in the special case $\Sigma={\mathbb R}\times S^1$. It has a generating brane whose endomorphism algebra ${\mathscr A}(\Sigma,\gamma)$ is computable explicitly, and defines the KLRW algebra of the surface $\Sigma$. This gives Theorem \ref{sige},
\beq
{\mathscr D}_{{\cal Y}(\Sigma)}\simeq{\mathscr D}_{{\mathscr A}(\Sigma,\gamma)},
\eeq
where ${\mathscr D}_{{\mathscr A}(\Sigma,\gamma)}$ is the derived category of modules of ${\mathscr A}(\Sigma,\gamma)$.

The correspondence in Theorem \ref{sige} is a statement of the Langlands correspondence: On the left-hand side, only the $\fg'$ symmetry is manifest, as ${\cal Y}(\Sigma)$ is the $G'$ Zastava space. The right-hand side has the Langlands dual $\fg$ symmetry manifest, categorifying $U_{\fq}(\fg)$-modules on $\Sigma$. It categorifies the quantum geometric Langlands correspondence.
\subsection{Organization}
The paper is organized as follows.

Section 2 reviews the essential content of Parts I and II \cite{A1,A2}, including the $\fgh_\kappa$ Knizhnik-Zamolodchikov equation and the relation of its monodromies to $U_{\fq}(\fg)$ link invariants. There are analogous equations of geometric origin, the quantum differential equation and its mirror. Their monodromy matrix elements get categorified by spaces of morphisms in the corresponding categories of $B$- and $A$-branes. Once the KZ equation is realized as a quantum differential equation for some $({\cal X},{\rm T})$, or in the mirror description $({\cal Y},{\cal W})$, its monodromy has a natural categorical lift. This provides a path to categorification of quantum link invariants. 

Section 3 recalls the IIB string origin of the $(2,0)$ little string theory and its self-dual string defects. The theories on self-dual strings in the abelianized phase of the $(2,0)$ theory are the quiver gauge theories described in \cite{DM,AH}. In an appropriate setting, the K-theoretic vertex functions of these gauge theories produce solutions of the qKZ equation, a q-difference deformation of the KZ equation.

Having understood the string-theory origin of the qKZ equation, obtaining the KZ equation amounts to taking an appropriate limit. The limit, taken in Section 4, is the same one that takes little string theory to the $(2,0)$ conformal field theory. In the limit, the gauge theory description of self-dual strings in the little string theory goes over to a pair of non-linear sigma models, related by T-duality. One has target $({\cal X},{\rm T})$, and the other has target $({\cal Y},{\cal W})$.  The pair of T-dualities explains the string-theoretic origin of the mirror symmetry relating them. 

Section 5 applies the results of the previous section to obtain invariants of a link $K$ in $\Sigma\times{\mathbb R}$, by considering heavy self-dual strings supported on $K\times{\mathbb R}$ in $M_6=\Sigma\times{\mathbb C}_{\fq}\times{\mathbb R}^2$. Presenting $K$ as a plat closure of a braid, the self-dual strings are supported on an interval times ${\mathbb R}$, with a pair of branes as boundary conditions. The section provides the six-dimensional construction of boundary conditions associated to plat closures. These arise from monopole bubbling and correspond to the ${\cal U}$ and ${\mathscr B}{\cal U}$ branes of $\MDX$ and $\MDY$. This identifies the space of supersymmetric ground states of the self-dual string theory with
\( {\cal H}^{*,*}(K)=Hom_{\mathscr{D}_*}^{*,*}({\mathscr B}{\cal U},{\cal U}) \).
Its six-dimensional origin explains why these vector spaces are invariants of links. This completes the derivation of theories studied in \cite{A1,A2}. In section 6 we recall some aspects of the theories in \cite{A1, A2} we will need; in particular how mirror symmetry is proven and used to solve the theories.

Sections 7 and 9 describe generalizations implied by the string construction. Section 7 describes the generalization to non-simply laced Lie algebras. Section 9 extends the construction to links in $M_{3, CS}=\Sigma\times{\mathbb R}$, where $\Sigma$ is an arbitrary open surface. Section 8 computes, as a simple example, the homology of the unknot colored by a minuscule representation $V$ of an arbitrary ${\fg}$. Outside of type A, the result is new. Section 9 explains how Theorem \ref{sige}, which solves the theory on $\Sigma$, categorifies the quantum Langlands correspondence.

\subsection{Acknowledgments}
Elise LePage, Peng Zhou and Vivek Shende have made essential contributions to the theory laid out in this series of works. The work derives from earlier collaborations with Andrei Okounkov, Edward Frenkel and Nathan Haouzi, for which I am grateful. I also benefited from discussions with Mohammed Abouzaid, Roman Bezrukavnikov, Sergei Cherkis, Ivan Danilenko,  Davide Gaiotto, Michael Finkelberg, Joel Kamnitzer, Spencer Tamagni, Ben Webster and Edward Witten. This work has been supported by the Simons Investigator and Simons Fellows Awards from the Simons Foundation, by the NSF Grants PHY-2112880 and PHY-2610058, and by the Miller Institute for Basic Science and the Leinweber Institute for Theoretical Physics at UC Berkeley.
\section{A tale of two equations}\label{sec2}

The Knizhnik-Zamolodchikov equation and the quantum differential equation are two equations of the same flavor, which arise in different parts of mathematical physics. The first leads to $U_{\fq}(\fg)$ invariants of links. The second has a natural categorical lift, through categories of branes and their monodromy functors. In this section, we recall these two parallel stories. In the next section, we will specialize to the situation in which the two equations coincide.

\subsection{The KZ equation}\label{ss:KZ}
Let
\[
M_{3, CS}={\Sigma}\times {\mathbb R}_{CS} ,
\]
where, until stated otherwise, $\Sigma$ is an infinite cylinder
\beq\label{cylc}
\Sigma={\mathbb R}\times S^1.
\eeq
After choosing a complex coordinate $y$, one may regard $\Sigma$ as the punctured complex plane $\mathbb C^\times={\mathbb C}\backslash 0$. 
\subsubsection{}
Choose a collection of interior marked points $a_i\in \Sigma={\mathbb C}\backslash 0$, labeled by finite-dimensional representations $V_i$ of $\fg$, $a_i\neq 0, \infty$. Label the punctures at $0$ and $\infty$ by generic highest-weight Verma modules $V_\lambda$ and $V_{\lambda'}$. The corresponding conformal blocks of the affine Lie algebra $\fgh_k$ have the form
\begin{equation}\label{confb}
{\mathcal V}=\langle \lambda'|\,\Phi_{V_1}(a_1)\cdots \Phi_{V_n}(a_n)\,|\lambda\rangle .
\end{equation}
They are flat sections of the Knizhnik-Zamolodchikov connection
\begin{equation}\label{KZ}
\kappa\,\partial_i {\mathcal V}=\sum_{j\neq i} r_{ij}(a_i/a_j)\,{\mathcal V},
\end{equation}
where $\kappa=k+h^\vee$, $\partial_i=a_i\frac{\partial}{\partial a_i}$, and $r_{ij}$ is the trigonometric classical $r$-matrix acting on $V_i\otimes V_j$. In the \eqref{KZ}, we include in the sum over $j$ the point at $0\in {\mathbb C}$, which is strictly deleted from $\Sigma$. 
\subsubsection{}
For generic $\kappa$, the space of conformal blocks of the form \eqref{confb} is the fixed weight subspace of the tensor product,
\begin{equation}\label{weight}
(V_1\otimes\cdots\otimes V_n)_{\lambda'-\lambda}
\end{equation}
 with weight $\lambda'-\lambda$. This space is the Hilbert space of Chern-Simons theory with gauge group $G$ and level $k$ on $\Sigma$. We will take $G$ to be a simply connected Lie group with Lie algebra ${\fg}$, so its character lattice is as large as possible. 

\subsubsection{}
A loop in the configuration space of the $n$ marked points on $\Sigma$ defines a braid $B$ of $n$ strands in $M_{3, CS}.$ 
Monodromy of the KZ equation along this loop acts on the space of conformal blocks on $\Sigma$, of the form in \eqref{confb}. By the Drinfeld-Kohno theorem, the monodromy matrix of the KZ equation is given in terms of products of $R$ matrices of the $U_{\fq}(\fg)$ quantum group. Correspondingly, monodromy assigns to a braid $B$ an element
\beq\label{bm}
{\mathfrak B} \in Hom_{U_{\fq}(\fg)}((V_1\otimes\cdots\otimes V_n)_{\lambda'-\lambda}).
\eeq
The quantum deformation parameter ${\fq}$ is related to $\kappa$ by
$$
\fq=\exp(2\pi i/\kappa).
$$
Since $\Sigma$ is an infinite cylinder, the KZ equation in \eqref{KZ} is of trigonometric type. Thus $B$ is an element of the affine braid group.
\subsubsection{}
The KZ equation thus relates Chern-Simons theory with gauge group $G$ and level $k$, and its Hilbert space on $\Sigma$ and $U_{\fq}(\fg)$ quantum groups. Since ${\fq}$ and hence $k$ are generic, Chern-Simons theory is analytically continued.

\subsection{$U_{\fq}(\fg)$ link invariants}

Any link 
$K\subset M_{3, CS}$ can be presented as the plat closure of some braid $B$ with an even number $2n$ of strands. The plat closure is by a collection of $n$ cups or caps, pairing nearby marked points on $\Sigma$.

\subsubsection{}\label{ss-qmat}
The $U_{\fq}(\fg)$ invariant of $K$ is the matrix element
\beq\label{PKq}
P_K=\langle {\mathfrak B}{\mathfrak U}|{\mathfrak U}\rangle,
\eeq
where ${\mathfrak B}$ is the braiding matrix corresponding to the braid $B$ from 
\eqref{bm}, $|{\mathfrak B}{\mathfrak U}\rangle={\mathfrak B}|{\mathfrak U}\rangle$, and 
\beq\label{cups}
|{\mathfrak U}\rangle
\eeq
is the state in Chern-Simons Hilbert space on $\Sigma$ describing the collection of $n$ cups. The conformal block ${\cal V}_{U}$ of $\fgh_k$ corresponding to the state $|{\mathfrak U}\rangle$ has $n$ pairs of ${\fgh}_k$ vertex operators, $\Phi_{V_i}(a_{2i-1})\otimes \Phi_{V_i^*}(a_{2i})$, fusing to a trivial representation.

\subsubsection{}
The quantum-group matrix element in \eqref{PKq} is invariant under braid moves by construction, as a consequence of Yang-Baxter equation satisfied by $R$-matrices of $U_{\fq}(\fg)$. Its Chern-Simons origin makes it manifest it is an invariant of a framed link $K$ in $M_{3, CS}$. 
\subsubsection{}
Specializing to links $K$ that fit into a three-ball in $M_{3,CS}$, $P_K=P_K(\fq)$ coincides with the invariant of a link $K\subset {\mathbb R}^3$ from \eqref{PKf}. The link categorification problem is describe the matrix elements in \eqref{PKq} as the graded Euler characteristic \eqref{EJ} of a bigraded vector space ${\cal H}(K) =\bigoplus_{i,j \in {\mathbb Z}} {\cal H}^{i,j}(K)$ which itself is a link invariant. 

For links in $M_{3,CS}$ for the more general $\Sigma$ in \eqref{cylc}, there are additional integral gradings associated to winding of $K$ around the $S^1$ in $\Sigma$. They are kept track of, from Chern-Simons theory perspective, by the holonomy of its gauge field around the $S^1$. The choice of holonomy is that of the highest weight of the Verma module $V_\lambda$. For links in a three-ball, the $\lambda$ dependence drops out.

\subsection{The quantum differential equation}\label{s-qde}
The Knizhnik-Zamolodchikov equation has a geometric counterpart. In mirror symmetry, one encounters a differential equation of the same general form,
\beq\label{qdifg}
\partial_i {\cal V}_\alpha-(C_i)^\beta_{\alpha}{\cal V}_\beta=0.
\eeq
The equation, its solutions and its monodromies all have a geometric interpretation. Moreover, the monodromy matrix elements have a manifest geometric categorification.
\subsubsection{}
Let ${\cal X}$ be a Kahler manifold, $\gamma_i\in H^2({\cal X})$ be a divisor class, and let $\partial_i$ be the tangent vector to the Kahler moduli space in the corresponding direction. The operator $C_i$ in \eqref{qdifg} is quantum multiplication by $\gamma_i$ acting on $H^*(\cal X)$. Thus
\[
(C_i)^\beta_{\alpha}=C_{\gamma_i\alpha\delta}\eta^{\delta\beta},
\]
where $\eta$ is the intersection pairing on $H^*(\cal X)$ and $C_{\alpha\beta\gamma}$ are the genus-zero three-point functions of the A-model on ${\cal X}$, with three observables corresponding to $\alpha, \beta, \gamma \in H^*(\cal X)$ inserted. Equivalently, the quantum product is
\[
\alpha*\beta=C_{\alpha\beta\delta}\eta^{\delta\gamma}\gamma .
\]
The associativity of the quantum product, $[C_i,C_j]=0$ and the fact the genus-zero potential satisfies $\partial_iC_j=\partial_jC_i$ ensure flatness of the connection.
\subsubsection{}\label{s-solutions}
Solutions of the quantum differential equation are labeled by $K$-theory classes of B-type branes on ${\cal X}$. For a B-type brane which we will denote by ${\cal L}$, its $K$-theory class $[{\cal L}]$ determines ${\cal V}_\alpha[{\cal L}]$, as follows.

The vector-valued flat section ${\cal V}_\alpha[{\cal L}]$ is the partition function of the ${\cal N}=2$ supersymmetric two dimensional sigma model with target ${\cal X}$ on ${\sf D}$, an infinite cigar. The sigma model path integral is over maps 
$${\sf D} \;\longrightarrow \;{\cal X}.$$
The theory in the interior of ${\sf D}$ has a topological A-type twist. The infinite length of the cigar projects to the space of supersymmetric ground states. We place a B-type boundary condition ${\cal L}$ at the circle boundary at infinity (see Chapter 39 of \cite{mirrorbook}, for detail). At the tip of the cigar we insert a chiral-ring operator $\Phi_\alpha\in H^*({\cal X})$. 

The cigar with the trivial insertion at the tip also plays an important role. We will call ${\cal Z}[{\cal L}]$ the scalar central charge, for reasons we will recall, and view ${\cal V}_{\alpha}[{\cal L}]$ as its vector-valued analogue. If $\Phi_{\slzero}=1$ denotes the identity operator, we have 
\beq {\cal Z}[{\cal L}]\equiv{\cal V}_{\slzero}[{\cal L}]. \eeq 
\subsubsection{}
Let the mirror of the sigma model to ${\cal X}$ be the Landau-Ginzburg model with target ${\cal Y}$, top holomorphic form $\Omega$ and superpotential ${\cal W}$. 
Mirror symmetry exchanges the A- and B-type twists and identifies the quantum differential equation of ${\cal X}$ with the Gauss-Manin system of the
mirror. In the mirror description, the same flat section is the partition function of the Landau-Ginsburg model on the cigar ${\sf D}$, with a B-type twist in the interior, and 
a mirror A-type brane, which we denote by the same letter ${\cal L}$, at infinity. 
It is computed by the period integral 
\beq\label{intV}
{\cal V}_\alpha[{\cal L}]=\int_{\cal L}\Phi_\alpha\,\Omega\,e^{-{\cal W}},
\eeq
where $\Phi_\alpha$ is inserted at the tip of the cigar. The scalar central charge, with identity insertion, is given by  
\beq\label{intZ} {\cal Z}[{\cal L}]=\int_{\cal L}\Omega\,e^{-{\cal W}}. \eeq

\subsubsection{}
The quantum differential equation of ${\cal X}$, and equivalently the Gauss-Manin system of ${\cal Y}$, have integral monodromy, assuming ${\cal W}$ is single valued. On the mirror side this is the familiar Picard-Lefschetz monodromy of cycles. For example, if ${\cal U}$ is a Lagrangian vanishing cycle, monodromy around its vanishing locus acts on any other Lagrangian ${\cal L}$ by
\beq\label{PL}
[{\cal L}]\longmapsto [{\cal L}] + \#({\cal U},{\cal L})[{\cal U}],
\eeq
where $\#({\cal U},{\cal L})$ is the intersection number, and $[{\cal L}]$ denotes the homology class of ${\cal L}$. This leads to the monodromy of solutions to the Gauss-Manin system, which is the geometric analogue of the braid-group monodromy of conformal blocks. When ${\cal W}$ is not single valued, the integer intersection numbers become graded, as we will explain in detail in section 9.

Monodromy acts on the K-theory class of the brane ${\cal L}$ at infinity of ${\sf D}$. ${\cal Z}[{\cal L}]$ and its vector-valued cousin ${\cal V}_\alpha[{\cal L}]$ have the same monodromy, since they differ only in the insertions at the origin of ${\sf D}$.

\subsection{Categorification from geometry}
The sigma-model description gives a natural categorical lift of the monodromy of the quantum differential equation and its mirror. The corresponding monodromy matrix elements are categorified by spaces of morphisms in the corresponding brane categories.

\subsubsection{}
B-type branes on ${\cal X}$ are objects of a category, the derived category of coherent sheaves, \beq {\mathscr D}_{\cal X}=D\operatorname{Coh}({\cal X}), \eeq or its equivariant version when ${\cal X}$ has a torus action. A-branes on $({\cal Y},{\cal W})$ are objects of the derived Fukaya-Seidel category \beq {\mathscr D}_{\cal Y}=D\operatorname{Fuk}({\cal Y},{\cal W}). \eeq Objects of ${\mathscr D}_{\cal X}$ are B-branes which are complexes of coherent sheaves. Objects of ${\mathscr D}_{\cal Y}$ include A-branes which are graded Lagrangians and their complexes. (See \cite{mirrorbook, Douglas} and especially \cite{Horiearly} for a review.)

\subsubsection{}
Homological mirror symmetry states that the two categories associated to the mirror pair ${\cal X}$ and $({\cal Y},{\cal W})$ are equivalent: \beq {\mathscr D}_{\cal X}\simeq{\mathscr D}_{\cal Y}. \eeq 
One consequence of mirror symmetry is a map of a flat section of quantum differential equation of ${\cal X}$ to that of the Gauss-Manin system of $({\cal Y},{\cal W})$.

\subsubsection{}\label{ss-susyg}
In addition to objects, a category has morphisms. Morphisms in either category are spaces of supersymmetric ground states of the theory on a strip, with a pair of branes as boundary conditions. In the category of A-branes, morphism spaces are constructed as follows (see \cite{A2} for more detail). Take the strip $[-1,1]\times{\mathbb R}$. The perturbative supersymmetric ground states with A-branes represented by graded Lagrangians ${\cal L}_{-1},{\cal L}_1\subset{\cal Y}$ as boundary conditions are generated by intersection points \beq {\cal P}\in{\cal L}_{-1}\cap{\cal L}_1. \eeq They form the vector space \beq \operatorname{hom}^*({\cal L}_{-1},{\cal L}_1)=\bigoplus_{{\cal P}\in{\cal L}_{-1}\cap{\cal L}_1}\mathbb C{\cal P}. \eeq The space is graded by the fermion number, or Maslov degree. Intersection points can appear or disappear in pairs, so the space of perturbative ground states is not an invariant. Instantons generate the Floer differential \beq \partial:\operatorname{hom}^*({\cal L}_{-1},{\cal L}_1)\longrightarrow\operatorname{hom}^{*+1}({\cal L}_{-1},{\cal L}_1), \eeq obtained by counting holomorphic strips with boundary on ${\cal L}_{-1}$ and ${\cal L}_1$ that interpolate between intersection points. In the situations relevant here, $\partial^2=0$. The space of exact supersymmetric ground states is the Floer cohomology \beq \operatorname{Hom}^*_{{\mathscr D}_{\cal Y}}({\cal L}_{-1},{\cal L}_1)=\ker\partial/\operatorname{im}\partial. \eeq The Floer cohomology is invariant under Hamiltonian isotopy of the branes. Its Euler characteristic is the signed intersection number \beq \#({\cal L}_{-1}\cap{\cal L}_1)=\sum_{{\cal P}\in{\cal L}_{-1}\cap{\cal L}_1}(-1)^{M({\cal P})}. \eeq Thus, Floer cohomology provides the geometric categorification of Lagrangian intersection theory.
\subsubsection{}
The space of morphisms ${Hom}^*_{{\mathscr D}_{\cal X}}({\cal L}_{-1},{\cal L}_1)$ on the B-side is defined by first finding a resolution of the brane ${\cal L}_{-1}$ in terms of a set of projective generators ${\cal T}_i$ of ${{\mathscr D}_{\cal X}}$. The resolution of ${\cal L}_{-1}$ is a complex of projectives. The morphism space ${Hom}^*_{{\mathscr D}_{\cal X}}({\cal L}_{-1},{\cal L}_1)$ is a cohomology of a complex whose terms involve ${Hom}^*_{{\mathscr D}_{\cal X}}({\cal T}_i,{\cal L}_1)$. 

Thus, both on the A-side and on the B-side, morphism spaces are cohomology groups, as is manifest in their supersymmetric quantum mechanics interpretation.
\subsubsection{}
Let ${\mathfrak B}$ denote the monodromy associated to a loop in the moduli space of the theory. On the A-side, symplectic parallel transport around the loop gives a symplectomorphism of ${\cal Y}$. It takes a Lagrangian brane ${\cal L}$ to a brane ${\mathscr B}{\cal L}$, and defines an autoequivalence ${\mathscr B}$ of ${\mathscr D}_{\cal Y}$. On $K$-theory, the action of ${\mathscr B}$ is the Picard-Lefschetz monodromy ${\mathfrak B}$ of the Gauss-Manin system. 

Mirror symmetry identifies this action on ${\mathscr D}_{\cal Y}$ with an autoequivalence of ${\mathscr D}_{\cal X}$. On the B-side, the action is harder to see geometrically, because it is associated to monodromy in complexified Kahler moduli, that can change the underlying manifold.

\subsubsection{}
Consider the path integral on an annulus $[-1,1]\times S^1$, with ${\cal L}_{-1}$ and ${\cal L}_1$ as boundary conditions on its two boundary components, and with the moduli varying with the coordinate on $[-1,1]$ along the path $B$ in the moduli space. The annulus partition function computes the matrix element of the monodromy ${\mathfrak B}$ of the quantum differential equation, or the mirror Gauss Manin system, between ${\cal V}_{-1}={\cal V}[{\cal L}_{-1}]$ and ${\cal V}_1={\cal V}[{\cal L}_{1}]$.

The theory on the annulus preserves two supercharges, of A- or B-type, picked out by the branes, under arbitrary variations of any of its moduli along the interval \cite{GMW2}. 

The annulus partition function can be shown to depend only on the homotopy class of the path in moduli space \cite{GMW2, A1}. This lets us move all the variation of the parameters to one or the other boundary -- thus keeping the moduli of the theory constant along the interval, and choosing the pair of branes to be either $({\mathscr B}{\cal L}_{-1}, {\cal L}_1)$ or $({\cal L}_{-1}, {\mathscr B}^{-1}{\cal L}_1)$. 
\subsubsection{}
Now view the same annulus in the open string channel, with time running around the circle. Cutting it open gives a strip with boundary conditions ${\cal L}_{-1}$ and ${\mathscr B}^{-1}{\cal L}_1$. Its space of exact supersymmetric ground states is \beq \operatorname{Hom}^*_{\mathscr D}({\cal L}_{-1},{\mathscr B}^{-1}{\cal L}_1), \eeq where ${\mathscr D}$ is either ${\mathscr D}_{\cal X}$ or ${\mathscr D}_{\cal Y}$, depending on the choice of description. Since ${\mathscr B}$ is an autoequivalence, \beq \operatorname{Hom}^*_{\mathscr D}({\cal L}_{-1},{\mathscr B}^{-1}{\cal L}_1)\simeq\operatorname{Hom}^*_{\mathscr D}({\mathscr B}{\cal L}_{-1},{\cal L}_1). \eeq 

\subsubsection{}
The annulus partition function in the open string channel is the Euler characteristic of this morphism space. Therefore, \beq \langle{\mathfrak B}{\cal V}_{-1}|{\cal V}_1\rangle=\chi({\mathscr B}{\cal L}_{-1},{\cal L}_1). \eeq The morphism space $\operatorname{Hom}^*_{\mathscr D}({\mathscr B}{\cal L}_{-1},{\cal L}_1)$ is therefore the categorical lift of the corresponding monodromy matrix element. 

\subsubsection{}

To summarise: whenever the KZ equation is realized as the quantum differential equation of ${\cal X}$, or equivalently as the Gauss-Manin system of $({\cal Y},{\cal W})$, the $U_q(\mathfrak g)$ braid matrices have natural categorical lifts. 

By construction, the resulting vector space will be a braid invariant. To obtain link invariants, one must identify the special ${\cal U}$ branes corresponding to cups and caps, leading to states $|{\mathfrak U}\rangle$ in \eqref{cups}. Then, one must show that the resulting morphism spaces are independent of the chosen plat presentation. 

In the next two sections, we will learn how, starting with IIB string theory, one discovers the mirror pair of manifolds which lead to the KZ equation. We will then use the same setting to identify the branes corresponding to cups and caps. Finally, we will be able to understand, a priori, why the theory leads to invariants of links.

\section{Little string theory and the qKZ equation}

To derive the geometric realization of the KZ equation from string theory, we begin with a step further back. We will derive from string theory a geometric realization of a difference counterpart of the KZ equation, the quantum Knizhnik-Zamolodchikov (qKZ) equation \cite{FIR}. Replacing the KZ equation by the qKZ equation corresponds to deforming the affine Lie algebra $\fgh$ to the quantum affine algebra $U_h(\fgh)$.

The qKZ equation arises from the ten-dimensional type IIB string theory on an ADE surface ${\sf R}$, in a decoupling limit in which it becomes the six-dimensional $(2,0)$ little string theory of type $\fg$. The D3-branes wrapping two-cycles of ${\sf R}$ become its self-dual strings. In a suitable background, the partition functions of these strings satisfy the qKZ equation. They are the $q$-conformal blocks of the quantum affine algebra $U_h(\fgh)$.

The little string theory has a conformal limit $m_s\to\infty$, in which it becomes the $(2,0)$ conformal field theory. The limit, taken in section~\ref{ss-limit}, sends $U_h(\fgh)$ to $\fgh_k$, degenerates qKZ to KZ, and turns the theory on self-dual strings into two-dimensional sigma models with targets $({\cal X},{\rm T})$ and $({\cal Y},{\cal W})$ that featured in \cite{A1} and in \cite{A2}. Thus, the present section derives the finite $m_s$ precursor of the geometric KZ theory from its ten-dimensional IIB origin.

\subsection{The little string theory}\label{LST}

We begin by recalling the type IIB origin of the $(2,0)$ little string theory of ADE type $\fg$, and of its self-dual strings.

Start with IIB string on a surface ${\sf R}$ which is a resolution of the ADE orbifold ${\mathbb C}^2/\Gamma$ singularity. Here, $\Gamma$ is a discrete subgroup of $SU(2)$ holonomy. Such $\Gamma$'s are in one to one correspondence with simply laced Lie algebras $\fg$. 

The geometry of the ADE surface ${\sf R}$ has a deep relation to the representation theory of the Lie algebra ${\fg}$. This relation is the origin of the geometric description of quantum groups we are about to discover.
\subsubsection{}\label{locg}

The ADE surface ${\sf R}$ encodes the root and weight lattices of $\fg$. The homology classes of vanishing two-cycles $[{\sf C}_a]\in H_2({\sf R},\mathbb Z)$ are in one-to-one correspondence with the simple positive roots $e_a$ of $\fg$:
\beq
[{\sf{C}}_a]=e_a.
\eeq
Their intersection form
\beq
\#([{\sf{C}}_a],[{\sf{C}}_b])=-\langle e_a,e_b\rangle
\eeq
is the negative of the Cartan matrix of $\fg$. The same data is equivalently encoded in the Dynkin diagram of $\fg$. It follows that the root lattice of $\fg$ is identified with $H_2({\sf R},\mathbb Z)$.
\subsubsection{}

The weight lattice of ${\fg}$ can be identified with $H_2({\sf R}, \partial {\sf R}, {\mathbb Z})$, which contains classes of two-cycles which may have a boundary at $\partial{\sf R}$. The group contains the $H_2({\sf R}, {\mathbb Z})$ as a sub-lattice, by restricting to compact 2-cycles. 
$H_2({\sf R}, \partial {\sf R}, {\mathbb Z})$ has a basis spanned by two-cycles ${\sf C}_a^{\vee}$ which satisfy
\beq\label{hsdual}
\#([{\sf C}_a^{\vee} ], [{\sf C}_b]) = \langle w_a, e_b\rangle =\delta_{ab},
\eeq
and which we identify with fundamental weights $w_a$ of ${\fg},$ 
\beq\label{weighti}
[{\sf C}_a^{\vee}] = w_a.
\eeq
The non-compact cycle ${\sf C}_a^{\vee}$ can be modeled as follows. Near a vanishing cycle ${\sf C}_a$, the local geometry of ${\sf R}$ is $T^*{\sf C}_a$. One gets ${\sf C}_a^{\vee}$ by picking a generic point on the ${\sf C}_a$, and taking the fiber over it.

The equation \eqref{hsdual} expresses the fact that for the surface ${\sf R}$, $H_2({\sf R}, \partial {\sf R}, {\mathbb Z})$ and $H_2({\sf R}, {\mathbb Z})$ are dual lattices. This is a geometric counterpart of the fact that, for a simply laced Lie algebra ${\fg}$, the root lattice is dual to the weight lattice (for a general Lie algebra ${\fg}$, the root lattice is dual to the co-weight lattice).

\subsubsection{}\label{srecall}

The $(2,0)$ little string theory is obtained from the ten dimensional IIB string on ${\sf R}$ by taking a decoupling limit. The limit sends the IIB string coupling $g_B$ to zero,
$$
g_B\rightarrow 0
$$
while keeping the string scale $m_s$ fixed, and the modes of supported at the singularity dynamical \cite{Witten:1995zh, Seiberg:1997zk, Losev:1997hx}.

In the limit, we keep fixed expectation values of scalar fields of the $(2,0)$ theory, normalized as
\beq\label{mod}
\int_{{\sf C}_a} {m_s^4\, \omega_{I,J,K}/g_B}, \qquad \int_{{\sf C}_a} m_s^2 \, B_{NS}/g_B, \qquad \int_{{\sf C}_a} m_s^2 \, B_{RR}.
\eeq
The bosonic fields in the $(2,0)$ tensor multiplet are scalar fields in \eqref{mod} and the self-dual two-form 
\beq\label{sdt}
\int_{{\sf C}_a} C^{(4)},
\eeq
where $ C^{(4)}$ is IIB RR four-form potential. A power of $g_B$ accompanies NS sector fields but not the RR sector ones since this is how they enter the low energy effective action of IIB string. Powers of $m_s$ are introduced to give the scalar fields of the $(2,0)$ tensor multiplet the natural mass dimension 2.
\subsubsection{}
The $(2,0)$ LST has four ${\mathbb R}$-valued scalar fields -- the first four in \eqref{mod} --  and one compact scalar. 
The two IIB string $B$-fields have periods $\int_{{\sf C}_a} B$, which have periodicity $2\pi$. In the little string decoupling limit, the scalar field coming from $B_{NS}$, normalized as in \eqref{mod} decompactifies, but not the scalar coming from $B_{RR}$. 
As a result, the 6d LST theory has an $SO(4)$ R-symmetry rotating the first four scalars in \eqref{mod}.

\subsubsection{}\label{strings}
The $(2,0)$ LST contains two kinds of strings. The strings that give it its name are inherited from the fundamental IIB string and have tension $m_s^2$.

It also contains self-dual strings, named so as they couple to the self-dual two-form in \eqref{sdt}. They originate from D3 branes of IIB string, wrapping compact 2-cycles in ${\sf R}$. Their tension is set by the expectation values of scalar fields in \eqref{mod}. 
The charges of self-dual strings are labeled by elements of $H_2({\sf R}, {\mathbb Z})$, the root lattice of ${\fg}$. At the origin of the moduli space, where all the expectation values of the scalar fields in \eqref{mod} vanish, the self-dual strings become tensionless, and the $(2,0)$ LST has an unbroken ${\fg}$ symmetry. We will always be in the phase where ${\fg}$ is broken to its Cartan, and the self-dual strings have finite tension.

The theory also has self-dual string defects, coming from D3 branes wrapping non-compact 2-cycles. Their charges lie in $H_2({\sf R},\partial {\sf R}, {\mathbb Z})$, the weight lattice of ${\fg}$. We will sometimes call these the heavy self-dual strings.

\subsection{The self-dual string theory in $(2,0)$ LST}

Given the origin of the self-dual strings of the $(2,0)$ LST as D3 branes of IIB string wrapping 2-cycles of $\sf{R}$, their two-dimensional world-sheet theory can be derived from first principles, using the work of Douglas and Moore \cite{DM} from 1996. The only additional input needed, besides \cite{DM}, is to take the $g_B\to0$ limit. We recall the answer below.

In the next section, we will use this as the starting point to derive the theory on self-dual strings of the $(2,0)$ CFT. This result is new.

\subsubsection{}\label{sscharge}

Take a collection of D3 branes wrapping compact 2-cycles $[\sf C]\in H_2({\sf R}, {\mathbb Z})$, whose class is

\beq\label{comp} [{\sf C}] = \sum_{a=1}^{\rm rk}  \,d_a\,e_a, \eeq 
where the non-negative integers $d_a$ count the compact wrapped branes and will determine the ranks of the gauge groups. Next, we add non-compact D3 branes, whose class in $H_2({\sf R}, \partial {\sf R}, {\mathbb Z})$ is \beq\label{ncomp} [{\sf C}^\vee] = \sum_{a=1}^{\rm rk} \, m_a \, w_a , \eeq with integer multiplicities $m_a$. The gauge groups on D3 branes supported on non-compact cycles are non-dynamical. They lead to flavor symmetry.

\subsubsection{}\label{os}

We will choose to work in the background where
$\int_{\sf C_*} {\vec \omega}$ and $\int_{\sf C_*}B_{RR}$ vanish and the $B_{NS}$ periods are real, positive
\beq\label{fbac}
\int_{\sf C_*} B_{NS}>0,
\eeq
and otherwise generic.
This corresponds to the orbifold point, where the analysis of \cite{DM} and \cite{AH} takes place. We will make the same choice here.
\subsubsection{}\label{susy}

The central charge of each wrapped D3 brane is a vector. 
The vector is equal to the period of the triplet of self-dual two forms ${\vec \omega} = (\omega_I, \omega_J, \omega_K)$ and the NSNS and RR B-fields $B_{NS}$ and $B_{RR}$, on 
the 2-cycle in ${\sf R}$ which the D3 branes wrap. 

In the background chosen, D3 branes wrapping different curve components all have their central charges  aligned. The self-dual strings preserve eight supercharges and have finite tension given by
\beq\label{tension}
M_a^2 \equiv \int_{{\sf C}_a} m_s^2 B_{NS}/{g_B }.
\eeq
At the same time, the bulk ${\fg}$ symmetry is broken to its Cartan.
\subsubsection{}\label{quiverQ} 
The theory on D3-branes with charges in \eqref{ncomp} and \eqref{comp}, in the background \eqref{fbac}, is a quiver gauge theory. Its quiver diagram ${\scQ}$ is based on the Dynkin diagram of ${\fg}$ \cite{DM}. The quiver gauge group $G_\scQ$ and flavor symmetry group $G_W$ are 

\beq\label{3dgauge} G_W = \prod_{a} U(m_a), \qquad G_\scQ=\prod_a U(d_a), 
\eeq 
Observe that, in the present setting, the quiver diagram is necessarily that of the Lie algebra ${\fg}$ and not its affine extension $\fgh$ as in \cite{DM}. This is due to the decoupling limit which defines the little string theory. The D-brane associated to the affine node is a D1 brane, whose tension $m_s^2/g_B$ goes to infinity as we send $g_B$ to zero.


\subsubsection{}\label{Coulomb3d} 
The charges above will determine the representation-theoretic data of the $q$-conformal blocks. We make this identification explicit before using the worldsheet theory to construct them.

Let $V_a$ be the fundamental representation of ${\fg}$ whose highest weight is the fundamental weight $w_a$, labeled by the $a$-th node of the Dynkin diagram. The representations assigned to the marked points determine a tensor product \beq\label{Vtensor} V=\bigotimes_i V_i=\bigotimes_a V_a^{\otimes m_a}. \eeq The integers $m_a$ are the multiplicities with which the fundamental representations $V_a$ occur.

The ranks $d_a$ determine the weight subspace of representation $V$ which $q$-conformal blocks live in. If the chosen weight is \beq\label{weighta} \mathrm{weight}=\sum_i w_i-\sum_a d_a e_a, \eeq where $e_a$ are the simple positive roots of ${\fg}$, then $d_a$ is the rank of the gauge group attached to the $a$-th node.

\subsubsection{}

The tension determines, from the perspective of the low energy theory on the wrapped D3 branes, the gauge couplings of the 2d quiver gauge theory. They are given by 
\beq\label{gaugec} 1/g_{2d,a}^2 = \int_{{\sf C}_a} B_{NS}/(g_B m_s^2 ) =M_a^2/m_s^4. 
\eeq At fixed finite $m_s$, the theory can be weakly coupled for sufficiently large nonzero tensions $M_a^2$, by \eqref{gaugec}.

The triplet of moduli $\int_{{\sf C}_a} {m_s^2\, \omega_{I,J,K}/g_B}$ from \eqref{mod}
correspond to the 2d Fayet-Iliopoulos parameters.  Throughout, we will turn the complex FI parameters off, 
\beq\label{csc}
 \int_{{\sf C}_*} \,\omega_{J,K}=0.
\eeq
The real FI parameters combine with the theta angles $\int_{\sf C_a} B_{RR}$ to $\rm{rk}\,\fg$ complex parameters 
\beq\label{FI}
\lambda_{a, 2d} = \int_{{\sf C}_a} (m_s^2 \omega_I/g_B+i B_{RR}),
\eeq
which we view as a perturbation about the fixed background where they vanish.

\subsection{Quantum Knizhnik-Zamolodchikov equation}

The Knizhnik-Zamolodchikov equation from section \ref{ss:KZ}
has a one parameter deformation, which is a difference equation called the quantum Knizhnik-Zamolodchikov equation \cite{FIR}. It is based on replacing the affine Lie algebra $\widehat{\fg}_k$
by the quantum affine algebra $U_{h}(\fgh).$ It is given by

\beq\label{qKZ}
\begin{aligned}
{\cal V}^K(a_1, \ldots p a_i, \ldots a_n) =&{\sf R}_{i i-1}(p a_{i}/a_{i-1}) \cdots {\sf R}_{i 1} (p a_{i}/a_{1})   (h^{\rho})_i\\
\times &{\sf R}_{i n}(a_{i}/a_n) \ldots {\sf R}_{i i+1}(a_{i}/a_{i+1}) {\cal V}^K(a_1, \ldots a_i, \ldots a_n) .
\end{aligned}
\eeq
The step $p$ of the difference equation is related to $h$, the quantization parameter in $U_h(\fgh)$, by
\beq\label{hp}
p = h^{-(k + h^{\vee})}.
\eeq
In the limit where $p$ and $h$ go to $1$ keeping $k$ in \eqref{hp} fixed, the qKZ equation in \eqref{qKZ} reduces to the KZ equation in \eqref{KZ}.

\subsubsection{}
The right hand side of \eqref{qKZ} is given in terms of ${\sf R}$-matrices of $U_{h}(\fgh)$. The $\sf R$-matrix 
$$
{\sf R}_{i j}(y/y') \in  {\rm Hom}_{U_{h}(\fgh)}({V}_i(y) \otimes {V}_j(y')).
$$
acts by exchanging a pair $V_i(y), V_j(y')$ of finite dimensional $U_{h}(\fgh)$ modules. The spectral parameters $y$ and $y'$ live on a copy of ${\mathbb C}^{\times}$, which should be identified with our $\Sigma.$ 
\subsubsection{}\label{confbr}
 Taking $h\neq 1$ breaks conformal invariance of an affine Lie algebra as $U_{h}(\fgh)$ does not contain the Virasoro algebra as a subalgebra. Before breaking conformal invariance, regarding $\Sigma$ as punctured complex plane and an infinite complex cylinder is equivalent. After breaking conformal invariance, this no longer is the case. To get the quantum affine deformation needed here, we take $\Sigma$ to be an infinite complex cylinder. Starting with the punctured complex plane, one gets the Yangian deformation instead \cite{FIR}. 
\subsubsection{}

 The deformation is remarkable because, despite breaking conformal invariance, one preserves structures provided by conformal field theory. 
In particular, the solutions of the qKZ equation may be obtained by vertex operators and sewing, and have the same form as in the conformal case,
\eqref{confb},
\beq\label{qconfb}
{\mathcal V}^K  = \langle \lambda'| \;{ \Phi}_{V_{w_1}} (a_{1}) \cdots { \Phi}_{V_{w_n}} (a_{n})    \;|\lambda\rangle,
\eeq
except that everything is $h$-deformed.

\subsection{The qKZ equation from self-dual strings}\label{bckg}

The $U_{h}(\fgh)$ qKZ equation and its solutions arise from the $(2,0)$ little string theory with a collection of D3 branes. The fact this is the case is a key result of \cite{AFO}. To this end, we will study the $(2,0)$ LST on the six manifold 
\beq\label{6mani0}
M_6= \Sigma  \times {\sf S}\times {\sf D},
\eeq
where ${\sf S}={\mathbb C}$, 
\beq\label{sigmaag}
\Sigma = {\mathbb R}\times S^1(R) 
\eeq
where $S^1(R)$ is a circle of radius $R$, and for the time being ${\sf D} = {\mathbb C}$. The little string theory on $M_6$ is a shortcut, as we reviewed above, for starting with IIB string on ${\sf R}\times M_6$ and taking the LST decoupling limit. In addition, we will turn on an $\Omega_B$ background on $M_6$, which we will define in \ref{ls11}.

\subsubsection{}
Take all the D3 branes to be at points on 
\beq\label{4mani0}
M_4 =\Sigma\times {\sf S},
\eeq
in \eqref{6mani0} and supported on ${\sf D}$. We will require the non-compact D3 branes to be at the origin of ${\sf S}={\mathbb C}$, and at the collection points on 
$\Sigma$ where the vertex operators in \eqref{qconfb} are. We also match the class of the curve with the highest weight of the representation coloring the vertex operator. Explicitly, to introduce the vertex operator $\Phi_{V_i}(a_i)$ in \eqref{qconfb}, place a D3 brane wrapping a non-compact 2-cycle in ${\sf R}$ whose homology class is the highest weight $w_i$ of representation $V_i$ at the point on $\Sigma$ with
with coordinates $y=a_i$.
The compact D3 branes, of total charge in \eqref{comp}, can apriori be anywhere on $M_4$.

\subsubsection{}\label{tdual}

Since the bulk theory is a string theory, and the $S^1(R)$ in $\Sigma$ is a compact circle transverse to D3 branes, the theory on the D-branes wrapping 2-cycles in ${\sf R}$ and supported on ${\sf D}$ is actually a three dimensional quiver gauge theory on ${\sf D}$ times a circle. The massive tower of strings winding around $S^1(R)$ circle is equivalent to the tower of momenta on the T-dual circle $S^1(R')$ of radius $R'$ related to $R$ by
\beq\label{dualR}
R \, m_s^2 = 1/R'.
\eeq
We can make this manifest by T-duality which trades IIB on $S^1(R)$ with D3 branes at points on the circle to IIA on $S^1(R')$ with D4 branes wrapping the circle. At low energies, the theory on these D4-branes is a three dimensional gauge theory with quiver ${\scQ}$ on $S^1(R')\times {\sf D}.$

The T-dual IIA is compactified on ${\sf R}$ times a six manifold $M_6'$ obtained from $M_6$ in \eqref{6mani0} by replacing $S^1(R)$ with $S^1(R')$. 
The decoupling limit from section \ref{srecall}, maps to the limit in which IIA on ${\sf R}\times M_6'$ becomes the $(1,1)$ little string theory on $M_6'$. The $(1,1)$ LST reduces to the six dimensional maximally supersymmetric YM theory at low energies. 
\subsubsection{}\label{ls11}
We will put the $(1,1)$ LST theory on $M_6'$ in the standard $\Omega_A$-background, defined as follows. Let the coordinates on ${\sf D}$ and ${\sf S}$ be $x$ and $w$, respectively. Accompany the rotation around $S^1(R')$, by a rotation\footnote{A-priori, $p$ and $t$ are phases, though the theory depends on them holomorphically.} of the two complex planes,
\beq\label{4dOmega}
(x, w) \rightarrow ({p} x, t^{-1} w).
\eeq
To preserve supersymmetry, we accompany this with the scaling of the holomorphic symplectic form on ${\sf R}$
\beq\label{ex}
\omega^{2,0}_{\sf R} \rightarrow h^{-1} \,\omega^{2,0}_{\sf R},
\eeq
where $p$, $h$ and $t$ are related by
\beq\label{pth}
h=p/t.
\eeq
The $\Omega_B$ background for the $(2,0)$ LST on $M_6$ is defined to be the $S^1(R')$ T-dual of $\Omega_A$.

\subsubsection{}

The partition function of $(2,0)$ LST on $M_6$ in $\Omega_B$ background computes the trace \cite{NO}
\beq\label{3dindex}
{\rm Tr} (-1)^{2S}\; p^{S-S_H} t^{S_H-S_V}.
\eeq
More directly, the \eqref{3dindex} is the partition function of the T-dual $(1,1)$ LST on $M_6'$ in $\Omega_A$ background.
Here, $S$ and $S_V$ generate the rotations of the ${\sf D}$ and the ${\sf S}$ planes in $M_6'$, respectively; $S_H$ generates the symmetry that rotates the phase of $\omega^{(2,0)}_{\sf R}$. The insertions of $p^\#$ and $t^\#$ are the effect of $\Omega_B$ background. 
\subsubsection{}
In the absence of D-branes, the index in \eqref{3dindex} is trivial. The gauge group of the 6d $(1,1)$ gauge theory is abelianized by the background \eqref{fbac} in which the scalar fields coming from $B_{NS}$ are turned on. To compute the partition function of the resulting abelian 6d gauge theory, first replace the (T-dual of $\Sigma$) with a torus of complex structure $\tau$. The resulting partition function is computed in section 6.2 of \cite{NO} (with mass of the adjoint hypermultiplet set to $0$). Then, send $\tau$ to $i\infty$ to decompactify the torus to $\mathbb{C}^\times$. In the limit, all the instanton corrections vanish, and the partition function is simply $1$, up to a constant prefactor depending on $p$ and $h$ which we will suppress.
\subsubsection{}

Since the index \eqref{3dindex} of the $(2,0)$ LST on $M_6$ is trivial in the absence of D3 branes, it follows that with D3 branes included, it coincides with the partition function of the theory on the branes. 

The index of $(2,0)$ LST on $M_6$ with D3 branes on ${\sf D}$ in $\Omega_B$ background thus equals the index of 3d ${\cal N}=4$ gauge theory on $S^1(R')\times {\sf D}$ from section \ref{quiverQ}. This is the gauge theory living on D4 branes in $(1,1)$ LST on $M_6'$, supported on ${\sf D} \times S^1(R')$. From perspective of the 3d gauge theory on D4 branes, $S_V$ and $S_H$ in \eqref{3dindex} generate the $U(1)_H\times U(1)_V$ subgroup of the $SU(2)_H\times SU(2)_V$ R-symmetry of the 3d theory, labeled by whether they act on hyper (H) or vector (V) multiplet moduli.
\subsubsection{}
The holonomies of background gauge fields complexify the real masses and the Fayet-Iliopoulos parameters.
 In the IIA frame, $\lambda_{a,2d}$ from \eqref{FI} is the real FI parameter of the effective 2d quiver gauge theory on ${\sf D}$ complexified with the theta angle:
\beq\label{lambda}
\lambda_{a, 2d} = \int_{\sf C_a\times S^1(R')} m_s^3  \omega_I/g_A+i \,C_{RR}^{(3)}.
\eeq
The 2d theory is obtained by compactifying the 3d gauge theory on $S^1(R')$, so ${\rm{Re}}\lambda_{a, 2d} = R' \lambda_{a, 3d}$, where $\lambda_{a, 3d}$ is the 3d FI parameter.
The imaginary part of $\lambda_{a, 2d}$ is the holonomy of the 6d gauge field around $S^1(R')$, which in turn is the period of the RR 3-form potential, $\int_{\sf C_a\times S^1(R')} \,C_{RR}^{(3)}$. 

Similarly, the positions of non-compact D3-branes on $\Sigma$ are the complexified real masses. In the dual D4 brane frame, the real mass is complexified by the holonomies of the $G_W$ gauge fields on $S^1(R')$.

\subsubsection{}

The index in \eqref{3dindex}, viewed as the partition function of the 3d ${\cal N}=4$ gauge theory,  has a precise mathematical formulation. As explained in \cite{Nekrasov, NO2}, for any $\fg$, the gauge theory index is the vertex function of quantum K-theory of the Higgs branch ${\cal X}^{\vee}$. The Higgs branch ${\cal X}^{\vee}$ is the Nakajima quiver variety corresponding to the quiver $\scQ$:
\beq\label{NV}
{\cal X}^{\vee}=T^*{{\rm Rep }\,\scQ} \rd G_{\scQ} 
\eeq
where 
\beq\label{repQ}
{{\rm Rep }\,\scQ}=
\bigoplus_{a\to b}{\rm Hom}(\mathbb C^{d_a},\mathbb C^{d_b})
\;\oplus\;
\bigoplus_a{\rm Hom}(\mathbb C^{d_a},\mathbb C^{m_a}).
\eeq
The quantum K-theory of Nakajima quiver varieties such as ${\cal X}^{\vee}$ was formulated in \cite{OK}. How this comes about is well known, see \cite{AFO} for review.
\subsubsection{}
 Briefly, for large real FI parameters, the supersymmetric path integral of the three dimensional theory on ${\mathbb R}\times {\sf D}$ localizes to the supersymmetric quantum mechanics on the moduli space of solutions to vortex equations on ${\sf D}$ \cite{Nekrasov, NO2}. 
Solutions to vortex equations on ${\sf D}$ are quasi-maps
$${\sf D} \mapsto {\cal X}^{\vee}.$$
The quasimap degree is the vortex charge.
The index of supersymmetric quantum mechanics on the vortex moduli space is the count of quasi-maps of a given degree as defined by quantum K-theory of ${\cal X}^{\vee}$ \cite{OK}. 

The insertions of $p^{S-S_V} h^{S_V-S_H}$ in the supertrace correspond to working equivariantly with respect to rotations of ${\sf D}$ by $p$, and global symmetries ${\bf T}^{\vee}$ of ${\cal X}^{\vee}$. The equivariant torus is 
\beq\label{dT}
{\bf T}^{\vee}= A \times {\mathbb C}_h^{\times},
\eeq
where $A$ comes from flavor symmetries of the 3d gauge theory, associated with fundamental hypermultiplets, and ${\mathbb C}_h^{\times}$ is the $U(1)_H$ $R$-symmetry. The latter scales the holomorphic symplectic form of ${\cal X}^{\vee}$, and is inherited from scaling of the holomorphic symplectic form of ${\sf R}$. 

To compute \eqref{3dindex} one sums over all possible degrees. The
 degree $k$ of the quasimap is kept track of by $z^k$, where 
$z = \exp(R' \lambda_{3d})$ and $\lambda_{3d}$ is the real FI parameter complexified by the theta angle.

\subsubsection{}

The K-theoretic vertex function of ${\cal X}^{\vee}$ is the K-theory analogue of Givental's $I$-function \cite{GivM} which we denoted by ${\cal Z}$ in \eqref{intZ}. Its vector-valued analogue, generalizing Givental's $J$-function \cite{G} is denoted by ${\cal V}_\alpha$ in \eqref{intV}. It is obtained by including a line operator at the origin of ${\sf D}$. A line operator is labeled by a K-theory class
\[
\alpha \in K_{{\bf T}^{\vee}}({\cal X}^{\vee}).
\]
We will denote the resulting K-theoretic vector vertex function by ${\cal V}_{\alpha}^{K,\vee}$.
\subsubsection{}
One of the key results of \cite{OK} is a theorem that 
$${\cal V}_{\alpha}^{K,\vee},$$
solves a difference equation in terms of the equivariant variables $a\in A$ which coincides with the quantum Knizhnik-Zamolodchikov equation in \eqref{qKZ}. It was proven in \cite{bethe} that one gets the qKZ equation in the standard form \eqref{qKZ} if the class inserted at the origin is given in terms of K-theoretic stable envelopes \cite{OK} of ${\cal X}^{\vee}$. These form a very special basis of  
${K}_{{\bf T}^\vee}({\cal X}^{\vee})$, which coincides with the natural basis for representations of $U_{h}(\fgh)$.

\subsubsection{}\label{nsv}

In addition to the line operator inserted at the origin of ${\sf D}$, the $K$-theoretic vertex functions depend on the choice of the supersymmetric vacuum at infinity of ${\sf D}$. 
The space of supersymmetric vacua of the three-dimensional theory on \(S^1(R')\times {\sf D}\) is the equivariant elliptic cohomology
\[
\operatorname{Ell}_{{\bf T}^{\vee}}({\cal X}^\vee),
\]
see \cite{Nikitaell}, for example. The elliptic curve is the boundary torus
\(
E=\partial\bigl(S^1(R')\times {\sf D}\bigr)\simeq \mathbb C^\times/p^{\mathbb Z}.
\)
For generic masses and FI parameters, a basis of $\operatorname{Ell}_{{\bf T}^{\vee}}({\cal X}^\vee)$ is obtained by choosing ``non-singular boundary conditions" at infinity of \(\sf D\), in terminology of \cite{OK}. These are labeled by isolated \({\bf T}^{\vee}\)-fixed points of ${\cal X}^{\vee}$: in the corresponding vacuum, the fields approach a fixed point at infinity of \(\sf D\), and the gauge field strength vanishes there. The reason the $\Omega$-background parameter $p$ gets related to the complex structure of the torus at infinity of ${\sf D}\times S^1$ is explained in \cite{NW}.

\subsubsection{}
All the ingredients of the qKZ equation have a gauge theory and string theory interpretation.
The step $p$ of the qKZ equation 
is the parameter by which the complex ${\sf D}$ plane rotates as we go around the $S^1(R')$. The parameter $h$ in $U_{h}(\fgh)$ is the holonomy around the $S^1$
of the off-diagonal $U(1) \in SU(2)_H\times SU(2)_V$, as seen in \eqref{3dindex}. The insertion $p$ and $h$ are related to the $\epsilon$-parameters of the $\Omega$ background (which have dimensions of mass), by
\beq\label{Omegab}
p= \exp(R'\epsilon_p),\qquad h = \exp(R'\epsilon_h).
\eeq
The complexified real masses $m_i$ of the 3d theory are related to positions $a_i$ of the vertex operators in \eqref{qconfb}, and the positions of D3 branes on $\Sigma$, by
\beq\label{ma}
a_i = \exp(R' m_i).
\eeq
Because the real masses get complexified by the holonomy of a global $U(1)$ gauge symmetry around the $S^1(R')$, that $a_i$'s are valued on $\Sigma \cong {\mathbb C}^{\times}$, where $\Sigma$ is the Riemann surface in $M_6$. 

The weights $\lambda_a$ of the Verma module inserted at $0\in \Sigma$ enter through
\beq\label{zp}
 z_a = \exp(R' \lambda_{a, 3d}).
\eeq
The parameters $\lambda_{a, 3d}$ are the complexified real FI parameters of the three dimensional quiver gauge theory, and $\lambda_{a, 3d}=\lambda_a\epsilon_p$.

\subsubsection{}\label{Kintegral}
 The K-theoretic vertex functions ${\cal V}^{K, \vee}$ are naturally a series in $z^k$ keeping track of vortex charge and quasimap degree. They also have an integral representation, found in \cite{AFO, bethe}, which we will need later. Existence of an integral representation of the Higgs branch index is well known in physics literature, see for example \cite{Nekrasov, Beem}.

The integral representation of  ${\cal V}^{K, \vee}$ comes \cite{AFO} from thinking about the 3d theory on $S^1\times {\sf D}$ as supersymmetric quantum mechanics on $S^1$. Its index can be computed in steps. Start by giving generic expectation values to real scalars in vector multiplets, complexified by the holonomy of the gauge field around the $S^1$. This places the D3 branes at a collection of generic points on $\Sigma={\mathbb C}^\times$. Since branes of the same charge are identical, this picks a point on 
\beq\label{Yfir}Y=\prod_a Sym^{d_a}(\Sigma).
\eeq
Next, integrate out contributions of massive charged matter fields. 
The resulting determinants give the holomorphic integrand in
\beq\label{intvK}
 {\cal V}_{\alpha}^{K, \vee}[L] = \int_{L}  \; \Phi_{\alpha}^K \,\Omega_Y\,\exp(-{{ W}^K}).
\eeq
Here 
\beq\label{tf}
\Omega_Y=\bigwedge_{a,\alpha}dy_{\alpha,a}/y_{\alpha,a}
\eeq
is the top holomorphic form on $Y$.
Finally, one projects to gauge-invariant contributions by integrating over holonomies. 
From a mathematics perspective \cite{AFO}, this computes
the count of quasi-maps from ${\sf D}$ to ${\cal X}^{\vee}=T^*{{\rm Rep }\,\scQ} \rd G_{\scQ}$ in geometric invariant theory terms: One studies quasi-maps to the pre-quotient, 
$T^*{{\rm Rep }\,\scQ}$ and then projects to gauge invariant configurations by integrating over the maximal torus of $G_{\scQ} =\prod_a GL({\mathbb C}^{d_a})$.

\subsubsection{}\label{WK}

The function ${{ W}^K}$ in \eqref{intvK} is derived in \cite{AFO, bethe} to be a sum of three terms,
\beq\label{Ksup}
{{W}^K} = {W}^K_{1}+{W}^K_{2}+{W}^K_{3}
\eeq
all of which can be read off from the quiver $\scQ$. The first term 
$W^K_1$ is classical, whereas ${W}^K_{2,3}$ are the results of a one loop computation, viewing the 3d theory on $S^1\times {\sf D}$ as an effective quantum mechanics on $S^1$. 

The first term is the contribution of Fayet-Iliopoulos terms in the classical Lagrangian, one for each node of the Dynkin diagram of $\fg$:
\beq\label{firsta}
{{W}^K_{1}}=  \sum_{a, \alpha} \lambda_a\, \ln(y_{\alpha, a}).
\eeq
The second term is the contribution of matter fields associated to framing nodes of the quiver:
\beq\label{seconda}
{{W}^K_{2}}=  \sum_{a, \alpha, i}  \langle e_a, w_i\rangle  \; f(y_{\alpha, a}/a_i),
\eeq
where $\langle e_a, w_i\rangle$ is $1$ if $w_i$ is the fundamental weight dual to the simple root $e_a$, and zero otherwise.
The third term is associated to the Dynkin diagram of ${\fg}$ itself
\beq\label{thirda}
 {W}^K_{3}= - \sum_{(a,\alpha)< (b, \beta)} \langle e_a, e_b\rangle \; f(y_{\alpha, a}/y_{\beta,b}),
\eeq 
where $\langle e_a,e_b\rangle$ is the Cartan matrix of ${\fg}$. In the above, 
$$f(y) \equiv \ln {\phi_p(y)\over  \phi_p(t y)},$$
where $\phi_p(y)=\prod_{n\geq 0} (1-p^n y) $ is the quantum dilogarithm.

\subsubsection{}
The functions $\Phi^K$ in \eqref{intvK} reflect insertion of a line operator supported at the origin of ${\sf D}$ and winding around $S^1(R')$. They are elements of $K_{\rm T^{\vee}}({\cal X}^{\vee}),$ the equivariant $K$-theory of ${\cal X}^{\vee}$. 

In \cite{AFO} it was explained that one may trade insertion of class in ${\rm Ell}_{{\rm T}^\vee}({\cal X}^{\vee})$, which records a choice of a boundary condition at $\partial{\sf D}$, and integration over the maximal torus of $G_{\scQ}$ which is canonical, for a integration of over a Lagrangian cycle $L \subset Y$ as we have done in \eqref{intvK}. 

\subsection{qKZ equation and 3d mirror symmetry}
Three-dimensional mirror symmetry exchanges two holomorphic symplectic varieties associated to a three-dimensional \(\mathcal N=4\) quiver gauge theory:
$$
{\cal X} \qquad \textup{and} \qquad {\cal X}^{\vee}.
$$ 
The Higgs branch ${\cal X}^{\vee}$ is the Nakajima quiver variety, and ${\cal X}$ is the Coulomb branch. Coulomb branches for quiver varieties were constructed explicitly in \cite{Nak13a, Nak13b}, see also \cite{GC, Spencer}.

Mirror symmetry exchanges ${\cal X}$ and ${\cal X}^{\vee}$ while also exchanging the mass and FI parameters:
\beq\label{LA}
\Lambda = A^{\vee}, \qquad A = \Lambda^{\vee}.
\eeq
$A$ is the flavor torus acting on the framing hypermultiplets of the quiver gauge theory and $\Lambda$ the torus of complexified FI parameters. Thus $A$ is the equivariant torus for ${\cal X}^{\vee}$ and the Kahler torus for ${\cal X}$, while $\Lambda$ is the Kahler torus for ${\cal X}^{\vee}$ and the equivariant torus for ${\cal X}$. Three-dimensional mirror symmetry also exchanges the two $R$-symmetry fugacities $h$ and $p/h$.

\subsubsection{}
The \(K\)-theoretic vertex functions of \({\cal X}^{\vee}\) give one basis of solutions of the \(q\)KZ equation, characterized by having a power series expansion in the \(z\)-variables, parameterizing the $\Lambda$ torus. 

There is another natural basis, given by \(q\)-conformal blocks of $U_h(\fgh)$, from \eqref{qconfb}, obtained by sewing vertex operators. The latter are characterized by analyticity in the \(a\)-variables, parameterizing the $A$-torus. If they have a geometric interpretation, it must be given as K-theoretic vertex functions of ${\cal X}$, the Coulomb branch.

Descriptions in terms of both ${\cal X}$ and ${\cal X}^{\vee}$
exist as the supersymmetric index of the quiver gauge theory in \eqref{3dindex} can be computed in either of its two infrared descriptions: as the sigma model to the Higgs branch \({\cal X}^{\vee}\), or as the sigma model to the Coulomb branch \({\cal X}\). Many examples are worked out in \cite{KapW2, KapW1} for theories on compact three manifolds.

\subsubsection{}

In the present problem the three-dimensional theory is placed on the noncompact space
$
S^1(R')\times {\sf D} .
$
The index therefore depends on the boundary condition imposed at the boundary torus
\(
E=\partial\bigl(S^1(R')\times {\sf D}\bigr).
\)
Boundary conditions chosen in the \({\cal X}^{\vee}\) description need not match those in the \({\cal X}\) description. Thus the corresponding solutions of the \(q\)KZ equation need not be the same. They are related by changing the boundary condition at infinity of \({\sf D}\).

Let ${\cal V}^{K,\vee}_i$ be basis of $z$-solutions to qKZ equation which come from vertex functions of ${\cal X}^{\vee}$, and ${\cal V}^{K}_i$ the $a$-solutions, which should come from ${\cal X}$. Any two fundamental solutions to the same $p$-difference equation are related by a linear change of basis. There exists a matrix 
${\mathfrak S}_i^{j}$ of constants with respect to $p$-shifts of both the $a$- and the $z$-variables which relates a fundamental $z$-solution to qKZ equation to the $a$-solution: 
\beq\label{stabrel1}
{\cal V}^K_i := \sum_j \;{\mathfrak S}_i^{j}  \;{\cal V}_j^{K,\vee} , 
\eeq
where the summation is over the basis vectors in the fundamental solution. 
The $a$-solutions on the left hand side are analytic in a chamber corresponding to a specific ordering of vertex operators along the cylinder ${\Sigma}$. They are the q-conformal blocks of $U_{h}(\fgh)$. The matrix ${\mathfrak S}$ is called the pole subtraction matrix in \cite{ese}. It provides a linear combination of $z$-solutions, so as to subtract their poles in the $a$-variables. 
\subsubsection{}
The change of basis in \eqref{stabrel1} corresponds to the change of boundary conditions at $\partial {\sf D}$. The matrix ${\mathfrak S}_i^{j}$ has a geometric interpretation.  By a theorem of  \cite{ese}, it provides the elliptic stable basis of ${\cal X}^{\vee}$, 
$
{\mathfrak S} \propto Stab^{ell}({\cal X}^{\vee}).
$
The elliptic stable basis is a basis of 
${\rm Ell}_{{\rm T}^{\vee}}({\cal X}^{\vee})$, the equivariant elliptic cohomology of ${\cal X}^{\vee}$,
consisting of classes with support on attracting manifolds to ${\rm T}^{\vee}$-fixed points of ${\cal X}^{\vee}$. It gives a map 
$
Stab^{ell}({\cal X}^{\vee}) : {\rm Ell}_{\rm T}({\cal X}^{\vee}_{{\rm T}^{\vee}}) \rightarrow {\rm Ell}_{{\rm T}^{\vee}}({\cal X}^{\vee}).
$
The theorem thus says that q-conformal blocks of $U_{h}(\fgh)$ are the K-theoretic vertex functions of ${\cal X}^{\vee}$, expressed in the elliptic stable basis for ${\rm Ell}_{{\rm T}^{\vee}}({\cal X}^{\vee})$.  

Physically, this change of boundary condition can be implemented by gluing an interval \(I\) to the boundary. This produces an interface theory on
\(
E\times I .
\)
In the language of \(q\)KZ solutions, the interface gives the change-of-basis matrix between the vertex-function basis and the \(q\)-conformal-block basis. This is the pole-subtraction matrix from \cite{ese} described below.

\subsubsection{}

It is natural to identify the $a$-solutions of the qKZ equation, produced by the change of basis in \eqref{stabrel1}, with K-theoretic vertex functions of ${\cal X}$. Then, we get a second geometric interpretation of the qKZ equation as the quantum difference equation of ${\cal X},$
where the latter is the K-theory generalization of the the quantum differential equation.

The identification of $a$-solutions of the qKZ equation with K-theoretic vertex functions of ${\cal X}$ is physically natural, but not mathematically rigorous. The issue is that, in K-theory, unlike in cohomology, the available definition of the vertex function relies on a compactification of moduli of maps from ${\sf D}$ to the target, which 
comes from thinking about holomorphic maps as quasi-maps. In physics terms, quasi-maps are vortex solutions of gauge theories, and as such they require a target space which is a GIT quotient.
From its very definition ${\cal X}^{\vee}$, the Nakajima variety is such a quotient, but the Coulomb branch ${\cal X}$ is not, outside of quivers of type $A$. 

\subsubsection{}
The fact that, for general ${\fg}$, there is no mathematical definition of quantum K-theory of ${\cal X}$, is not an obstacle for us. We are interested not in the qKZ equation and its solutions, but in its conformal limit, the KZ equation. As we will explain, what the conformal limit is simple from perspective of ${\cal X}$ -- it trades quantum K theory for quantum cohomology, and K-theoretic vertex functions of ${\cal X}$ for cohomological ones.

In quantum cohomology, a theorem of \cite{Danilenko} identifies the quantum differential equation of ${\cal X}$ with the KZ equation. In section~\ref{ss-limit}, the conformal limit will give the physical origin of this identification; its explicit geometric form is described in section~\ref{Dan}.

\section{$(2,0)$ conformal theory and the KZ equation}\label{ss-limit}

In this section we derive the two geometric realizations of the KZ equation from the theory on self-dual strings of the six-dimensional $(2,0)$ CFT. We obtain the world-sheet theory by taking the conformal limit of $(2,0)$ LST. The same limit takes 
\beq\label{alglim}
U_h(\widehat{\mathfrak g})\longrightarrow\widehat{\mathfrak g}_k, 
\eeq restoring the Virasoro symmetry of the affine theory.

The two geometric realizations originate from two mirror descriptions of the theory on the self-dual string probing
\beq\label{M4}M_4 = \Sigma \times {\sf S},
\eeq
with ${\Sigma} = {\mathbb R}\times S^1(R)$ and ${\sf S}={\mathbb C}$, in the $\Omega_B$ background from section \ref{tdual}. 

One description is the \({\rm T}\)-equivariant sigma model on \({\cal X}\), from \cite{A1}, whose quantum differential equation is the ${\mathfrak g}_k$ KZ equation. The other is the Landau–Ginzburg model with target \({\cal Y}\) and superpotential \({\cal W}\), used in \cite{A2}, where the KZ equation is the Gauss-Manin equation. 

The targets \({\cal X}\) and \({\cal Y}\) are Coulomb branches, or equivalently, monopole moduli spaces. The monopole description is used later to construct link invariants and to extend the theory to general non-compact Riemann surfaces.

The conformal blocks of $\widehat{\mathfrak g}_k$ are partition functions of the self-dual string supported on ${\sf D}={\mathbb C}$, viewed as an infinitely long cigar, in $(2,0)$ CFT on $M_6$ from \eqref{6mani0}, which is
\beq\label{6mani}
M_{6} = M_4\times {\sf D},
\eeq 
with $M_4$ as in \eqref{4mani0}.
The \(\Omega_B\) background on \(M_6\) induces the \({\mathbb C}^{\times}_{\mathfrak q}\subset{\rm T}\) action on \({\cal X}\) and, in the mirror description, the equivariant part of the superpotential \({\cal W}\) on \({\cal Y}\).The equivariant parameter of the action on ${\cal X}$ is identified with 
\beq\label{qdef}
{\fq} = \exp(-2\pi i \epsilon_h/\epsilon_p),
\eeq
where $-\epsilon_h$ is the parameter of ${\Omega}_B$ background on ${\sf R}$ and $\epsilon_p$ on ${\sf D}$.


\subsection{Conformal limit of $(2,0)$ little string}

The conformal limit of the $(2,0)$ little string is to take 
\beq\label{mtoi}
m_s \rightarrow \infty, 
\eeq
keeping the moduli of the $(2,0)$ theory in \eqref{mod} fixed. 
Also fixed is the background LST is placed on: the metric on the six manifold $M_6$ \eqref{6mani}, and the $\Omega_B$-background on it.

Since the limit takes the tension $m_s^2$ of strings inherited from the fundamental IIB strings to infinity, they decouple. By contrast, the tensions $M_a^2$ of the self-dual strings, given in \eqref{tension}, remain fixed because they are set by scalar expectation values of the six-dimensional theory. These strings survive as the self-dual strings of the $(2,0)$ CFT.

\subsubsection{}
In LST, theory on self-dual strings probing $\Sigma={\mathbb R}\times S^1(R)$ and supported on ${\sf D}$ was a three-dimensional quiver gauge theory on $S^1(R')\times {\sf D}$, where $S^1(R')$ is the circle T-dual to $S^1(R)$.
Keeping the metric on $\Sigma$ fixed in the limit \eqref{mtoi}, keeps the radius $R$ of its circle fixed. Hence the radius of the T-dual circle goes to zero, \beq\label{radlimit} R'={1\over m_s^2R}\longrightarrow0. \eeq 
In the limit, the string winding modes around $S^1(R)$, which supplied the momentum modes on $S^1(R')$, become infinitely heavy. 

In the conformal limit, the world-sheet theory on the self-dual strings in the $(2,0)$ CFT becomes two-dimensional, supported on ${\sf D}$.

\subsubsection{}
The conformal limit determines the scaling of the masses and FI parameters of the 3d quiver gauge theory on the self-dual strings of LST. The positions of the heavy strings on $\Sigma$ remain fixed, since $M_6$, and hence $\Sigma$ is fixed. Viewing $\Sigma$ as the complex $y$ plane with $0, \infty$ deleted, their positions $y=a_i$ are related to the three-dimensional masses $m_i$ of fundamental matter fields by 
\beq\label{afixed} a_i=\exp(R'm_i). 
\eeq Thus $R'm_i$, rather than the three-dimensional masses $m_i$, remain fixed in the limit.

The 3d FI parameters $\lambda_{3d,a}$ are determined by the expectation values of scalar fields in the $(2,0)$ theory, and hence they remain fixed. As a consequence,
$\lambda_{a, 2d} = R'\lambda_{3d, a}$ from \eqref{lambda} vanishes in the limit: 
\beq\label{llim} R'\lambda_{3d,a}=\int_{{\sf C}_a}\left({m_s^2\omega_I\over g_B}+iB_{RR}\right)\longrightarrow0. 
\eeq
 Another way to see the vanishing in \eqref{llim} is that $R'\lambda_{3d,a}$ is $m_s^{-2}$ times a combination of the six-dimensional scalar expectation values in \eqref{mod} that remains fixed, it vanishes in the limit. In rewriting \eqref{lambda} as in \eqref{llim} we used that $R'm_s/g_A=1/g_B$.  Consequently, \beq\label{thelimit2} z_a=\exp(R'\lambda_{3d,a})\longrightarrow1. \eeq
Keeping the $\Omega$ background fixed means that $\epsilon_p$ and $\epsilon_h$ remain fixed. The corresponding fugacities therefore satisfy \beq\label{thelimit} h=\exp(R'\epsilon_h),\qquad p=\exp(R'\epsilon_p)\longrightarrow1. \eeq 
\subsubsection{}\label{clim}

In previous section, we identified the partition functions of the LST self-dual string theory on $S^1(R')$ times the cigar with q-conformal blocks of $U_h(\fgh)$.
We just established that the conformal limit of $(2,0)$ LST takes 
\beq\label{cftlimit} h,\qquad p=h^{-(k+h^\vee)},\qquad z_a=p^{\lambda_a}\longrightarrow1, \eeq 
with fixed $a_i$ and the dimensionless ratios 
\beq\label{fixedd} 
{\epsilon_h\over\epsilon_p}\equiv-{1\over k+h^\vee},\qquad {\lambda_{3d,a}\over\epsilon_p}\equiv\lambda_a.
\eeq 
For the future, it will be useful to have $\lambda_a$, $a=1, \ldots, {\rm rk}{\fg}$ and $k$ at the same footing, so define
\beq\label{l0def}
\lambda_0 \equiv {1\over k+h^\vee}.
\eeq

\subsubsection{}
The limit taken in \eqref{clim} is precisely the one in which the qKZ equation becomes the KZ equation, and the $q$-conformal blocks of $U_h(\fgh)$ become conformal blocks of $\fgh_k$: The dimensionless rations in \eqref{fixedd} determine the level $k$ of the affine Lie algebra, and the highest weight $\lambda_a$ of the Verma module inserted at $0\in\Sigma$.

It follows that conformal blocks of $\fgh_k$ are the partition functions of the $(2,0)$ CFT self-dual string on a cigar. Our next task is to identify their two-dimensional world-sheet theory.

\subsection{A description of the self-dual string theory}\label{kV}

The conformal limit is a two-dimensional $R'\rightarrow 0$ limit of the theory on the self-dual string in LST. One description of the world-sheet theory on the self-dual strings with charges from is section \ref{sscharge} is a 2d sigma model to the 3d Coulomb branch 
$${\cal X},$$
with equivariant ${\rm T}$-torus action. This is the geometric realization used in \cite{A1}.

\subsubsection{}\label{rightl}

The conformal limit keeps fixed the product $R'm_i$ of $R'$ and the masses of 3d fundamental matter fields and the 3d Fayet-Iliopoulos parameters $\lambda_{a,3d}$. Moreover, in the limit, the two-dimensional gauge couplings in \eqref{gaugec} diverge: 
\beq\label{gaugeclimit} 
1/g_{2d,a}^2=\int_{{\sf C}_a}{B_{NS}\over g_Bm_s^2}={M_a^2\over m_s^4}\longrightarrow0, 
\eeq 
since we are keeping the self-dual-string tensions $M_a^2$ fixed. 

The conformal limit is therefore not the one in which the original 3d quiver gauge theory becomes a 2d gauge theory with finite masses and gauge couplings. The limit that would do that would have kept fixed $m_i$ and $R' \lambda_{a,3d}$ and the two-dimensional gauge coupling.
\subsubsection{}
 A three dimensional gauge theory of the kind that arises on the LST self-dual string has different two-dimensional limits, depending what one keeps fixed \cite{AHKT, Seibergrecent}. 
 
The two limits, the limit we need and the two-dimensional gauge theory limit, are related by 3d mirror symmetry, since one limit is obtained from the other by exchanging the mass and FI parameters keeping the $\Omega_B$ background parameters $\epsilon_t, \epsilon_p$ and $\epsilon_h$ fixed.

In the 2d gauge theory limit, the resulting 2d theory would have a low energy description as a sigma model to the Higgs branch ${\cal X}^{\vee}$. Since 3d mirror symmetry exchanges ${\cal X}$ and ${\cal X}^{\vee}$, it follows the theory on the self-dual string has a description as the sigma model to the 3d Coulomb branch ${\cal X}$. 

\subsubsection{}

The Coulomb branch ${\cal X}$ of 3d quiver gauge theory on D3 branes, with charges as in \eqref{sscharge} has complex dimension \[ \dim_{\mathbb C}{\cal X}=2\sum_a d_a. \]  The parameters $a_i$, which are the positions of heavy strings on ${\cal \Sigma}$, become the complexified K\"ahler parameters of ${\cal X}$. For their generic values, ${\cal X}$ is smooth provided all representations $V_i$ are minuscule.
The sigma model also inherits from the 3d theory the equivariant 
\beq\label{equivT} T=\Lambda\times{\mathbb C}^{\times}_{\fq}, \eeq
action, as the equivariant parameters of its action on ${\cal X}$ are kept fixed. 
 The ${\mathbb C}^{\times}_{\fq}$ action, coming from $\Omega_B$ background on $M_6$, scales the holomorphic symplectic form 
$\omega^{2,0}_{\cal X}$ of ${\cal X}$
with weight
$${\fq} = exp(2\pi i \lambda_0).
$$The $\Lambda$ action, with with equivariant parameters $\lambda_{a}$, $a=1, \ldots, {\rm rk \fg}$ preserves it.

\subsubsection{}\label{Dan}
We can now identify the cigar amplitudes geometrically. By the conformal limit \eqref{cftlimit} of the qKZ equation, they are conformal blocks of $\fgh_k$.

The same amplitudes are the cohomological vector vertex functions ${\cal V}({\cal X})$, or Givental's J-functions of ${\cal X}$. The limit keeps Kahler variables of ${\cal X}$ fixed, while taking ${R'}$, the radius of the circle that deforms cohomology to $K$-theory to zero.

The cohomological vector vertex functions ${\cal V}$ are the solutions of the quantum differential equation of ${\cal X}$ from section~\ref{sec2}. 
When K-theoretic vertex function ${\cal V}^K={\cal V}^K({\cal X})$ is defined, ${\cal V}({\cal X})$ arises as its limit
\beq\label{SXb} {\cal V}=\lim_{p,z\to1}{\cal V}^K. \eeq 
\subsubsection{}
In the present case, the connection in \eqref{qdifg} comes from the ${\rm T}$-equivariant Gromov-Witten theory of ${\cal X}$, where ${\rm T}$ is the torus in  \eqref{equivT}.  Working equivariantly with respect to ${\mathbb C}_{\fq}^{\times}\subset T$ action is necessary to ensure the quantum part of the quantum connection does not vanish. Since ${\cal X}$ is holomorphic-symplectic, and not a generic Calabi-Yau, turning off the ${\mathbb C}_{\fq}^{\times}$ action that scales the holomorphic symplectic form $\omega^{2,0}_{\cal X}$ of ${\cal X}$ with weight ${\fq}$ would have reduced the quantum connection to classical.

\subsubsection{}
The classical and quantum contributions to the quantum differential equation of ${\cal X}$ can be recognized directly in the KZ connection. The trigonometric $r$-matrix which enters the KZ equation in \eqref{KZ} is 
\beq\label{ra}r_{ij}(a_i/a_j)={a_i r_{ij}+a_j r_{ji}\over a_i-a_j},
\eeq
see \cite{EFK}.
Here $r_{ij}$ denotes this operator acting on $V_i\otimes V_j$, $r={1\over2}\sum_a { h}_a\otimes { h}^a+\sum_{\alpha>0}{ e}_\alpha\otimes { f}_\alpha,$ where $\{h_a\}$ and $\{h^a\}$ are dual bases of the Cartan subalgebra, and ${e}_\alpha,{f}_\alpha$ are root vectors for the positive and negative roots. Consider a chamber in the ordering of vertex operators on $\Sigma$, such as 
\beq\label{cham} |a_1|\ll|a_2|\ll\cdots\ll|a_n|. \eeq
The chamber \eqref{cham} is simultaneously a chamber in Kahler moduli of ${\cal X}$. In the chamber, the $r$-matrix expression in \eqref{ra} separates into terms that remain finite at the boundary of the chamber and terms that are exponentially suppressed there. 
On the quantum-cohomology side, these are respectively ordinary equivariant multiplication by divisors and the contributions of nonzero-degree rational curves in ${\cal X}$. 

\subsubsection{}
The string theory prediction, therefore, is that the $\fgh_k$ conformal blocks of are cohomological vertex functions of ${\rm T}$-equivariant Gromov-Witten theory of ${\cal X}$. The sigma model to ${\cal X}$, working equivariantly with respect to ${\rm T}$ is one description of self-dual strings of $(2,0)$ theory on $M_6$. It also follows that the KZ-equation from \eqref{KZ} is the quantum differential equation of ${\cal X}$. 

The fact that the KZ equation from \eqref{KZ}, solved by \eqref{confb}, coincides with the quantum differential equation of ${\cal X}$ is in fact a theorem due to I. Danilenko \cite{Danilenko} from 2022.  The identification of quantum cohomology with the $r$-matrix in its canonical form is made in the stable-envelope basis of $H_{\rm T}^*({\cal X})$ \cite{Danilenko}. 

\subsubsection{}
Before we go on, observe that, the fact the $B$-fields in the sigma model to ${\cal X}$ are periodic identifies the complexified K\"ahler moduli space of ${\cal X}$ with the configuration space of the marked points on $\Sigma={\mathbb R}\times S^1$, modulo overall translations.

\subsection{Towards the second description}

In the next section, we will find the exact dual description of the world-sheet theory as the 2d mirror of $({\cal X}, {\rm T})$. 
Here, we will obtain an effective description of mirror theory, based on a target $Y$ with superpotential $W$. As we will see in section \ref{EHMS}, $(Y,W)$ is the equivariant (as opposed to ordinary) mirror of $({\cal X}, {\rm T})$.  Ordinary 2d mirror symmetry will provide the additional degrees of freedom, integrated out in the description based on $(Y, W)$, needed for the full mirror description.

\subsubsection{}\label{Ksimpler}
While ${\cal X}$ and ${\cal X}^{\vee}$ enter symmetrically in three dimensions, the conformal limit breaks the symmetry. The description based on ${\cal X}$ survives, while the description in terms of ${\cal X}^{\vee}$ cannot. In the conformal limit, the Kahler variables of ${\cal X}^{\vee}$ are taken to $1$ and as a result, interpretation of solutions to KZ equation in terms of enumerative geometry of ${\cal X}^{\vee}$ is lost. Loss of description in terms of ${\cal X}^\vee$ means the limit
\beq\label{SXao}
{\cal V}=\lim_{p,z, h \rightarrow 1}{\cal V}^{K, \vee},
\eeq
loses enumerative interpretation in terms of ${\cal X}^{\vee}$.  Here, ${\cal V}^{K, \vee}={\cal V}^{K}({\cal X}^{\vee})$ are the K-theoretic vector vertex functions of ${\cal X}^\vee$. At the same time, the limit in \eqref{SXao}
must lead conformal blocks of $\fgh_k$. The resolution of the tension is as follows.
\subsubsection{}
Recall from section \ref{Kintegral} that K-theoretic vertex functions ${\cal V}^{K, \vee}$ have an integral representation given by
 \beq\label{intvag}
 {\cal V}_{\alpha}^{K, \vee}[L] = \int_{L}   \Phi_{\alpha}^K \,\Omega_Y\,\exp(-{{ W}^K}),
\eeq
where $W^K$ is defined in \eqref{Ksup} and the equations that follow. 
The integral representation has to survive the $z\rightarrow 1$ limit as it does not rely on the existence of a power series expansion in terms of quasimaps to ${\cal X}^{\vee}$. 

\subsubsection{}

 The limit \eqref{SXao} takes \eqref{intvag} to an integral of the same form, 
\beq\label{SXa}
{\cal V}_{\alpha}[L] =  \int_{L} \Phi_{\alpha} \;\Omega_Y  \exp(-{ W}),
\eeq
with $\Phi_{\alpha}$ and $W$ one can simply compute \cite{AFO}. One gets:
\beq\label{Wbase} W=  \lim_{q,z \rightarrow 1} W^K=\sum_{a, \alpha_a} \Bigl(\lambda_a \log(y_{\alpha_a}) + \lambda_0 \log(f_{a, \alpha_a}(y))\Bigr), \eeq where: \beq\label{fdef} f_{a, \alpha_a}(y) =\frac{\prod_{(b,\alpha_b) \to (a,\alpha_a)}\left(1 - \frac{y_{b,\alpha_b}}{y_{a,\alpha_a }} \right) \prod_{[a,i_a] \to (a,\alpha_a)}\left(1 - \frac{a_{a,i_a}}{y_{a, \alpha_a}} \right)}{\prod_{(a,\alpha_a) \neq (a,\beta_a)} (1-\frac{y_{a,\alpha_a}}{ y_{a,\beta_a}})}. \eeq

The result in \eqref{SXa}
is an integral solution of the KZ equation from \eqref{KZ}, for any $L$ on which the integral converges. It reproduces integral solutions to the $\fgh_k$ KZ equation, discovered earlier in the works of Feigin and Frenkel \cite{FF} and Schechtman and Varchenko \cite{SV, SV1}.
\subsubsection{}
The integral representation of solutions to the KZ equation has the form of integral solutions to the Gauss-Manin system from Section 2, associated to the Landau-Ginsburg model 
$$(Y,W),$$
with target 
\beq\label{Yfirst}
Y=\prod_a Sym^{d_a}(\Sigma),
\eeq 
superpotential $W$ from \eqref{Wbase}, and top holomorphic form $\Omega_Y$ from \eqref{tf}.
Both $Y$ and $W$ have a physical interpretation in terms of the self-dual strings in the 6d $(2,0)$ theory on $M_6$.
\subsubsection{}
As noted in section 3, $Y$ is the configuration space of self-dual strings on $\Sigma$. The modes parameterizing positions of strings on ${\sf S}$ in the transverse $\Sigma \times {\sf S} \subset M_6$, are massive due to $\Omega_B$ background on $M_6$.

$W$ is the effective superpotential of the 2d world-sheet theory with target $Y$, with fluctuations along ${\sf S}$ integrated out. It is obtained from the effective super-potential at finite $m_s$ by taking the conformal limit. It can be derived as follows.
\subsubsection{}
%
 The three dimensional gauge theory on the self dual string of LST, supported on $S^1(R')\times{\sf D}$, is a two-dimensional theory on ${\sf D}$, with infinitely many massive modes. Make all charged fields massive, by giving expectation values to complexified real scalars in vector multiplets. There are additional contributions to their masses due to KK momenta on $S^1(R')$.

Integrating out all massive charged fields generates a superpotential which is
a sum of three terms, analogous to \eqref{Ksup}, but different:
\beq\label{3ds}
W_{3d} = W_{3d, 1} + W_{3d, 2} + W_{3d, 3}.
\eeq
The reason for the analogy is that both are obtained by a one loop computation.

The first term, $W_{3d, 1}$ is the same as in \eqref{firsta}, and is the classical twisted superpotential.
The remaining terms $W_{3d, 2}$ and $W_{3d, 3}$ are analogous to \eqref{seconda} and \eqref{thirda}, with the analogous origin. The function $f$ in  \eqref{seconda} and \eqref{thirda}, which is given in terms of the quantum dilogarithms, is replaced with 
$$f(x) ={1\over R' } ({\rm Li}_2(x) - {\rm Li}_2(tx)),$$
where ${\rm Li}_2(x)=\sum_{n=1}^{\infty} {x^n/ n^2}$ is the classical dilogarithm, and $t=p/h$ is the holonomy around the $S^1$ of the $U
(1)_V$ symmetry.

The conformal limit in \eqref{cftlimit}
sends $f(x)$ 
$$
f(x) \;\rightarrow\; \epsilon_t\, \log(1-x),
$$
since $t=\exp(R'\epsilon_t)$. 
The limit thus takes 
$$
W_{3d}/\epsilon_p  \;\;\longrightarrow \;\; W.
$$
The factor of $\epsilon_p$ is a familiar effect of the $\Omega$ background on ${\sf D}$, in two dimensions \cite{Nekrasov}. 

\subsubsection{}
The descriptions of the theory based on 
\beq\label{equim}({\cal X}, {\rm T}), \qquad (Y, W),
\eeq
are not equivalent.
Unlike the description in terms of $({\cal X}, {\rm T})$, the Landau-Ginsburg model 
$(Y, W),$ is not an exact description of the theory on self-dual strings probing $M_4$ in the $(2,0)$ CFT on $M_6$ in \eqref{6mani}. Rather, it is an effective one, in which the massive fluctuations of the string in ${\sf S} = {\mathbb C}$ directions of $M_4$ from \eqref{4mani0} are integrated out.

We will next identify a second, exact description of the self-dual string theory which is equivalent to that based on $({\cal X}, {\rm T}).$ The duality relating them is ordinary 2d mirror symmetry.

\subsection{SYZ mirror symmetry in self-dual string theory}

In this section we will use T-duality in the self-dual string theory to find the 2d mirror description to the sigma model on ${\cal X}$.

\subsubsection{}

The self-dual strings on $M_6$ in \eqref{6mani} probe $M_4$ from \eqref{4mani0}, 
\beq\label{4mania} M_4=\Sigma \times {\sf S} \subset M_6,
\eeq
where $\Sigma= {\mathbb R}\times S^1(R)$, with $S^1$ whose radius $R$ we make explicit, and ${\sf S}={\mathbb C}$. 
We will view ${\sf S}$ as an infinitely long cigar with an asymptotic radius $S^1(R^{\vee})$.
The torus
\beq\label{ttf}
S^1(R)\times S^1(R^{\vee})
\eeq
in the target space translates into a product of $d=\sum_a d_a$ tori in 2d sigma model on the self-dual string.

\subsubsection{}
Self-dual string being a string, it has T-duality symmetry 
that exchanges the transverse circle of radius ${\widetilde R}$, with the T-dual circle of radius $1/M_a^2 {\widetilde  R}$, 
$$
M_a \widetilde{R} \;\;\longleftrightarrow \;\; 1/M_a {\widetilde R}.
$$ 
Here, $M_a^2$ is the string tension of the self-dual string associated to the $a$-th node of the Dynkin diagram of ${\fg}$, see \eqref{tension}. 

 T-dualizing $S^1(R)$ and $S^1(R^{\vee})$ will give us two dual descriptions of the theory on the self-dual string. T-duality of the 2d sigma model is 2d mirror symmetry, so the descriptions we will obtain will be mirror to each other.

\subsubsection{}

The description based on 
$$({\cal X}, {\rm T}),$$
comes about by T-dualizing $S^1(R)\subset \Sigma$. ${\cal X}$ being a Coulomb branch, it is parameterized by triplets of real scalar fields, which originate from positions of D3 branes on ${\mathbb R} \times {\sf S}$, and by the dual photons. The dual photons are defined by scalar-vector duality in 3d. In a gauge theory compactified to 2d on a circle of radius $R'$, this becomes a scalar-scalar duality, which relates the dual photon and the Wilson line of the 3d gauge theory on $S^1(R')$. The latter is the position of a D3 brane on $S^1(R)$. 

\subsubsection{}
Strominger, Yau and Zaslow (SYZ) \cite{SYZ} explained that 2d mirror symmetry is T-duality. The mirror pair of Calabi-Yau $d$-folds are torus fibrations by real tori over the base of same real dimension. If some fibers of ${\cal X}$ degenerate, this equips its mirror ${\cal Y}$ with a superpotential ${\cal W}$ \cite{SYZ, HoriICM,AurouxX}.

The origin of the superpotential is as follows. SYZ mirror ${\cal Y}$ of a Calabi-Yau ${\cal X}$ is the moduli space of an A-brane wrapping the torus fiber of ${\cal X}$. The brane is mirror to a point-like B-brane probing all of ${\cal Y}$. By definition, their moduli spaces agree. The superpotential is generated by disk instantons ending on the torus wrapped A-brane. These can be present if ${\cal X}$ has degenerate fibers so some $S^1$'s in the A-brane are contractible in ${\cal X}$. 
\subsubsection{}

T-dualizing $S^1(R^\vee) \subset {\sf S}$ instead of $S^1(R)\subset \Sigma$, we will find the 2d mirror of ${\cal X}$. 
Since $S^1(R^{\vee}) \subset {\sf S}$ is contractible, this is an instance where ${\cal X}$ contains contractible fibers, and the mirror ${\cal Y}$ will be equipped with a superpotential.

A way to compute both ${\cal Y}$ and the superpotential ${\cal W}$ starting from ${\cal X}$ was developed in \cite{HoriICM, AurouxX}, in the course of studying mirror symmetry. Start by finding a divisor $D^{\vee}$ in ${\cal X}$ so that, working in its complement
\beq\label{Yvee}
{\cal Y}^{\vee} = {\cal X}\backslash D^{\vee},
\eeq
all the disk instantons that would contribute to ${\cal W}$ vanish. Removing the divisor is the same as giving the holomorphic disks strictly infinite area. 
Simultaneously, replacing ${\cal X}$ by ${\cal Y}^{\vee}$ which has no singular fibers, the torus fibration becomes effectively trivial.  

Mirror ${\cal Y}$ is obtained from ${\cal Y}^{\vee}$ by simply replacing its torus fibers by their duals. Having found the mirror ${\cal Y}$, compute the superpotential ${\cal W}$ by putting the divisor $D^{\vee}$ back in, and counting disks that pass through it. 

\subsubsection{}
We will implement the strategy of \cite{HoriICM, AurouxX} as follows. We will choose 
the divisor $D^{\vee}$ in ${\cal X}$ to correspond to the locus where a self-dual string is at the origin of ${\sf S}$. Removing the divisor $D^{\vee}$ from ${\cal X}$ is equivalent to replacing the six manifold $M_6$ in \eqref{6mani0} by
\beq\label{6manim}
M_6^{\times}=\Sigma \times {\sf S}^{\times } \times {\sf D},
\eeq
where ${\sf S}^{\times }=\mathbb C^\times$. In fact, we can take ${\sf S}^{\times} = {\mathbb R} \times S^1(R^\vee)$ with a circle of constant radius. The theories we are describing do not depend on the metric, but assuming a particular one makes the analysis simpler.
\subsubsection{}
At finite $m_s$, replacing $M_6$ by $M_6^{\times}$ takes the theory on the D3 branes probing $M_4$ in \eqref{4mani0} from a three dimensional quiver gauge theory on a circle $S^1(R')$ to a four dimensional gauge theory on the torus
\beq\label{twotorus}
S^1(R')\times S^1({R'^\vee}).
\eeq
whose radii are related to $R$ and $R^{\vee}$ by
 \beq\label{dR}R' = 1/(m_s^2 {R}), \qquad {R'^\vee} = 1/(m_s^2 {R}^{\vee}).
 \eeq This is the theory on D5 branes wrapping \eqref{twotorus} in a T-dual IIB string.
To obtain the self-dual string theory of the $(2,0)$ CFT, we need to take $m_s$ to infinity, keeping the $(2,0)$ theory background fixed.
Both radii of the torus shrink to zero, while \(R\), \(R^\vee\) and the self-dual-string tensions \(M_a^2\) remain fixed.

The four dimensional theory on a torus \eqref{twotorus} can be viewed as a three dimensional gauge theory on a circle in two different ways, depending on the value of the ratio 
$${R'\over R'^\vee} = {R^\vee\over R}.
$$
\subsubsection{}

The description based on $\mathcal Y^\vee$ is associated with the regime
\begin{equation}\frac{R^\vee}{R} \geq 1.\end{equation}
 It is obtained by T-dualizing $S^1(R)$. The target $\mathcal Y^\vee$ is the Coulomb branch of the four-dimensional quiver gauge theory compactified on a circle, or a K-theoretic Coulomb branch.

The two sets of compact scalars on $\mathcal Y^\vee$ have radii
\begin{equation}\label{yv}\frac{1}{M_aR},\qquad M_aR^\vee.\end{equation}
When both radii are large, the ratio $R^\vee/R$ is large as well.

The first set of compact scalars is inherited from $\mathcal X$ and comes from the dual photons of the three-dimensional gauge theory. The radius of a dual photon is the inverse of the radius of its dual Wilson scalar. The Wilson line around $S^1(R')$ has radius squared
${R'/g_{3d,a}^2(R')^2}.$
The gauge coupling is related to the self-dual-string tension by
${R'}/{g_{3d,a}^2}={1}/{g_{2d,a}^2}={M_a^2}/{m_s^4}.$
Thus the dual-photon radius squared is
\[\frac{g_{3d,a}^2(R')^2}{R'}={1\over (M_aR)^2},\]
giving the first radius above.

The second set of compact scalars describes the positions of the D4 branes on $S^1(R^\vee)$. In the D5-brane description, these are Wilson lines around $S^1(R^{\prime\vee})$. Their radius squared is
$${R' \over g_{4d,a}^2R^{\prime\vee}}={R' \over g_{3d,a}^2(R^{\prime\vee})^2}=(M_aR^\vee)^2,$$
giving the second entry in \eqref{yv}.

\subsubsection{}
The description based on $\mathcal Y$ is obtained in the opposite regime,
\[\frac{R^\vee}{R}\leq 1.\]
T-dualizing $S^1(R^\vee)$ maps the D3 branes to D4 branes wrapping $S^1(R^{\prime\vee})$ in the other IIA description. Exchanging $R$ and $R^\vee$ in the preceding calculation gives Wilson scalars and dual photons of radii
\begin{equation}M_aR,\qquad\frac{1}{M_aR^\vee},\end{equation}
respectively.

\subsubsection{}

We learn that the theory on self-dual strings probing $M_6^{\times}$ in \eqref{6manim} has two exact mirror descriptions as a pair of 2d sigma models to 
\beq\label{4dc}
{\cal Y}^{\vee} \;\;\xleftrightarrow{\rm{2d\; mirror}} \;\; {\cal Y}
\eeq
both of which are Coulomb branches of 4d gauge theories on $S^1$, or K-theoretic Coulomb branches. By contrast, the bulk $(2,0)$ theory has no explicit description.

\subsection{The mirror description}

We next describe $\mathcal Y$ explicitly and determine the superpotential ${\cal W}$ generated when we restore the origin of ${\sf S}$.

\subsubsection{}\label{s-heavyinf}
Before we begin, we need to account for one effect of mirror symmetry. The $\mathbb C_q^\times\subset{\rm T}$ symmetry of $\mathcal X$ exists because all the non-compact D3 branes are at $0\in {\sf S}$. Per construction above, the origin $0\in {\sf S}$ is sent by T-duality to infinity in the dual ${\mathbb C}^\times$ factor. The T-dual non-compact D4 branes are sent to infinity as well. In quiver gauge theory terms, the fundamental matter acquires infinite real mass. 

Thus, $\mathcal Y$ is the Coulomb branch of the four-dimensional quiver gauge theory without fundamental matter. The positions of the non-compact D3 and D4 branes on $\Sigma$ are remembered through the superpotential ${\cal W}$. This effect remains whether or not we work equivariantly.

\subsubsection{}\label{ybf}
The target $\mathcal Y$ fibers over the configuration space
\begin{equation}Y=\prod_a\operatorname{Sym}^{d_a}(\Sigma),\end{equation}
of $d=\sum_a d_a$ points on $\Sigma=\mathbb C^\times$, colored by simple roots of $\mathfrak g$ and otherwise identical. The colors encode the charges of the D4 branes wrapping $S^1(R^{\prime\vee})$, and the base records their positions on $\Sigma$.

Over a generic point in $Y$, the fiber of ${\cal Y}$ is
\begin{equation}Y_F=\prod_a(\mathbb C^\times)^{d_a}.\end{equation}
It parameterizes the positions of the D4 branes on $\mathbb R\subset{\sf S}^\times$, complexified by the dual photons. The dual photons are T-dual to the Wilson lines of the D4-brane gauge fields around $S^1(R^{\prime\vee})$.

Over $\Delta\subset Y$, there are special fibers. The divisor $\Delta$ is where either D4 branes of charges $e_a$ and $e_b$, with $\langle e_a,e_b\rangle\ne0$, collide on $\Sigma$, or a D4 brane of charge $e_b$ collides with a D4 brane of charge $w_i$ with $\langle e_a,w_i\rangle\ne0$.
\subsubsection{}\label{coord}

Write $\mathcal Y^o$ for the restriction of the fibration to $Y\setminus\Delta$. It has an open embedding
\begin{equation}\mathcal Y^o\subset\prod_a\operatorname{Sym}^{d_a}(\mathbb C^\times\times\Sigma),\end{equation}
corresponding to the Coulomb branch of the abelianized theory. 
Give $\prod_a Sym^{d_a}({\mathbb C}^\times \times\Sigma)$ explicit coordinates 
\beq\label{coords}
(u_{\alpha, a}, y_{\alpha, a}), \qquad \alpha=1, \ldots ,d_a, \;\; a=1, \ldots, \rm{rk}\,\fg.
\eeq
The coordinates $y$ parameterizes the positions of D4 branes on $\Sigma$; $u$ is the coordinate on mirror of ${\sf S}^\times$. 

\subsubsection{}\label{ccoord}
The full K-theoretic Coulomb branch $\mathcal Y$ specifies how ${\cal Y}^o$ gets extended across $\Delta$.
There are three cases:

\begin{itemize} \item[--]{When two light branes colored by the same root $e_a$ with coordinates $y_a$ and $y_a'$ approach one another, the fiber coordinates have simple poles:
\[u_a=\frac{x_a}{y_a-y_a'},\qquad u_a'=\frac{x_a'}{y_a'-y_a},\]
where $x_a,x_a'\in\mathbb C^\times$, while
\beq\label{cc1}
u_a+u_a'=\frac{x_a-x_a'}{y_a-y_a'}
\eeq
remains finite.}
\item[--]{When branes colored by different roots $e_a$ and $e_b$ collide, and   $\langle e_a,e_b\rangle=-1$, their fiber coordinates satisfy
\beq\label{cc2}
u_au_b=x_{ab}(y_a-y_b),\qquad x_{ab}\in\mathbb C^\times.
\eeq
If $\langle e_a,e_b\rangle=0$, nothing special happens at the collision.}
\item[--]{When a brane of charge $e_b$ approach a brane at $y=a$ whose charge is the fundamental weight $w_i$, the local coordinates satisfy
\beq\label{cc3}
u_b=x_b(y_b-a),\qquad x_b\in\mathbb C^\times,
\eeq
if $\langle e_b,w_i\rangle=1$. If $\langle e_b,w_i\rangle=0$, the fiber coordinate is unchanged. }
\end{itemize}

The relation between local descriptions of this kind and the global coordinates on the Coulomb branch is reviewed in \cite{ADZ}.

\subsubsection{}\label{upsuperpotential}
The superpotential ${\cal W}$ on ${\cal Y}$ is the sum of a topological, and equivariant superpotential: 
\beq\label{supUPba} {\cal W} = {\cal W}_{top}+{\cal W}_{equiv}. 
\eeq 
The topological term \beq\label{Wtop} {\cal W}_{top}= \sum_{a, \alpha_a} u_{a, \alpha_a} \eeq is generated when we trade $M_6^{\times}$ in \eqref{6manim} by $M_6$ in \eqref{6mani}. Equivalently, it is generated by filling in the divisor $D^{\vee}$ in
${\cal Y}^{\vee} = {\cal X}\backslash D^{\vee}$, trading ${\cal Y}^{\vee}$ for ${\cal X}$. The divisor $D^{\vee}$ corresponds to the locus in ${\cal X}$ where at least one light self-dual string is at the origin of ${\sf S}$.

Each $\mathbb C^\times$ factor of $Y_F$ is mirror to ${\sf S}^\times=\mathbb C^\times$. Filling in the origin to obtain ${\sf S}=\mathbb C$ generates the superpotential $u$, where $u$ is the coordinate on the mirror $\mathbb C^\times$ (see the Appendix of \cite{A2}). The  $u\to 0$ end of ${\mathbb C}^{\times}$ mirrors the open end of ${\sf S}$, while $u\to\infty$ end, where the superpotential diverges, corresponds to its origin. Summing over all $d$ factors gives $\mathcal W_{\rm top}$.

The equivariant term is
\begin{equation}\mathcal W_{\rm equiv}=\sum_{a,\alpha_a}\lambda_a\log(y_{\alpha_a})+\lambda_0\sum_{a,\alpha_a}\log\bigl(u_{a,\alpha_a}^{-1}f_{a,\alpha_a}(y)\bigr),\end{equation}
where the functions $f_{a,\alpha_a}(y)$ are those in \eqref{fdef}. It contains the logarithmic, multivalued terms in $\mathcal W$ and mirrors the equivariant ${\rm T}$-action on $\mathcal X$.

\subsubsection{}\label{upsuperpotentiald}
The superpotential ${\cal W}$ on ${\cal Y}$ also splits as the sum of the fiber and the base superpotential
\beq\label{supUPb} {\cal W} = W_F + W. \eeq 
The base superpotential $W$ is the one obtained in \eqref{Wbase}--\eqref{fdef}.
The fiberwise superpotential is
\beq\label{WF} W_F = \sum_{{a, \alpha_a}}(\lambda_0\log(u_{a, \alpha_a}^{-1}) +u_{a, \alpha_a}). \eeq
$W_F$ is the superpotential on the mirror to $d=\sum_a d_a$ copies ${\sf S}={\mathbb C}$ together with an equivariant action taking $z\rightarrow \fq z$, with $\fq = e^{2\pi i \lambda_0}$.

\subsubsection{}\label{downsuperpotential}
We can now make a connection with the effective Landau-Ginsburg model $(Y,W)$. Away from $\Delta$ and for $\lambda_0\neq 0$, the fiberwise theory $(Y_F, W_F)$ is massive. The critical-point equations are
\[\frac{\partial W_F}{\partial\log u_{a,\alpha_a}}=-\lambda_0+u_{a,\alpha_a}=0,\]
so $W_F$ has a unique nondegenerate critical point, given by $u_{a,\alpha_a}=\lambda_0$
for all $a, \alpha_a$. 
The effective value of the superpotential at the critical point is
\[{\mathcal W}|_{\rm crit.}=W+\rm{const}.\]
The additional constant term is independent of the positions on $\Sigma$. 
Thus, as long as equivariant parameter $\lambda_0\neq 0$, we recover the effective Landau-Ginzburg model $(Y,W)$.
The full mirror $(\mathcal Y,\mathcal W)$ retains the fiber degrees of freedom, even though they are generically massive.

\subsection{Monopole moduli spaces and self-dual strings}\label{s-monopoles}
We identified ${\cal X}$ and ${\cal Y}$ as Coulomb branches. They are also monopole moduli spaces, a fact which we will use extensively in the next section, and again in section 9, where we generalize the theory to arbitrary non-compact Riemann surface $\Sigma$. Here, we recall why this is the case. 

\subsubsection{}

At finite $m_s$, T-duality relates the $(2,0)$ little string on a circle $S^1({\widetilde R})$ to the $(1,1)$ little string on the T-dual circle $S^1({\widetilde R}')$. Unlike the $(2,0)$ little string, the $(1,1)$ theory has a gauge-theory description at energies below $m_s$. It is a six-dimensional gauge theory, based on Lie algebra ${\fg}$, with gauge coupling ${g_{6d}}$ given by  
$$
\frac{1}{g_{6d}^2}=m_s^2.
$$
It is abelianized by expectation values of the real scalar fields 
\beq\label{monopolefield} \Psi_a=\int_{{\sf C}_a}\frac{m_sB_{NS}}{g_A}>0, 
\eeq 
where $g_A$ is the IIA string coupling. The background carries over from IIB string in \eqref{fbac}.
The massless gauge fields, valued in the Cartan torus, come from the periods of the RR 3-form potential $C^{(3)}_{RR}$ on the compact cycles of the ADE surface ${\sf R}$.

\subsubsection{}

A D3 brane wrapping a two-cycle ${\sf C}\subset{\sf R}$ becomes a D4 brane wrapping ${\sf C}\times S^1({\widetilde R}')$. The D4 brane is a magnetic source for the RR four form field strength $F_{RR}^{(4)}=dC^{(3)}_{RR}$, 
$$dF_{RR}^{(4)}=*j^{(5)},
$$
where $j^{(5)}$ is a current with a delta function support on the D4 brane (the RR 1-form potential vanishes throughout).
The source $*j^{(5)}$ is the product of a delta-function three-form in the transverse three-manifold and a two-form $\omega_{{\sf C}}$ Poincare dual to the wrapped cycle ${\sf C}_*\subset{\sf R}$. Expand the RR field strength as 
$$
F^{(4)}=\sum_{b=1}^{{\rm rk}\,{\fg}}F_b\,\omega_b,
$$
where the $\omega_b$ form a basis of integral compactly supported two-forms, 
$$
\omega_b\in H^2_{\rm cmpct}({\sf R},{\mathbb Z}).
$$
The monopole charge is the flux of the abelianized field strengths $F_b$ through a small sphere surrounding the monopole. Integrating the source equation gives 
\beq 
\sum_{b=1}^{{\rm rk}\,{\fg}}\oint_{S^2}F_b\; [\omega_b]=[\omega_{{\sf C}}], 
\eeq 
where the D4-brane charge density has been set to one. Thus the magnetic charge of a D4 brane is determined by the cohomology class of the two-form $\omega_{{\sf C}}$ with delta-function support on the cycle ${\sf C}$ the brane wraps.

\subsubsection{}
For a brane wrapping the non-compact cycle ${\sf C}_a^\vee$, the two-form with delta-function support on that cycle is
\[
\omega_a^\vee\in H^2({\sf R},{\mathbb Z}).
\]
The forms $\omega_a^\vee$ give a basis of $H^2({\sf R},{\mathbb Z})$. The integral pairing $H^2({\sf R},{\mathbb Z})\times H_2({\sf R},{\mathbb Z})\longrightarrow{\mathbb Z}$ identifies $H^2({\sf R},{\mathbb Z})$ with the dual lattice to $H_2({\sf R},{\mathbb Z})$. Since $H_2({\sf R},{\mathbb Z})$ is identified with the root $\Lambda'_{{\rm rt}}$ lattice of $\fg$, its dual is the coweight lattice $\Lambda'_{{\rm cowt}}$. Thus, the magnetic charge of the brane on ${\sf C}_a^\vee$ is therefore the fundamental coweight $w_a^\vee$.
\subsubsection{}
For a brane wrapping a compact cycle ${\sf C}_a$, the magnetic charge is represented by its compactly supported dual two-form, \[ \omega_a\in H^2_{\rm cmpct}({\sf R},{\mathbb Z}). \] This cohomology group is canonically dual to $H_2({\sf R},\partial{\sf R},{\mathbb Z})$, which we identified with the weight lattice. Its dual is the coroot lattice, where the charges of smooth 't Hooft-Polyakov monopoles take values.

\subsubsection{}\label{globalform}
The Dirac monopole charges normally live in the cocharacter lattice $\Lambda_{\rm cochar}'$ of the monopole group $^LG$. The cocharacter lattice lives between the coroot and coweight lattices 
$$\Lambda_{\rm cort}'\subset \Lambda_{\rm cochar}'\subset \Lambda_{\rm cowt}',$$
and is determined by the global form of the group $^LG$ (whose Lie algebra $^L\fg$ happens to coincide with ${\fg}$, since it is of ADE type). The coweight and cocharacter lattices coincide when $^LG$ is of adjoint type, which ours must therefore be.

In summary, a D4 branes are monopoles, Dirac and 't Hooft-Polyakov, in the three dimensional space transverse to them of the Lie group of adjoint type $^LG$, whose Lie algebra $^L\fg=\fg$ is simply laced.

\subsubsection{}

The D4 branes supported on two-cycles in ${\sf R}$, with only \eqref{monopolefield} turned on are mutually BPS, preserving the same half the supersymmetry of the background, per discussion in section \ref{susy} in the D3 brane context. In particular, forces between them cancel, and the positions in the transverse 3-space of compact D4 branes are moduli.
As a result of supersymmetry, for $d_a$ branes wrapping each compact cycle ${\sf C}_a\times S^1({\widetilde R}')$, their string tension is 
\beq\label{massD} M^2(d)= \sum_a d_a M_a^2,
\eeq
where $M_a^2=  {\widetilde R}' \int_{{\sf C}_a}{m_s^3 B_{NS}/g_A}$. This is a constant, independent of the brane positions. It is equal to the tension of the dual D3 branes, $M_a^2=  \int_{{\sf C}_a}{m_s^2 B_{NS}}/{g_B}$ given in \eqref{tension}, using the relation of string couplings, $ {\widetilde R}' m_s/g_A=1/g_B$.
The string tension depends only on the vacuum, which determines $M_a$, and total brane charge $d_a$. 
\subsubsection{}\label{exist}

In the IIA product background with constant string coupling, these D4 branes are BPS monopoles. Their fields satisfy the Bogomolny equation 
\beq\label{bogomolny} F=\ast D\Psi. 
\eeq 
The D4-brane moduli space is the moduli space of solutions of the Bogomolny equation. It is helpful to briefly recall how this comes about.

\subsubsection{}

 The relevant terms in the type IIA string-frame action are 
$$S_{\rm IIA} = -\frac{1}{4\kappa_{10}^{2}}\int d^{10}x \; \sqrt{g_{10}}  \;\bigl(g_A^{-2}|dB_{\rm NS}|^{2}+|F_{\rm RR}^{(4)}|^{2}\bigr)\,+\ldots,
$$
with constant string coupling $g_A$. Reducing on $\sf R $ gives the Cartan gauge and scalar kinetic terms of the six-dimensional theory. Their non-abelian completion is
$$\frac{1}{2g_{6d}^{2}}\int_{M_6}\int d^{6}x \; \sqrt{g_{6}}  \bigl(|F|^{2}+|D\Psi|^{2}\bigr),
$$
which are the bosonic kinetic terms of 6d maximally supersymmetric Yang-Mills with gauge group $^LG$. Take $M_6$ to be a product $S^1({\widetilde R}') \times M_3 \times {\sf D}$, and consider configurations that depend only on $M_3$. 
Completing the square gives the energy per unit length of the theory on ${\sf D}$ as 
$$M^2=\frac{{\widetilde R}'}{2g_{6d}^{2}}\int_{M_3}d^{3}x \; \sqrt{g_{3}}   |F-*D\Psi|^{2} \;+\frac{{\widetilde R}'}{g_{6d}^{2}}\int_{\partial M_3}\langle\Psi,F\rangle,
$$
where the second equality uses the Bianchi identity. The boundary term is the magnetic charge contribution to the supersymmetry algebra, so the square measures the energy above the BPS bound. For supersymmetry parameters preserved by the D4 branes, 
the gaugino variation is proportional to the Clifford contraction of $F-*D\Psi$. Thus, solving the Bogomolny equation \eqref{bogomolny} is precisely the condition for preserving supersymmetry. 
\subsubsection{}
The monopole interpretation survives the conformal limit. The dual $(1,1)$ little string becomes a six-dimensional gauge theory compactified on the shrinking $S^1( {\widetilde R}')$ circle, since we are keeping the radius $ {\widetilde R}$ fixed. The result is a five-dimensional gauge theory with gauge coupling 
$$
\frac{1}{g_{5d}^2}= \frac{{\widetilde R}'}{g_{6d}^2}=\frac{1}{ {\widetilde R}}.
$$
This describes the conformal $(2,0)$ theory, compactified on $S^1( {\widetilde R})$, at energies below the Kaluza-Klein scale. The magnetic charges and the monopole interpretation of the wrapped branes survive, even though the bulk gauge-theory description becomes five-dimensional.
\subsubsection{}
We can now apply this construction to the theory on D3 branes on $M_6$ in \eqref{6mani} probing $M_4\subset M_6$ given by \eqref{4mani0},
$$
M_4= \Sigma \times {\sf S}.
$$
By T-dualizing  $S^1(R)\subset\Sigma=\mathbb R_{\sigma} \times S^1(R)$, we found a sigma model on the Coulomb branch ${\cal X}$ of a three-dimensional quiver gauge theory. This T-duality maps D3 branes to D4 branes wrapping $S^1(R')$. Thus, ${\cal X}$ is the moduli space of monopoles on the transverse space 
\beq\label{xsplit}
{\mathbb R}^3={\mathbb R}_{\sigma}\times {\sf S},
\eeq
with the ${\sf S} ={\mathbb C}$.

T-dualizing $S^1(R^\vee)\subset{\sf S}$, we get instead ${\cal Y}$. We do this in steps, as in section \ref{susy}, by first replacing ${\sf S}$ by ${\sf S}^{\times} = {\mathbb R}_t\times S^1(R^{\vee})$, and then filling the origin $0\in {\sf S}$ back in. The D4 branes wrap $S^1(R'^{\vee})$, and their transverse three-manifold is $\Sigma\times{\mathbb R}_t$. Thus ${\cal Y}$ is the moduli space of monopoles on 
\beq\label{ysplit}
{\mathbb R}^2 \times S^1(R) =\Sigma\times{\mathbb R}_t.
\eeq
 Filling in the origin equips ${\cal W}$ with topological superpotential ${\cal W}_{top}$ in \eqref{Wtop}. Turning on equivariant ${\rm T}$-action on ${\cal X}$ adds the equivariant superpotential ${\cal W}_{equiv}.$

\subsubsection{}
The splittings of the three manifold as in \eqref{xsplit} and \eqref{ysplit} pick out the complex structures on the monopole moduli spaces, see e.g. \cite{KW}. 
For example, on $\mathbb R_t\times\Sigma$, with the product metric and orientation $dt\wedge d{\rm vol}_\Sigma$, define $\mathcal D_t=D_t-i\Psi$. The Bogomolny equation splits into a complex equation and a real equation, 
\beq\label{twoe} [D_{\bar y},\mathcal D_t]=0,\qquad \ast_\Sigma F_\Sigma=D_t\Psi, \eeq 
where $F_\Sigma$ is the curvature along $\Sigma$. The complex equation says that $\mathcal D_t$ preserves the holomorphic structure defined by $D_{\bar y}$, and is invariant under complex gauge transformations. 

Solving the complex equation, modulo complex gauge transformations gives the monopole moduli space as a complex manifold we called ${\cal Y}$. 

The real equation is the moment-map condition; for fixed holomorphic data, it becomes an equation for the Hermitian metric. Apriori, the real equation may not have any solutions. In our case, we know solutions exist, from the existence of the supersymmetric D4-brane construction from section \ref{exist}.

For $\mathcal X$, the same construction uses ${\sf S}=\mathbb C$ in place of $\Sigma$.

\subsubsection{} The complex equation lets us describe a monopole by holomorphic scattering data.
The simplest example is that of smooth framed $G'=SU(2)$ monopoles of charge $k$ on $\mathbb{C}\times\mathbb{R}$.

The description by based rational maps was given by Donaldson \cite{DonaldsonMonopoles}, see also the account of Atiyah and Hitchin \cite{AtiyahH}. 
For each $y\in\mathbb C$, solve 
\[ 
\mathcal D_tV=0 
\] 
along the line $\{y\}\times\mathbb R_t$, where $V$ is a section of the rank-two bundle associated to the fundamental representation of $SU(2)$. Let $V_\pm$ denote nonvanishing solutions decaying as $t\to\pm\infty$, and choose an independent solution $W_-$ growing as $t\to-\infty$. With the leading normalizations fixed by the framing, write 
\[ 
V_+=a(y)V_-+b(y)W_-,\qquad S(y)=\frac{a(y)}{b(y)}. 
\]
The complex equation says $a$ and $b$ are holomorphic and so $S$ is meromorphic. When its poles are distinct, 
\[ 
S(y)=\sum_{i=1}^{k}\frac{u_i}{y-y_i},\qquad u_i\ne0. 
\] 
Changing the growing solution by a holomorphic multiple of the decaying solution changes $S$ by a holomorphic function, leaving its principal parts unchanged.

\subsubsection{}

The full rational-map space, including coincident poles, is the open Zastava space for ${\fg}'={\mathfrak{su}_2}$: The pairs $(u_i, y_i)$ are $\mathfrak{su}_2$ specializations of the local coordinates $(u_{\alpha,a},y_{\alpha,a})$ introduced in section \ref{coord}. 

The poles $y_i$ record the D4-brane positions on $\Sigma$, here $\Sigma=\mathbb C$, while the nonzero coefficients $u_i$ encode their positions along $\mathbb R$ and their gauge phases. The unordered pole positions therefore give the base $Y=\operatorname{Sym}^k(\Sigma)$, and, over the distinct-point locus, the coefficients give the fiber $Y_F=(\mathbb C^\times)^k$, as in section \eqref{ybf}. The base and fiber coordinates obtained there from the branes are thus the classical holomorphic scattering coordinates of monopoles. 

Restricting the colored divisor on $\mathbb{C}$ to $\mathbb{C}^{\times}$ gives the trigonometric open Zastava space, identified with the K-theoretic Coulomb branch ${\cal Y}$. In section 9, we will explain the geometric meaning of the local scattering data and how they assemble into ${\cal Y}(\Sigma)$ for a general open Riemann surface $\Sigma$.
\subsubsection{}

For general $G'$, the analogous scattering construction uses flags rather than lines. The asymptotic Higgs field $\Psi_a$, which we assume breaks $G'$ to the maximal torus, orders the growth rates of the transported solutions. Although an individual solution may be changed by adding solutions of faster decay, their successive spans are well defined and form a flag \cite{GC}.

This leads to the adjacent-color collision in \eqref{cc2}. Two adjacent simple roots $e_a,e_b$ of a simply laced algebra generate an $A_2$ root sub-system.  Let $V_1,V_2$ be transported solutions of $\mathcal D_tV=0$ whose successive spans define the scattering line and plane. In the first and second fundamental representation of $A_2$, these are represented by 
\[ 
V_1\in\mathbb C^3,\qquad \widetilde V_1=V_1\wedge V_2\in\bigwedge\nolimits^2\mathbb C^3. 
\] 
The latter satisfies the induced transport equation, with the connection acting on both factors. 

Choose a holomorphic basis $E_1,E_2,E_3$ of transported solutions adapted to the reference flag: its line is spanned by $E_1$, and its plane by $E_1,E_2$. The color-$a$ subgroup acts on the pair $E_1,E_2$. The color-$b$ subgroup mixes $E_2,E_3$, and therefore acts on the pair $E_1\wedge E_2,E_1\wedge E_3$ in the second representation. With one monopole of each color, the ${\mathfrak{su}}(2)$ description in the two pairs therefore gives \[ V_1=(y-y_a)E_1+u_aE_2+cE_3, \] \[ \widetilde V_1=(y-y_b)E_1\wedge E_2+u_bE_1\wedge E_3+dE_2\wedge E_3, \] where $c,d$ are additional constant coefficients. Using that 
$$V_1\wedge\widetilde V_1=0,
$$
which comes from $\widetilde V_1=V_1\wedge V_2$, we have that 
$$d(y-y_a)-u_au_b+c(y-y_b)=0.$$
The coefficient of $y$ gives $c=-d$. Setting $c=-d=x_{ab}$, the constant term gives $ u_au_b=x_{ab}(y_a-y_b)$, which is \eqref{cc2}. 

For $y_a$ and $y_b$ distinct, $x_{ab}\ne0$. Otherwise $u_au_b=0$, and either $V_1$ vanishes at $y_a$ or $\widetilde V_1$ vanishes at $y_b$, so the polynomial data fail to define a flag there. As $y_a$ and $y_b$ collide, $x_{ab}$ remains a nonzero coefficient of the scattering data while $u_au_b$ vanishes. This is precisely the adjacent-color collision condition in \eqref{cc2}. It expresses compatibility of the two fundamental-representation solutions, or equivalently, that the scattering line lies in the scattering plane.
\subsubsection{}
The third collision \eqref{cc3} follows from the $A_2$ model by taking the color-$a$ monopole infinitely heavy at $y_a=a$, so it becomes a fixed Dirac defect for the remaining smooth monopole. On the chart $u_a\ne0$, the adjacent-color relation becomes $u_b=x_b(y_b-a)$, with $x_b=-x_{ab}/u_a\ne0$.

\section{Knot invariants from self-dual strings}

In the previous section, we saw how the KZ equations and conformal blocks of $\fgh_k$ arise from self-dual strings of the $(2,0)$ CFT. It follows, as explained in section 2, that $U_{\fq}(\fg)$ braiding matrix elements get categorified by spaces of morphisms in the category of B-type branes in the sigma model $({\cal X}, {\rm T})$,
$${\MDX} =DCoh_{\rm T}({\cal X}),$$
for which the KZ equation is the quantum differential equation, or the category 
$$\MDY = DFuk({\cal Y}, {\cal W}),$$
of A-type branes on $({\cal Y}, {\cal W})$, for which the KZ equation arises as the Gauss-Manin system. The branes which are objects of these categories are boundary conditions for the worldsheet theory on self-dual strings needed to obtain the solutions to the KZ equation.

In this section, we identify the branes of the self-dual string world sheet theory which correspond to cups and caps and thus lead to $U_{\fq}(\fg)$ link invariants and their categorification. We also explain why the resulting amplitudes and morphism spaces depend only on the link. The bulk $(2,0)$ CFT perspective makes the invariance manifest. 

\subsection{A 4d topological theory with surface defects}

Start with the 6d $(2,0)$ CFT on $M_6$ in \eqref{6mani},
\beq\label{k6mani} M_6= \Sigma \times {\sf S} \times {\sf D}.
\eeq
In section \ref{ss-limit}, we viewed ${\sf S}\times {\sf D}$ as a product of two infinite cigars, with the $\Omega_B$-background parameters $-\epsilon_t$ and $\epsilon_p$, respectively. To obtain the KZ equation and its solutions, we placed self-dual strings on ${\sf D}$, with heavy strings at the origin of ${\sf S}.$
In this section, we will want to get matrix elements of the $U_{\fq}(\fg)$ monodromy matrix, and eventually, their categorification. Per section \ref{sec2}, we want to replace the cigar ${\sf D}$ of section \ref{ss-limit} with its region near infinity:
\beq\label{sD}
{\sf D}= {\mathbb R}_{CS} \times S^1(R_\tau).
\eeq
This has two effects, found by Nekrasov and Witten in \cite{NW}. First, the supersymmetries preserved by the theory on $M_6$ change relative to section \ref{ss-limit}, as we explain in section \ref{2dsusy}. Second, the radius of the circle we get in this way becomes $R_{\tau} =1/\epsilon_p$. To select the monodromy matrix element associated to a link $K$ in 
$M_{3, CS} = \Sigma \times {\mathbb R}_{CS}$, we will want to find a pair of boundary conditions for the theory on the light self-dual strings, determined by the plat presentation of $K$. The strings are then supported on 
$$[-1,1]\times S^1(R_\tau) \subset {\sf D},$$ with those boundary conditions at the ends of the interval. Topological invariance along ${\sf D}$ will make the interval's length immaterial. The quantum deformation parameter ${\fq}=  e^{2\pi i/(k+h^{\vee})}$ is the $\Omega_B$-background parameter
$$
{\fq} = \exp(-2\pi i R_\tau \epsilon_h)=\exp(-2\pi i \epsilon_h/\epsilon_p),
$$
using \eqref{fixedd}.

Cutting the circle \(S^1(R_\tau)\) in \eqref{sD} open modifies ${\sf D}$ once again, to give
\beq\label{kC}
{\sf D} = {\mathbb R}_{CS}\times {\mathbb R}_{\tau}.
\eeq
This realizes the automatic, geometric categorification from section \ref{sec2}. In the present context, it lifts the monodromy matrix elements to spaces of supersymmetric ground states of the self-dual string associated to the link $K$. With the circle cut open, the theory no longer depends on ${\fq}$; it is remembered only through the associated grading on the spaces of states.

\subsubsection{}\label{firsttwist}

The $(2,0)$ theory on $M_6$ in \eqref{k6mani} is partly topologically twisted \cite{WF}. It has the holomorphic twist on ${\sf S}$, so the amplitudes will depend holomorphically on ${\fq}$, the $\Omega_B$-background parameter associated to it. It gives rise to a topological field theory on the four manifold
\beq\label{fourmani}
M_{4, CS}  = \Sigma \times {\sf D}.
\eeq

One can derive this, following \cite{WF} in steps. First, viewing ${\sf S}={\mathbb C}$ as a cigar, one has to topologically twist to preserve supersymmetry. The twist makes use of a $U(1)_R\subset SO(5)_R$ R-symmetry, which acts by rotating the pair of scalars in \eqref{mod} coming from periods of $\omega_{\sf R}^{(2,0)}$, the holomorphic $(2,0)$ form on ${\sf R}$. Due to the twist, the holomorphic three form 
\beq\label{omegatwist}\omega_{\sf R}^{(2,0)}dw,
\eeq
where $w$ is the coordinate on ${\mathbb C}$, is invariant under the symmetry that scales $w\rightarrow {\fq} w$. This symmetry is also what we need for the ${\fq}$ grading to exist. This is a special case of a twist we would need were we to replace ${\sf S}={\mathbb C}$ by a more general Riemann surface. Upon twisting, eight of the original 16 supercharges become scalars on ${\sf S}$ and act as supercharges of the 4d ${\cal N}=2$ theory we get in the remaining four directions. The theory on $M_{4, CS}$ depends on the complex structure of the surface ${\sf S}$ which enters the definition of the theory; in this case, it depends on ${\fq}$. There is a unique way to twist the 4d ${\cal N}=2$ theory on $M_{4, CS}$ to get a topological theory, which is the Donaldson twist.

\subsubsection{}\label{tt}
For a general four manifold $M_{4, CS}$, the Donaldson twist preserves a single supercharge, so there is a unique BRST operator $Q$ which squares to $0$.
In our case, $M_{4, CS}$ is a product
\beq\label{fourmaniprod}
M_{4, CS}  = M_{3, CS}\times {\mathbb R}_\tau, 
\eeq
where $M_{3, CS}$ is the three manifold where Chern-Simons theory lives.  
The resulting theory preserves two supercharges of the original theory. The supercharges are Hermitian conjugates of each other. They are the pair of supercharges of the effective supersymmetric quantum mechanics on ${\mathbb R}_\tau.$ In addition, for us
\beq\label{cs3mani}
M_{3, CS}=\Sigma \times {\mathbb R}_{CS},
\eeq
so ${\sf D}={\mathbb R}_{CS}\times{\mathbb R}_{\tau}$ is flat, and the Donaldson twist becomes a twist on $\Sigma$ alone. As a result, we get an effective two dimensional theory on ${\mathbb R}_{CS}\times{\mathbb R}_{\tau}$ with four supercharges and $(2,2)$ supersymmetry. For $\Sigma={\mathbb R}\times S^1$, the twist on $\Sigma$ is trivial. It will become essential in section 9, where we will take $\Sigma$ to be a more general Riemann surface.

\subsubsection{}

We will abelianize the bulk theory, by turning on expectation values of the scalar field in \eqref{tension}. The bulk theory then has no interesting dynamics of its own, which we used repeatedly. Rather, the dynamics of the theory comes from its self-dual strings -- the light ones, whose tension $M_a^2$ is set by the expectation value of the 6d bulk scalar field in \eqref{tension},
\beq\label{tensiona}
M_a^2 =\int_{\sf C_a} m_s^2 B_{NS}/{g_B }.
\eeq
In addition, to get invariants of knots and links, we need to include self-dual string defects. These are the heavy strings, whose tension is in fact infinite.

The conditions under which including the heavy strings preserves topological invariance were studied in section 5.1 of \cite{WF}. We need all the strings to break $SO(5)_R$ to the $SO(4)_R$ symmetry compatible with the twist. This is the same as saying that only the expectation value of $B_{NS}$ restricted to the compact and non-compact D3 branes is not zero in the background. This is the setup we started with in section \ref{os}.

\subsubsection{}
To get a theory that has a ${\fq}$ grading we need the heavy self-dual strings to be at the origin of the ${\sf S} ={\mathbb C}$ plane in
\beq\label{trans}
M_{4}= \Sigma \times {\sf S}.
\eeq
The light strings are dynamical, unlike the heavy ones. Rather than asking the theory of the light strings to have the same properties, it suffices that its vacua do. For this it suffices to let the light strings be pointlike on $M_{4}$.
\subsubsection{}\label{2dsusy}
As long as we maintain 2d Lorentz invariance in ${\sf D}$-directions, adding the strings preserves all the supercharges the bulk does. This is still the case when we take ${\sf D}$ to be the long cigar with $\Omega_B$ background on it; however, out of the 4 supercharges from section \ref{tt}, only two are preserved near the tip of the cigar, and a different two near its infinity. This is the effect from \cite{NW}.

In the next subsection, we will impose boundary conditions in the self-dual string theory, and then the configuration preserves only two of the supercharges; the same two as near the infinity of the cigar. The two preserved supercharges are those coming from the Donaldson twist, which make the theory topological on $M_{4, CS}$.

\subsubsection{}

Take the heavy strings to have charges in \eqref{ncomp}. Take the charges of light strings to be given in \eqref{comp}. If
\beq\label{RS}
\Sigma = {\mathbb R}_{\Sigma} \times S^1(R),
\eeq
we get two explicit descriptions of the theory on self-dual strings with these charges.
The description based on $({\cal X}, {\rm T})$, the sigma model on ${\cal X}$ with equivariant ${\rm T}$ action, comes by T-dualizing $S^1(R)\subset \Sigma$. 
The description based on $({\cal Y}, {\cal W})$, the sigma model on ${\cal Y}$ with superpotential ${\cal W}$, comes from T-dualizing the circle $S^1(R^{\vee})$ in ${\sf S}={\mathbb C}$ viewed as a long cigar.

\subsection{Links and self-dual strings}

Take an oriented link $K$, 
$$
K\subset M_{3, CS} $$
colored by a minuscule representation $V_i$ of ${\fg}.$ Such a link corresponds to the heavy self-dual string, whose charge is the highest weight $w_i$ of $V_i$, supported on
\beq\label{surface}
K\times {\mathbb R}_{\tau} \subset M_{4, CS},
\eeq
and at the origin of ${\sf S}$. 

The minuscule representations of a simple Lie algebra are those for which all the weights are in a single orbit of the Weyl group. For the ADE Lie algebras, they are the $n$ fundamental representations of $A_n$; the defining representation and the two spinor representations of $D_n$; the two 27 dimensional representations of $E_6$; and the 56 dimensional representation of $E_7$.

\subsubsection{}

Present the link $K$ as a plat closure of a braid $B$
with $2n$ strands. As in section \ref{ss:KZ}, this can be done for any $K$. 
For every $s\in [-1,1]$ the braid $B$ determines $2n$ marked points on $\Sigma$, whose positions vary with $s$. The closure is by canonical collections of $n$ cups at $s=-1$ and $n$ caps at $s=1$. By topological invariance, we take instead the braid to be trivial inside $[-1, 1]$, moving all the braiding into the collection of caps at $s=1$. The two marked points joined by the $i$-th cup are labeled by $V_i$ and $V_i^*$, the representation coloring the cup and its conjugate.

 \subsubsection{}

 Let \(w_i\) and \(w_i^*\) be the highest weights of \(V_i\) and \(V_i^*\). The ranks \(d_a\) corresponding to the zero-weight subspace of \(V=\otimes_{i=1}^n(V_i\otimes V_i^*)\) are determined by
\[
\sum_{i=1}^n(w_i+w_i^*)=\sum_{a=1}^{\operatorname{rk}\mathfrak g}d_a e_a.
\]
As in \eqref{comp}, the \(d_a\) are the charges of the light self-dual strings. The total charge of the heavy defect strings is \(\sum_i(w_i+w_i^*)\), as a special case of \eqref{ncomp}. The theory on the light self-dual strings on 
$$[-1, 1] \times {\mathbb R}_{\tau}
$$
has two descriptions, in terms of  
$$({\cal X}, {\rm T}), \;\; {\textup{or}} \;\; ({\cal Y}, {\cal W}),$$
depending on the duality frame. The positions of heavy strings on $\Sigma$, for $s\in (-1,1
)$ determine their moduli.

\subsubsection{}
The collections of cups at $s=-1$ and braided cups at $s=1$ determine a pair of branes 
\beq\label{ccup}
{\cal U}, \;\; {\textup{and}} \;\; {\mathscr B}{\cal U},
\eeq
as the boundary conditions for the light-string theory. Here \({\mathscr B}\) denotes the action of the braid \(B\) on branes.

The boundary conditions break half of the ${\cal N}=(2,2)$ supersymmetry, preserving a pair of supercharges of B-type on $({\cal X}, {\rm T})$, and of A-type on $({\cal Y}, {\cal W})$. The two branes are objects of the categories of branes, $\MDX$ in the first duality frame and $\MDY$ in the second, preserving B- and A-type supersymmetry, respectively. 

Before any T-dualities, the self-dual string has mixed A/B type boundary conditions: A-type along $\Sigma$, and B-type along ${\sf S}$. Since the theory defined by $Q$ is topological along $\Sigma$ and holomorphic along ${\sf S}$, 
these are the boundary conditions compatible with the supercharge $Q$ of section \ref{tt} and its Hermitian conjugate.

\subsubsection{}
The space of supersymmetric ground states of the self dual string on $[-1, 1]\subset {\mathbb R}_{CS}$ with the pairs of branes in \eqref{ccup} as boundary conditions
coincides, per discussion of section \ref{ss-susyg}, with the space of morphisms  between the branes in the corresponding categories:
\beq\label{knotK}
{\cal H}^{*,*}(K)\equiv Hom^{*,*}({\mathscr B}{\cal U}, {\cal U}).
\eeq
For links in a three-ball in $M_{3,CS}$, the space is graded by the fermion number (cohomological, or Maslov) grading $i$ and a grading $j$ associated to ${\fq}$,
$$
{\cal H}^{*,*}(K) =\bigoplus_{i,j\in {\mathbb Z}} {\cal H}^{i,j}(K).$$
For more general links in $M_{3,CS}=\Sigma\times {\mathbb R}_{CS}$, there are additional gradings, associated to the generators of $H^1(\Sigma, {\mathbb Z})$, as we will elaborate on in section 9.
\subsubsection{}

The space ${\cal H}^{*,*}(K)$ becomes the space of BPS states of the 6d $(2,0)$ theory on $M_6$, in the presence of the heavy strings supported on the surface \eqref{surface},
once we can realize the boundary conditions in \eqref{ccup} in the six-dimensional theory.
Topological invariance of the 6d theory along $M_{4,CS}$ then makes it manifest  that ${\cal H}^{*,*}(K)$ is an invariant of the link $K$, depending on $\fg$ and the representations $V_i$ coloring its components. As we will see below, ${\cal H}^{*,*}(K)$ is in fact an invariant of a framed link.

\subsubsection{}

The link invariant ${\cal H}^{*,*}(K)$ has the equivariant Euler characteristic, 
\beq\label{PKqa}
P_K({\fq}) = \sum_{i,j}(-1)^i {\fq}^j {\rm dim} {\cal H}^{i,j}(K) =\langle {\mathfrak B}{\mathfrak U}|{\mathfrak U}\rangle 
\eeq
which is the $U_{\fq}(\fg)$ invariant of the link $K$, the matrix element of ${\mathfrak B}$ from \eqref{PKq}.

As in \eqref{cups}, $|{\mathfrak U}\rangle$ is the state in Chern-Simons Hilbert space on $\Sigma$ corresponding to the brane \({\cal U}\). It is the conformal block of \(\widehat{\mathfrak g}_k\) in which the punctures paired by each cup fuse to the trivial representation. The braided cap brane \({\mathscr B}{\cal U}\) corresponds to
\[
|{\mathfrak B}{\mathfrak U}\rangle={\mathfrak B}|{\mathfrak U}\rangle.
\]
In previous geometric approaches to categorification, due to Seidel and Smith \cite{SS} or Kamnitzer and Cautis \cite{CK1, CK2}, there was no a priori reason why the Euler characteristic should be the quantum group invariant of the link. There is no systematic way to identify what the invariant decategorifies to. For us, it is manifest, by relation of $\int_{\cal L} \Omega e^{-{\cal W}}$ to conformal blocks of $\fgh_k$, explained in section 4. It remains to explain the six-dimensional origin of the cup and cap boundary conditions.

\subsection{Cups and caps from self-dual strings}\label{ccs}

We now derive from the six-dimensional theory the boundary conditions corresponding
to cups and caps. In the two-dimensional descriptions, these are the ${\cal U}$-branes of
$\MDX$ and $\MDY$, together with their braid transforms ${\mathscr B}{\cal U}$. The same
construction also has a direct interpretation in Chern-Simons theory on
$M_{3,{\rm CS}}$.

\subsubsection{}
A boundary condition for a theory may be viewed as a domain wall between that
theory and a trivial one. For us, the trivial theory is any theory without light,
dynamical self-dual strings. In the notation of \eqref{comp}, it has
\beq
d_a=0
\eeq
for all $a$.

Across such a domain wall, the total self-dual-string charge must be conserved.
We will discover the boundary conditions needed using the fact ${\cal X}$ and ${\cal Y}$ are monopole moduli spaces and the phenomenon of monopole
bubbling, first described in \cite{KW}. The bubbling automatically preserves the
total monopole charge. 

\subsubsection{}
Monopole bubbling occurs when the monopole moduli space becomes singular as a
pair of singular monopoles, or equivalently heavy strings, come together. We will describe it in the $({\cal X},{\rm T})$ theory, corresponding to $G'$ monopoles on $\mathbb R^3$. Since
the phenomenon is local, it is independent of the three-manifold on which the
monopoles live. The
closely related construction exists for $({\cal Y},{\cal W})$, associated to monopoles on $\Sigma\times {\mathbb R}$.

In the \(({\cal X},{\rm T})\) description, monopole bubbling categorifies fusion in the category of representations of \(U_q({\mathfrak g})\) [A1]. In section \ref{cups}, we used fusion to construct the conformal blocks associated to cups and caps. We will now use categorified fusion to obtain the corresponding boundary conditions for the light self-dual strings.

\subsubsection{}\label{onlyt}

First, recall monopole bubbling from \cite{KW} (see \cite{A1} for a review). Take a pair of singular, Dirac monopoles of charges $w_b$ and $w_c$, which we identify with the highest weights of representations $V_b$ and $V_c$ of $G$, the Langlands dual of the monopole group.

As the two Dirac monopoles come together, the monopole moduli space can
develop a singularity at which a fixed number of smooth 't Hooft-Polyakov
monopoles coincide with the singular monopoles and bubble off. If smooth monopoles of charge
\beq
\sum_a d^a_{k,bc}e_a
\eeq
bubble off, they leave behind a single singular monopole of charge
\beq\label{bubble}
w_k=w_b+w_c-\sum_a d^a_{k,bc}e_a.
\eeq
The possible charges $w_k$ are precisely the highest weights of the irreducible representations $V_k$ appearing in the tensor product
\beq\label{tensora}
V_b\otimes V_c =\bigoplus_k V_{k}.
\eeq
Because $V_b$ and $V_c$ are minuscule, each $V_k$ appears with multiplicity
either zero or one.

\subsubsection{}

Geometrically, the singularity arises because a collection of compact cycles
$F_k \subset {\cal X}$, labeled by the representations $V_k$ appearing in \eqref{tensora}, shrink
to a point as the two singular monopoles come together.

Briefly, the geometry is as follows. When the two singular monopoles coincide,
${\cal X}$ degenerates to a singular space
\beq
{\cal X}^{*}=\bigcup_k {\cal X}^{*}_k .
\eeq
On the component ${\cal X}^{*}_k$, the smooth monopoles of total charge
\beq\label{cchar}
w_b+w_c-w_k=\sum_a d^a_{k,bc}e_a
\eeq
have bubbled off. The top-dimensional component, on which no smooth
monopoles bubble off, is noncompact. The lower-dimensional bubbling strata
compactify it by allowing some of the smooth monopoles to disappear into
the singular monopoles.

The moduli of the monopoles whose positions must be tuned for bubbling of type $V_k$ to occur form the transverse geometry $T^*F_k$ to ${\cal X}_k^*$. The cycle $F_k$ has complex dimension
\beq
d_{k,bc}=\sum_a d^a_{k,bc},
\eeq
and shrinks to zero as the two singular monopoles come together. The monopoles left over after bubbling are parametrized by ${\cal X}_k^*$, and locally ${\cal X}$ is a fibration over ${\cal X}_k^*$ with fiber $T^*F_k$.

\subsubsection{}\label{basic}

Monopole bubbling leads to the boundary conditions for the theory on light self-dual strings, as follows.
Start with the $(2,0)$ CFT on $M_6$, with a pair of heavy self-dual strings of charges $w_b$ and $w_c$, located at points on $\Sigma$, at the origin of ${\mathbb C}$, and  
supported on ${\sf D} = {\mathbb R}_{CS}\times {\mathbb R}_{\tau}$. 

Let the positions of heavy strings on $\Sigma$ vary as function of $s \in {\mathbb R}_{CS}$ 
\beq\label{vpath}
a_b=a_b(s), \qquad a_c=a_c(s),
\eeq
so that they interpolate from being at distinct points $a_b(0)\neq a_c(0)$ for $s\leq 0$ to coincident for $s\geq 1$. Projected to $\Sigma$, we get a path which begins at $a_b(0)$ and ends at $a_c(0)$. 
We will call this path a ``matching" of $a_b(0)$ and $a_c(0)$. 

The heavy strings are non-dynamical and do not interact with one another: without light strings, the world-sheet theory is trivial. Thus, although the path in \eqref{vpath} brings the two heavy strings together for $s\geq 1$, in the absence of light strings there is nothing to distinguish this path from the constant one -- for every finite $s$, they are homotopic. This is the physical origin of the trivial braid-group action on the weight $w_b+w_c$ subspace of $V_b\otimes V_c$. 
\subsubsection{}\label{HL}
Next, add light strings of total charge
\beq
\sum_a d^a_{k,bc}e_a,
\eeq
as in \eqref{cchar}. Initially supported on all of ${\sf D}$, they are described by the sigma model with target
\beq
{\cal X}=T^*F_k,
\eeq
where $F_k$ is the vanishing cycle associated with the fusion channel $V_k$.  The positions of heavy strings on $\Sigma$, modulo overall translation, are complexified Kahler moduli of ${\cal X}$. The path \eqref{vpath} is thus a path in complexified Kahler moduli of ${\cal X},$ which vary along the string.

In the presence of these light strings, the two heavy strings can bind for $s\geq 1$ into a single heavy string of charge $w_k$. Charge conservation then requires the light strings to bubble off for $s\geq  1$. The resulting domain wall is a boundary condition for the theory of light strings on $s<1$.
\subsubsection{}
For the light strings to bubble off, their fields at $s=1$ must lie on the vanishing cycle $F_k(1)$. The boundary condition is therefore Neumann along $F_k(1)$ and Dirichlet in the cotangent-fiber directions of $T^*F_k(1)$. In the construction above, $F_k(1)$ has zero volume. By perturbing the path in \eqref{vpath} within its homotopy class relative to the boundary, one can give $F_k(1)$ finite volume.

More generally, one may begin with any number of heavy strings and introduce a sequence of domain walls near $s=1$. At each wall, some light strings bubble off and some heavy strings fuse. Iterating this process, one obtains a boundary condition -- as opposed to a domain wall -- whenever all light strings have disappeared for sufficiently large $s$.
 \subsubsection{}
The most important example for our purposes is the boundary condition associated to a product of $n$ cups colored by minuscule representations $V_i$. We begin with $2n$ heavy strings, with charges $w_i$ and $w_i^*$, the highest weights of $V_i$ and $V_i^*$. The strings fuse in pairs and disappear for $s\gg 1$. 

Each cup gives a partial boundary condition, a domain wall, for light strings of charge $\sum_a d_{i,a}e_a$, where
\beq\label{chargew}
w_i+w_i^*=\sum_a d_{i,a}e_a.
\eeq
Thus, the total charges of the heavy and light strings are, respectively, $\sum_i(w_i+w_i^*)$ and $\sum_{i,a}d_{i,a}e_a$.
 \subsubsection{}\label{minimal}
Near the tip of a cup colored by $V_i$, the relevant region of ${\cal X}$ is a fibration with fiber 
\beq\label{cotU}T^*U_i
\eeq
over ${\cal X}_i^*$, where 
\beq\label{UGP}U_i=G'/P_i
\eeq
is the minuscule Grassmannian corresponding to $V_i$ \cite{A1}. The base ${\cal X}_i^*$ parametrizes the remaining singular and smooth monopoles, while the fiber describes the $d_{i,a}$ smooth monopoles of charge $e_a$ near the point where the singular monopoles of charges $w_i$ and $w_i^*$ come together. 

Above, $P_i$ is the maximal parabolic subgroup of $G'$, associated to the co-weight which gets mapped, by Langlands correspondence, to the highest weight of representation $V_i$ of $G$. $P_i$ is the Lie group obtained from $G'$ by deleting the node of the Dynkin diagram associated to the representation $V_i$ and replacing it with a $U(1)$. The simplest instance of it is ${\mathbb P}^1=SO(3)/U(1)$ corresponding to $G'=SO(3)$ and a single pair of singular monopoles coming together.
 \subsubsection{}
 The domain wall corresponding to a cup colored by $V_i$ imposes Neumann boundary conditions along $U_i$, Dirichlet boundary conditions in the cotangent directions of $T^*U_i$, and no boundary condition along ${\cal X}_i^*$. Topological invariance allows the cup domain walls to slide past one another, so their ordering is immaterial. Bringing them together gives a single boundary condition at $s=1$. In the corresponding local neighborhood,
\beq
T^*U=\prod_i T^*U_i.
\eeq
The resulting brane is
\beq\label{Ubrane}
{\cal U}={\cal O}_U\in{\MDX}.
\eeq
The brane is the structure sheaf of the vanishing cycle $U\subset {\cal X}$, and corresponds to having Neumann boundary conditions along $U$ and Dirichlet boundary conditions in cotangent directions.
\subsubsection{}

The boundary condition in \eqref{Ubrane} corresponds to a vanishing path on $\Sigma$ joining the points $a_b(0)$ and $a_c(0)$. In the terminology of section~\ref{basic}, this is a matching of minimal volume on $\Sigma$. As the matched points come together, $U$ shrinks to zero size.

Any two matchings \eqref{vpath} with the same colored endpoints are related by transport around a loop in the configuration space of \(2n\) marked points on \(\Sigma\). This configuration space is the complexified Kähler moduli space of \({\cal X}\). The loop defines a colored braid \(B\) in \(M_{3,{\rm CS}}\), and transport around it acts on $\MDX$, the category of B-branes on ${\cal X}$, by the braiding functor \({\mathscr B}\).  The functor maps the brane ${\cal U}$ corresponding to the first matching to
$$
{\mathscr B}{\cal U} \in \MDX,
$$
the brane corresponding to the second matching.
\subsubsection{}
The four-dimensional theory on $M_{4,{\rm CS}}=M_{3,{\rm CS}}\times{\mathbb R}_{\tau}$ becomes Chern-Simons theory with gauge group $G$ on $M{3,{\rm CS}}=\Sigma\times{\mathbb R}_{\rm CS}$ upon compactifying ${\mathbb R}_{\tau}$ to a circle $S^1(R_{\tau})$, with periodic boundary conditions for the fermions and holonomies around $S^1$ to keep track of the gradings associated to the torus ${\rm T}$.

This takes heavy self-dual strings wrapping the $S^1$ to Wilson lines in Chern-Simons theory,
colored by $V_b$ and $V_c$. 
For $s<1$, the Wilson lines are distinct, with positions in
$\Sigma$ specified by the path \eqref{vpath}.
For $s\geq 1$, they fuse into a single Wilson line colored by
\[
V_k\subset V_b\otimes V_c,
\]
whose highest weight is $w_k$.
Thus the monopole-bubbling boundary domain walls, constructed above, are
the six-dimensional origin of the fusion vertices in 
Chern-Simons theory.

\subsection{Link homology from self-dual strings}\label{lsl}
We now use the boundary conditions constructed above to explain why
\beq
H^{*,*}(K)=\operatorname{Hom}^{*,*}_{{\mathscr D}^*}( {\mathscr B}{\cal U}, {\cal U})
\eeq
depends only on the (framed) link $K$, and not on its chosen plat presentation.

\subsubsection{}\label{firstst}

Start with the $6d$ $(2,0)$ CFT on $M_6=M_{4,{\rm CS}}\times {\sf S}$, with $2n$ heavy strings of total charge
\beq
\sum_i(w_i+w_i^*),
\eeq
where $w_i$ and $w_i^*$ are the highest weights of the representations $V_i$ and $V_i^*$ coloring the strands of the braid $B$. Embed the strings in $M_{4,{\rm CS}}=\Sigma\times {\sf D}$ so that, as before, each is supported on ${\sf D}={\mathbb R}_{\rm CS}\times {\mathbb R}_\tau$ and lies at the origin of ${\sf S}={\mathbb C}$.

Let the positions of the heavy strings on $\Sigma$ vary with $s\in {\mathbb R}_{\rm CS}$. At $s=0$, they define $2n$ distinct marked points. For $s\geq 1$ and $s\leq -1$, the points come together in pairs colored by $w_i$ and $w_i^*$. The pairing for $s\geq 1$ defines one matching of the marked points, corresponding to the brane ${\cal U}$ we constructed above. Choose the matching for $s\leq -1$ to be obtained from it by reflection, followed by the braid $B$. The marked points together with these two matchings are precisely the data of the chosen plat presentation of $K$.

\subsubsection{}
While the data above are those of a plat presentation of $K$, the heavy strings alone do not form the link. In the absence of light strings, their support in
$
M_{3,{\rm CS}}=\Sigma\times {\mathbb R}_{\rm CS}
$
is an infinite, highly degenerate graph. Since the heavy strings do not interact, this graph can be disentangled. To obtain heavy strings with compact support on the link 
$$K\subset M_{3,{\rm CS}},
$$
add light strings of total charge
\beq\label{dranks}
\sum_{a, i}d_{a, i}e_a = \sum_i(w_i+w_i^*),
\eeq
related to the total heavy-string charge by the charge-conservation relation above. The index $i$ runs over $n$ cups, and $a$ over the $\rm{rk}\,\fg$ simple positive roots.

Initially, both the light and the heavy strings are supported on all of ${\sf D}$.
Using a sequence of monopole bubblings, we can remove all the light and heavy strings for $s\gg 1$ and $s\ll -1$.  Monopole bubbling uses light strings of charge $\sum_{i,a}d_{i,a}e_a $ to erase the equal and opposite heavy string charge.
This modifies the configuration of $2n$ heavy strings we started with from an infinite (degenerate) graph to the link $K$ we were after. 

\subsubsection{}
In summary, using monopole bubbling, we obtained a configuration of heavy strings supported on a surface 
\beq\label{Ksurface}
K\times {\mathbb R}_{\tau}\subset M_{4, CS},
\eeq
and labeled by the highest weights $w_b$ of representations $V_b$ of $G$.

The theory on $M_{4, CS}$ is topological, and inserting such a  heavy string preserves topological invariance. In particular, when the surface is
$K\times {\mathbb R}_\tau$, topological invariance guarantees that ambient isotopic links
$K,K'\subset M_{3,\rm CS}$, continuously deformed into one another without allowing
strands to cross, give rise to the same topological invariants.

We next identify this invariant with the supersymmetric ground states of the light-string theory.
\subsubsection{}

Monopole-bubbling produces boundary conditions for the sigma model to $({\cal X},{\rm T})$ preserving B-type supersymmetry. The space of supersymmetric ground states of the theory on the interval is the space of morphisms 
$$Hom_{\MDX}^{*,*}({\mathscr B}{\cal U},{\cal U}),$$ 
between the corresponding branes in the category $\MDX$ of ${\rm T}$-equivariant B-type branes on ${\cal X}$. We showed that this space is the invariant of the heavy-string configuration supported on $K\times {\mathbb R}_\tau$ from \eqref{Ksurface}. It depends only on the framed link $K\subset M_{3,{\rm CS}}$, and not on its projection to $\Sigma$, or the pair of matchings underlying ${\cal U}$ and ${\mathscr B}{\cal U}$.

\subsection{Self-dual strings and ribbon graphs}

The construction above leads to homological invariants of framed links, as expected from \cite{Jones}. Furthermore, it extends to give homological invariants of arbitrary ribbon graphs from \cite{RT1, RT2}.

\subsubsection{}
We get graphs as opposed to links by replacing the cup branes ${\cal U}$ by boundary conditions that correspond to more general fusion sequences, not necessarily fusion to a trivial representation of $G$. This in fact is the general case we started section \ref{ccs} with.

For the invariant to admit a description as a space of morphisms between branes in the derived category of a smooth ${\cal X}$, we need to fuse minuscule representations at each step. For ${\fg}={\mathfrak{su}_n}$, one can fuse $V_i\otimes V_j$ to $V_{i+j}$, where $V_i$ is the $i$'th anti-symmetric representation, all of which are minuscule. 

The invariants obtained are those of ribbon graphs in $M_{3,\rm CS}$, in blackboard framing, as we now explain.

\subsubsection{}

Recall the matching on $\Sigma$ from section~\ref{basic}. It is a path on $\Sigma$ at $s=0$, interpolating from the marked point $a_b(0)$ to $a_c(0)$. Its canonical framing is given by the vector field $\partial/\partial s$ along the matching. The ribbon can be pictured by pushing the interior of the matching into the $s>0$ half space. The original matching and its push-off form the boundaries of the ribbon.

When $w_k=0$ and $V_k$ is the trivial representation, we have $V_b=V_c^*$. The matching then defines a cap colored by $V_b$, and the construction above gives its canonical ribbon. More generally, when $V_k$ is nontrivial, the construction above gives a ribbon graph extending this local framing without additional twists.
\subsubsection{}
The fact that the graphs we obtain carry blackboard framing is consistent with the fact that, for each object $L$ of $\MDX$, we get a vector vertex function ${\cal V}[L]$. It computes a conformal block of $\fgh_k$, a solution of the KZ equation \eqref{KZ}, which is the quantum differential equation of ${\cal X}$, working equivariantly with respect to ${\rm T}$. The conformal block is a solution of the KZ equation obtained in free field formalism, which leads to link invariants in blackboard framing, see e.g. \cite{ALR}.

\subsection{Relation to the work of Witten}\label{s-WF}
Recall that nothing in the theory depends on the values of the tensions $M_a^2$ in \eqref{tensiona}. An approach to describing the theory without abelianization, at $M_a^2=0$, is laid out by Witten in \cite{WF}, and developed further in \cite{GW, GMW2}. It starts from the same setting -- the six-dimensional $(2,0)$ theory on $M_6$ with knots coming from the heavy self-dual strings --  and aims to obtain the resulting space of BPS states by analysis of bulk physics.

\subsubsection{}
Compactified on the circle of ${\sf S}$, viewed as a cigar, the $(2,0)$ theory becomes the five-dimensional gauge theory with gauge group $G'$ from section \ref{s-monopoles}, on $M_{4,CS}\times{\mathbb R}_+$. Knots, coming from the heavy strings, are at the boundary. The space of BPS states is the cohomology of a complex generated by solutions of the supersymmetry equations of the five-dimensional theory, with the differential obtained by counting its instantons. The theory can be viewed, roughly, as a category of A-branes on an infinite dimensional target, the space of $G_{\mathbb C}'$ connections on $\Sigma\times{\mathbb R}_+$, with the Chern-Simons functional as the superpotential \cite{WF, GW, GMW1, GMW2}, and boundary conditions that depend on the link. The challenging aspect of that approach is that, unlike the categories studied in the context of mirror symmetry, it requires new analytical tools to develop.

\subsubsection{}
The approach taken here uses $M_a^2\neq 0$ to describe the theory from the perspective of the self-dual strings. Once the bulk theory is abelianized, the space of BPS states is the space of supersymmetric ground states of a two-dimensional sigma model with a finite dimensional target, ${\cal X}$ or ${\cal Y}$. Its category of branes is of the kind studied in mirror symmetry, and homological mirror symmetry is what makes the theory explicitly solvable, as described in the next section.

\section{Homological $U_{\fq}(\fg)$ invariants from geometry}\label{sec3}

The construction of the preceding section identifies link homology with spaces of morphisms between branes of the worldsheet theory on the self-dual string. We now show how homological mirror symmetry makes the spaces of morphisms, and hence link homologies, explicitly computable. Equivariant homological mirror symmetry allows the link homology to be computed from $({Y},{W})$, where both the branes and the action of braiding are geometric. We continue to take $\fg$ simply laced, $\Sigma={\mathbb R}\times S^1$, and the link colors to be minuscule representations.

\subsection{Homological mirror symmetry}\label{HMS}

A model way to prove homological mirror symmetry relating $\MDX$ and $\MDY$ is to show both categories have a generating set of objects $\{{\cal T}_i\}$ whose endomorphism algebras 
\beq\label{ao}{\mathscr A} = \bigoplus_{i, j}hom^{*}({\cal T}_i,{\cal T}_j) = hom^*({\cal T}, {\cal T}),
\eeq
computed in the respective categories, are the same. Above, we denoted by ${\cal T} = \bigoplus_i {\cal T}_i$. 
The endomorphism algebra is the algebra of open strings beginning and ending on the generating branes. The algebra ${\mathscr A}$ is in general an $A_{\infty}$ algebra.
This translates mirror symmetry to a pair of equivalences 
\beq\label{firsthf}
{\mathscr D}_{\mathcal X} \cong  {\mathscr D}_{\mathscr A} \cong {\mathscr D}_{\mathcal Y},
\eeq
where 
${\mathscr D}_{\mathscr A}= D({\mathscr A}{\rm-mod})
$
is the derived category of graded ${\mathscr A}$-modules.
 In our setting, ${\mathscr A}$ is graded not only by homological degrees, as in \eqref{ao}, but also equivariant ones.

\subsubsection{}
An elementary example \cite{Auroux} is mirror symmetry relating a pair of infinite cylinders,
$$ {\cal X}={\mathbb C}^{\times},\qquad {\cal Y}={\mathbb R}\times S^1, $$
whose SYZ torus fibers are a pair of dual $S^1$'s. Both $\MDX$, the derived category of coherent sheaves on ${\cal X}$, and $\MDY$, the derived wrapped Fukaya category of ${\cal Y}$, are generated by a single object, which we denote by the same letter ${\cal T}$. The brane is the trivial line bundle on ${\cal X}$ and a real-line Lagrangian on ${\cal Y}$. Their algebras of open strings are the same:
$$ {\mathscr A}={\mathbb C}[x^{\pm1}], $$
the algebra of regular functions on the algebraic cylinder. On the A-side, the powers of $x$ record winding around the cylinder.
\begin{figure}[h]
\centering
\includegraphics[scale=0.45]{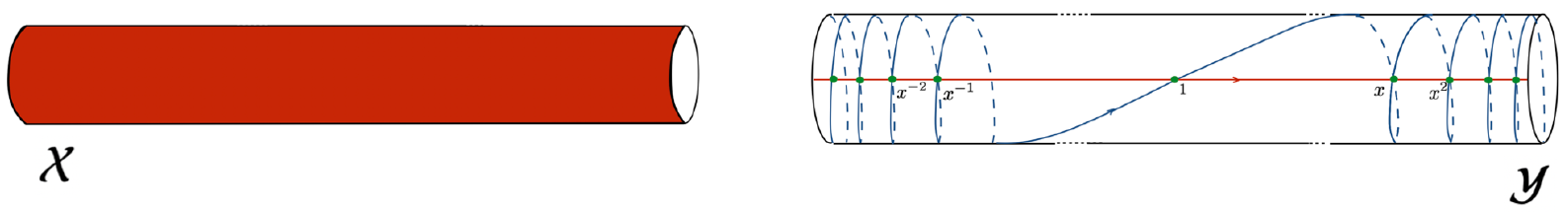}
\caption{A simple example of manifest mirror symmetry.}\label{f_cyl}
\end{figure}
The new examples of homological mirror symmetry which arise in \cite{A1,A2}, while more complicated than this elementary example, are its canonical generalizations.

\subsubsection{}\label{why}
The source of equivalence in \eqref{firsthf} is as follows. The fact that ${\cal T}$-branes generate means that any object of the category can be described as a complex, or a bound state, of ${\cal T}$-branes. A complex is a direct sum of ${\cal T}_{i}$ branes, in various degrees, which we will denote by ${\cal L}({\cal T})$,
 together with a differential $\delta_{\cal L}$:
 \beq\label{dg}
{\cal L} \cong ({\cal L}({\cal T}), \delta_{\cal L})
\eeq
The differential is a degree one operator, 
$$
\delta_{\cal L}: {\cal L}({\cal T}) \rightarrow {\cal L}({\cal T})[1],
$$
valued in ${\mathscr A}$, which encodes how to obtain ${\cal L}$ by taking iterated cones of ${\cal T}$-branes in ${\cal L}({\cal T})$. The differential describes open string tachyons, which condense to bind the ${\cal T}_i$ branes to recover ${\cal L}$, or rather a brane equivalent to it. See \cite{Douglas} and \cite{Horiearly} for reviews.

On the other hand, a ${\cal T}_{i}$ brane is a module for ${\mathscr A}$ given by $Hom^*_{\MDX}({\mathcal T}, {\mathcal T}_{i})$. The resulting module is called projective, as it is a direct summand of the regular module ${\mathscr A}$. The ${\cal T}$-brane itself corresponds to this regular module. This identifies a resolution of any object ${\cal L}$ of the category with a complex of projective modules for ${\mathscr A}$, which is an object of ${\mathscr D}_{\mathscr A}$.
\subsubsection{}
Implicit in this is the fact that categories for which one can attempt to prove homological mirror symmetry in this way are very special: When the category of branes admits a generator ${\cal T}=\bigoplus_i {\cal T}_i$, it behaves like the category of modules of a quiver, whose path algebra is ${\mathscr A}$, see \cite{A2} for a review.

\subsubsection{KLRW algebras from $(2,0)$ theory}

For $\MDX$ and $\MDY$ of previous section, mirror symmetry can be established in this model way. The corresponding algebra ${\mathscr A}$ turns out to itself be an algebra well known to representation theorists who study categorification of quantum groups. 
\subsubsection{}\label{s-BK}
Bezrukavnikov and Kaledin proved that the category of B-type branes of any symplectic resolution ${\mathcal X}$ \cite{kaledin, kaledinR, BK1, BK2} admits a special kind of a generator, the tilting bundle ${\cal T}$. The proof uses quantization in characteristic $p$. The tilting bundle is an honest vector bundle on ${\cal X}$, whose endomorphism algebra
\beq\label{AT}
{\mathscr A}=Hom^*_{\MDX}({\cal T},{\cal T})
\eeq
is an ordinary associative algebra, concentrated in homological degree zero. It is graded only by equivariant degrees. Our ${\cal X}$ is a canonical example of a symplectic resolution, being the Coulomb branch of a 3d ${\cal N}=4$ gauge theory, or the moduli space of monopoles on ${\mathbb R}^3$. The proof of \cite{BK1} does not supply the tilting vector bundle; they are not known in any generality. The virtue of characteristic $p$ methods is that they allow one to identify the algebra without knowing ${\cal T}$ itself.

\subsubsection{}

Webster computed ${\mathscr A}$ explicitly for any ${\cal X}$ which is a Coulomb branch of a 3d ${\cal N}=4$ quiver gauge theory, as is ours. He identified ${\mathscr A}$ with a cylindrical KLRW algebra \cite{W1,W2}. 

The original KLR algebras were introduced by Khovanov and Lauda \cite{KhL} and by Rouquier \cite{Rh}. They categorify (the positive half) of the $U_{\fq}(\fg)$ quantum group. Webster generalized the algebra in \cite{Webster12,webster} to the KLRW algebra in order to define categorified $U_{\fq}(\fg)$ quantum group invariants. 

The KLRW algebra of \cite{KhL, Rh, webster} has a diagrammatic representation in terms of strand diagrams on a plane, satisfying certain relations. The cylindrical generalization which leads to ${\mathscr A}$ in \eqref{AT} from \cite{W1,W2} corresponds to the same strand diagrams and relations, just drawn on the surface of a cylinder, rather than a plane. Mirror symmetry will explain why this is the case, and in particular show that strand diagrams represent open strings.

\subsubsection{}\label{TA}

The mirror of the tilting bundle ${\cal T}$ is a sum of A-branes ${\cal T}_i$ on ${\cal Y}$, which we denote by the same names. They project to
$Y=\prod_a Sym^{d_a}(\Sigma)$
as products of straight real-line Lagrangians on the cylinder, colored by simple roots $e_a$. Each real line Lagrangian on $\Sigma$ is paired with a real-line Lagrangians in the ${\mathbb C}^{\times}$ fiber over it.

For an A-brane ${\cal L}\in\MDY$ to be well defined, the superpotential ${\cal W}$ must restrict to a single valued function on the brane (see section 9 for more detail). Given the multi-valued superpotential ${\cal W}$ in \eqref{supUPba}, it follows that the real-line components can be isotoped past each other if they are colored by simple roots $e_a$ and $e_b$ with $\langle e_a,e_b\rangle=0$, so that the corresponding nodes of the Dynkin diagram are not linked. A component colored by $e_a$ can also be isotoped past a marked point on $\Sigma$ colored by a weight $w$ if $\langle e_a,w\rangle=0$. 

Non-isotopic orderings of products of real line Lagrangians on $\Sigma$ give different ${\cal T}_i$-branes. To get ${\cal T} = \bigoplus_i {\cal T}_i$, we sum over all inequivalent such branes.

\subsubsection{}\label{approx}
The endomorphism algebra of the ${\cal T}$-brane in $\MDY$ was computed in \cite{ADZ}. It was shown to be the cylindrical KLRW algebra from \cite{W1,W2}. Its strand diagrams in terms of which KLRW algebras are defined represent  open strings stretching between ${\cal T}_i$-branes. Its relations come from disk counts. 

The computation in \cite{ADZ} proceeds in two steps. First, remove the divisor $\Delta\subset {\cal Y}$ from section \ref{ybf}. In ${\cal Y}^o$, the complement of $\Delta$
in ${\cal Y}$, the only disks to count are products of one dimensional triangles and constant maps. They are essentially the same as in the model example we started this section with. The algebra ${\mathscr A}^o$ can be written down by inspection. It is naturally written in terms of strand diagrams on a cylinder which label intersection points of ${\cal T}$-branes.

Filling the divisor leaves the intersection points of ${\cal T}$-branes, and hence the generators of the algebra, the same, but changes the algebra structure: it leads to a deformation of the algebra which comes from products involving disks passing through $\Delta$. There is a vast simplification here as well, in that one ends up needing to count only a few such non-trivial disks; the rest of the algebra follows by associativity and degree counting.

\subsubsection{}

The results of \cite{ADZ}, combined with those of Webster \cite{W1,W2}, identify $\MDX$ with the subcategory of $D{\rm Fuk}({\cal Y},{\cal W})$ split-generated by ${\cal T}$. Full homological mirror symmetry would follow from the expected, but still unproved, statement that ${\cal T}$ split-generates all of $D{\rm Fuk}({\cal Y},{\cal W})$. The link homology theory does not hinge on this assumption: the algorithm of \cite{ALR} computes ${\cal H}^{*,*}(K)$ using ${\cal T}$-brane resolutions of ${\cal U}$ and ${\mathscr B}{\cal U}$, so per definition one stays within the subcategory where the equivalence is already established. Throughout, we will identify $\MDY$ with the said subcategory of the actual Fukaya category.
\subsection{Equivariant homological mirror symmetry}\label{EHMS}

The manifold ${\cal X}$ and its mirror ${\cal Y}$ fit into the following diagram: 
 \begin{center}
\begin{tikzcd}[sep=  large]
{\mathcal X}\ar[r,<->,"\textup{mirror}"] \arrow[<->]{rd}[description]{\textup{equiv. mirror}}&{\mathcal Y}\ar[d, ->,""] \\
X\ar[r,<->,"\textup{mirror}"]\ar[u,->,""] & Y
\end{tikzcd}
\end{center}\label{core2}
\vskip 0.2cm
The downstairs $X$ and $Y$ are targets of the worldsheet theory after integrating out modes made massive by the ${\mathbb C}^{\times}_{\fq}$ action on ${\sf S}\subset M_6$. 
$X$ is a holomorphic Lagrangian in ${\cal X}$, the locus where all the monopoles, singular or not, are at the origin of ${\sf S}={\mathbb C}$. 

${\cal Y}$ fibers over $Y$ with generic holomorphic Lagrangian fibers $Y_F=({\mathbb C}^{\times})^d$. The fibers $Y_F$ are equipped with the fiberwise superpotential $W_F$ from section \ref{upsuperpotentiald}, with a unique critical point for every generic point in $Y$. The total superpotential ${\cal W}$
on ${\cal Y}$, restricted to the critical point of $W_F,$ gives the downstairs superpotential $W$, as in section \ref{downsuperpotential}.

\subsubsection{}
The downstairs and upstairs categories are related by pairs of adjoint functors \cite{A2}. The functor
\beq\label{fwant}
k_*:\MDy\; \rightarrow\;\MDY
\eeq
pairs every point on a Lagrangian in $Y$ with a product of $d$ $U$-shaped branes in the ${\mathbb C}^{\times}$ factors of the fiber. Each such $U$-shaped brane mirrors a 0-brane at the origin of ${\mathbb C}$ with its ${\mathbb C}^{\times}_{\fq}$ action. Let $$k^*:\MDY\;\rightarrow\; \MDy,$$ 
be the left adjoint of $k_*$. If $k_*$ pairs the downstairs brane with $U$-branes in fibers over it, $k^*$ takes a fiberwise $hom$. Then for $L_F,L_G\in\MDy$,
\beq\label{model}
Hom^{*,*}_{\MDY}(k_*L_F,k_*L_G)=Hom^{*,*}_{\MDy}(k^*k_*L_F,L_G).
\eeq
This recovers the required morphism spaces  on ${\cal Y}$ from a calculation on $Y$, a space of half the complex dimension. The functors are compatible with braid transport.
\subsubsection{}
$(X, {\rm T})$ and $(Y, {W})$ are themselves related by symmetry \cite{A2}. The corresponding categories $\MDx$ of B-branes on $X$ and ${\MDy}$, the A-branes on $Y$ satisfy 
$${\MDx\cong {\mathscr D}_A\cong \MDy},
$$
where ${\mathscr D}_A={\mathscr D}(A-{\rm mod})$. Here, $A$ is the endomorphism algebra of $k^*{\cal T}_i = T_i$ branes. Viewed as A-branes on $Y$, $T_i$'s are the products of real line Lagrangians on $\Sigma$ without fibers. The algebra $A=Hom_*^*(T,T)$ is a quotient of ${\mathscr A}=Hom_*^*({\cal T}, {\cal T})$ by a two-sided ideal ${\cal I}$, $A={\mathscr A}/ {\cal I}.$
\subsubsection{}
For any simply laced Lie algebra $\fg$, the cup branes constructed in section \ref{ccs} are structure sheaves ${\cal U}={\cal O}_U$ of products of minuscule Grassmannians, supported on the core $X$. Their mirrors are
$$ {\cal U}=k_*I_{\cal U}\in\MDY, $$
where $I_{\cal U}\in\MDy$ is an interval brane. For a cup joining marked points colored by $V_i$ and $V_i^*$, the $I_{\cal U}$-brane is the configuration space of colored points on an interval ending on these marked points. Let $\mu_i$ and $\mu_i^*$ be the highest weights of $V_i$ and $V_i^*$; the points are colored by the simple roots in
$$\mu_i+\mu_i^*=\sum_a d_{a,i}e_a.$$
They are ordered by a sequence of simple-root lowerings from the highest weight $\mu_i$ to the lowest weight $-\mu_i^*$ of $V_i$. Since the representations are minuscule, different lowering sequences are related by exchanges of orthogonal roots, which do not change the brane. For several cups, $I_{\cal U}$ is the corresponding product of interval branes.
\subsubsection{}
The adjoint functors equate the link homology $Hom^{*,*}(K)$ with morphisms of branes in $\MDy$:
\beq\label{MPud}
Hom^{*,*}(K)= Hom^{*,*}_{\MDY}({\mathscr B}{\cal U},{\cal U})=Hom^{*,*}_{\MDy}({\mathscr B}E_{\cal U},I_{\cal U}).
\eeq
The second downstairs brane is
\beq\label{df}
E_{\cal U}=k^*k_*I_{\cal U}=k^*{\cal U}.
\eeq
The $E_{\cal U}$ brane can be computed explicitly for ${\mathfrak g}={\mathfrak{su}}_2$ as the image of $I_{\cal U}$ under $k^*k_*$. The result is the figure-eight brane [A2]. More generally, $E_{\cal U}$ is a product of nested figure-eight branes, with all self-intersections placed near one or the other infinity of $\Sigma$; see the figure below. The ${\cal U}$ brane has two equivalent descriptions as an object of ${\mathscr D}_{\cal Y}$: as $k_*I_{\cal U}$, obtained by pairing $I_{\cal U}$ with $U$-shaped branes in the ${\mathbb C}^{\times}$ fibers over it, or as $E_{\cal U}$ paired with real line Lagrangians in the ${\mathbb C}^{\times}$ fibers. Since $U$-shaped branes are canonically dual to the real line Lagrangians, the adjoint functor $k^*$ takes the second description of ${\cal U}$ to $k^*{\cal U}=E_{\cal U}$.

\begin{figure}[h]
\begin{center}
  \includegraphics[scale=0.37]{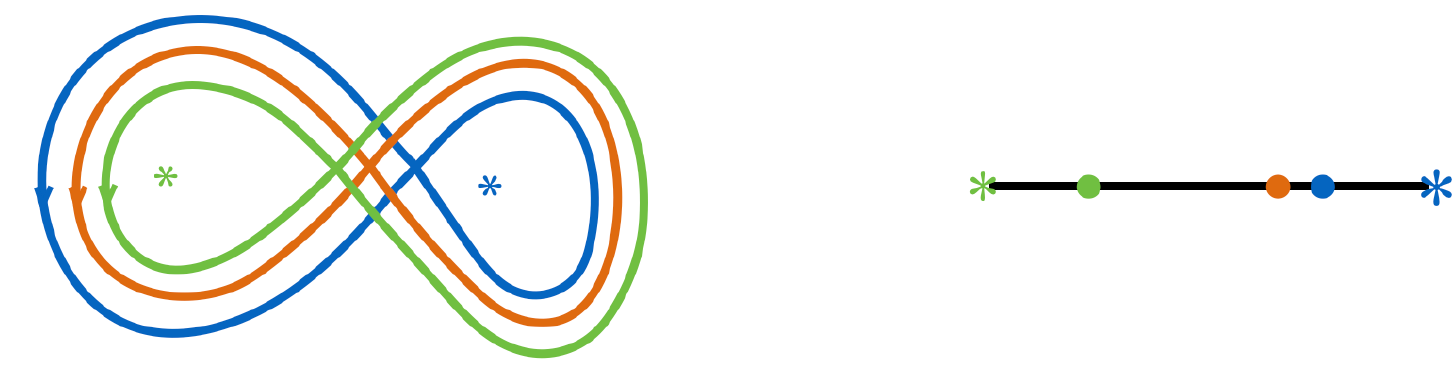}
 \caption{The $E_{\cal U}$ cup and $I_{\cal U}$ cap A-branes corresponding to the defining representation of ${\fg} ={\mathfrak{su}}_4$, colored by its three simple roots; they are equivariant mirror to a B-brane supported on a ${\mathbb P}^3$ as its structure sheaf.}\label{f-Eight}
\end{center}
\end{figure}

\subsection{Link invariants from geometry}
In the description based on $Y$, both the branes, and the action of braiding on them is geometric, so we can simply start with a link $K$ in $M_{3, CS}=\Sigma\times {\mathbb R}_{CS}$ and a choice of projection to the surface ${\Sigma}$.
\begin{figure}[h!]
\begin{center}
     \includegraphics[scale=0.25]{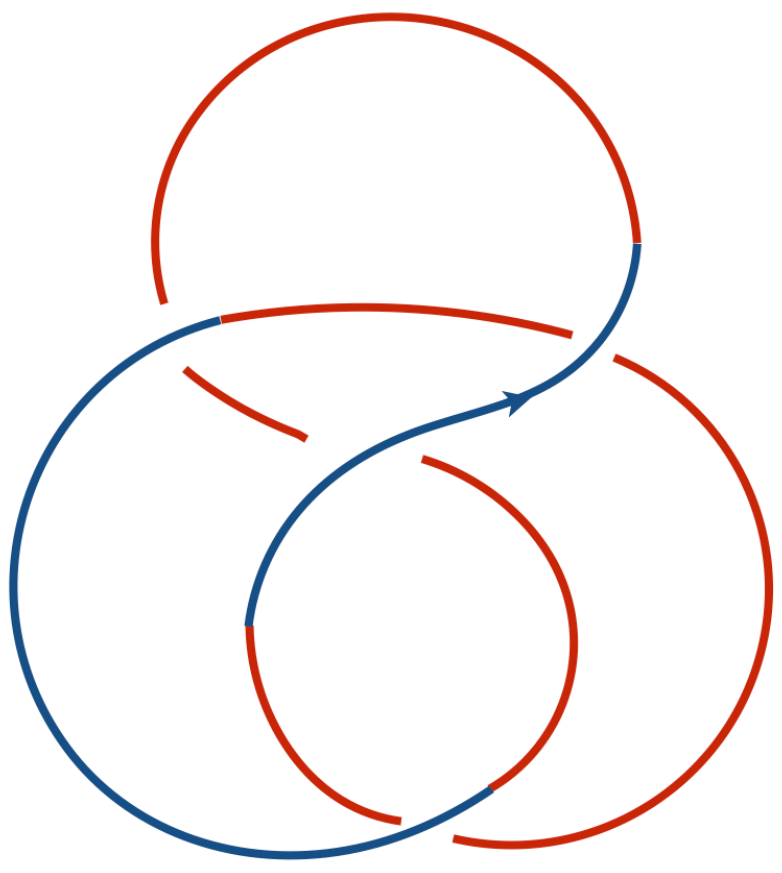}
 \caption{A bi-coloring of the left handed trefoil.  }\label{fT-1}
\end{center}
\end{figure}
\subsubsection{}
To translate the link to a pair of A-branes, start by choosing bi-coloring of the link, such that each of its components has an equal number of red and blue segments, and the red always underpass the blue.

For a link component colored by a representation $V_i$ of ${\fg}$, place a marked point on $\Sigma$ colored by its highest weight $w_i$ where a blue segment begins and its conjugate $w_i^*$ where it ends; the orientation of the link component distinguishes the two. The marked points and their weights determine $Y$ and $W$. The integers in $Y=\prod_a Sym^{d_a}(\Sigma)$ are $d_a=\sum_i d_{a,i}$, where $i$ runs over the cups associated to the blue segments, as in \eqref{dranks}. 

Replace the blue segments by interval branes to get the $I_{\cal U}$ brane, and the red segments by figure-eight branes to get the ${\mathscr B}E_{\cal U}$ brane.

\begin{figure}[h!]
\centering
\includegraphics[scale=0.25]{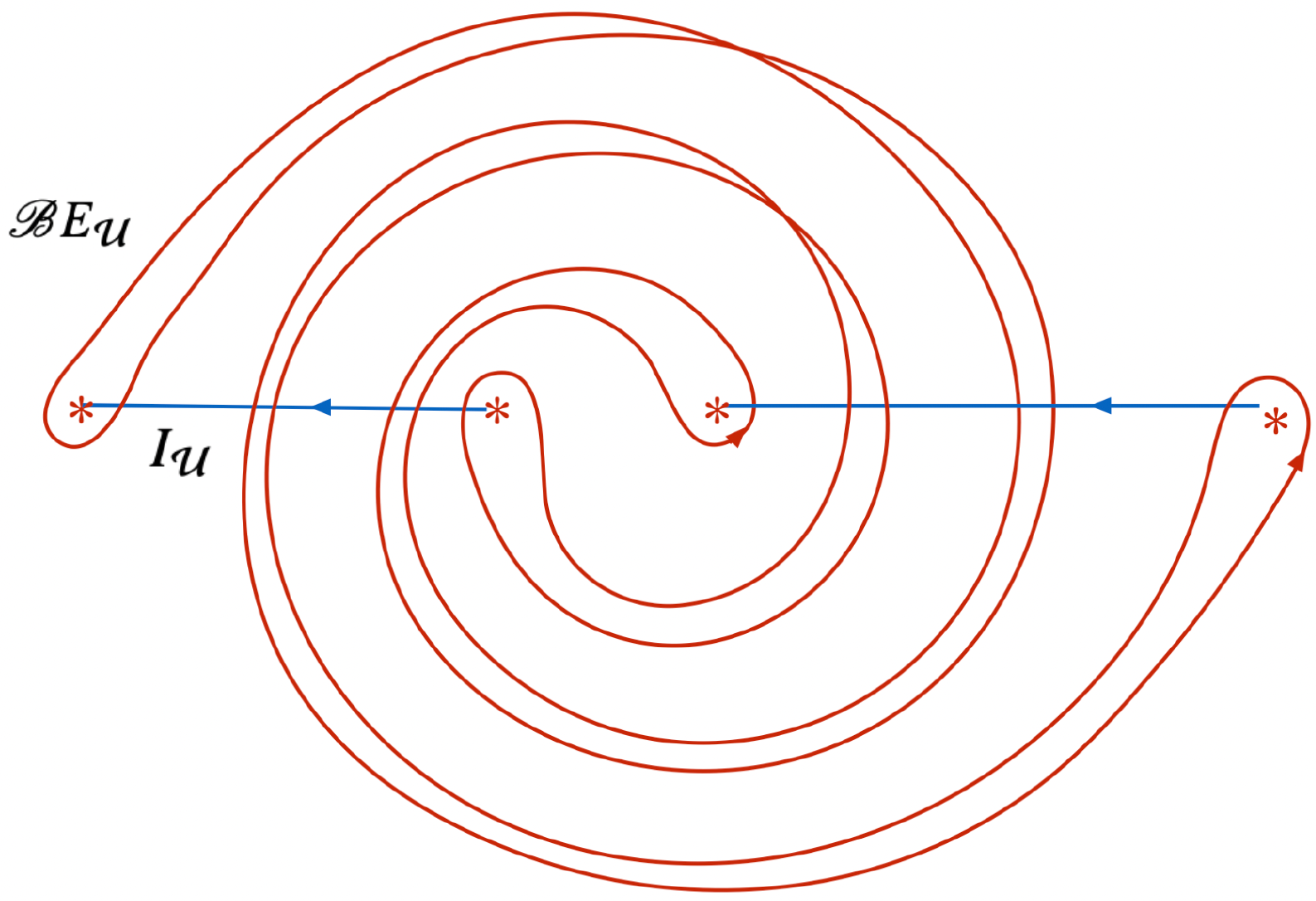}
\caption{The branes for the left-handed trefoil with $\fg={\mathfrak{su}}_2$. The knot has been isotoped relative to Figure \ref{fT-1}.}\label{f_T1}
\end{figure}
The link homology is the space of morphisms in \eqref{MPud}. Keeping all gradings,
\beq\label{MP}
Hom^{*,*}_{\MDy}({\mathscr B}E_{\cal U},I_{\cal U})=\bigoplus_{k\in{\mathbb Z},\,\vec d\in{\mathbb Z}^{\operatorname{rk}\fg+1}}Hom_{\MDy}({\mathscr B}E_{\cal U},I_{\cal U}[k]\{\vec d\}),
\eeq
where $k$ is the homological degree and $\vec d$ the equivariant degree. This is the cohomology of the Floer complex of these two branes.

\subsubsection{}

There is no general algorithm to compute the Floer complex or its cohomology. The algorithm of \cite{ALR} computes $Hom^{*,*}_{\MDY}({\mathscr B}{\cal U},{\cal U})=
Hom^{*,*}_{\MDy}({\mathscr B}E_{\cal U},I_{\cal U})$ by finding a resolution of the braided brane ${\mathscr B}{\cal U}$ brane in terms of ${\cal T}$-branes:
\beq\label{keya}
{\mathscr B}{\cal U}\cong({\mathscr B}{\cal U}({\cal T}),\delta_{{\mathscr B}{\cal U}}).
\eeq
The resolutions of ${\mathscr B}{\cal U}$ and ${\mathscr B}E_{\cal U}$ are related by the same functor that relates the branes. Since $k^*{\cal U}=E_{\cal U}$ and $k^*$ commutes with braid transport, applying it to the resolution in \eqref{keya} gives 
\beq\label{keyb}
{\mathscr B}E_{\cal U}\cong\bigl(k^*{\mathscr B}{\cal U}({\cal T}),\,k^*\delta_{{\mathscr B}{\cal U}}\bigr).
\eeq 
Here $k^*{\cal T}_i=T_i$, so the downstairs complex has the same multiplicities and grading shifts of its summands, with each ${\cal T}_i$ replaced by $T_i$. Its differential is obtained by reducing the matrix entries of $\delta_{{\mathscr B}{\cal U}}$ under the quotient ${\mathscr A}\to A={\mathscr A}/{\cal I}$.

Finding resolutions like \eqref{keya} and \eqref{keyb} is itself a very hard problem in general. Here, there is an algorithm to solve it. The algorithm of \cite{ALR} constructs these two resolutions together, starting from a common geometric decomposition. 

\subsubsection{}
To find the ${\mathscr B}{\cal U}({\cal T})$ terms in the complex \eqref{keya}, isotope the ${\mathscr B}E_{\cal U}$ brane on $\Sigma$, until it breaks at the two infinities of $\Sigma$. The brane that results is $k^*{\mathscr B}{\cal U}({\cal T})$; pairing it with real line Lagrangians in ${\mathbb C}^*$ fibers over it, gives ${\mathscr B}{\cal U}({\cal T})$.
The differential $\delta_{{\mathscr B}{\cal U}}$ is a matrix, mapping ${\mathscr B}{\cal U}({\cal T})$ to ${\mathscr B}{\cal U}({\cal T})[1]$. Its terms are elements of ${\mathscr A}$ which encode how to take connected sums to recover the ${\mathscr B}{\cal U}$ brane back. 

The full algorithm to $\delta_{{\mathscr B}{\cal U}}$ is in \cite{ALR}; it proceeds by finding $\delta^o_{{\mathscr B}{\cal U}}$, corresponding to thinking about the branes as living in ${\cal Y}^o = {\cal Y}/\Delta$; its terms are quite literally encoded in ${\mathscr B}E$, isotoped to near the breaking point. The maps come entirely from intersections on $\Sigma$, so the same initial matrix defines a differential both upstairs and downstairs. Upstairs, its entries lie in ${\mathscr A}^o$, the endomorphism algebra of ${\cal T}$-branes in ${\cal Y}^o$ from section \eqref{approx}, and $(\delta^o_{{\mathscr B}{\cal U}})^2=0$ there.

Restoring $\Delta$ deforms the algebra product from ${\mathscr A}^o$ to ${\mathscr A}$. The algorithm corrects $\delta^o_{{\mathscr B}{\cal U}}$ to a differential $\delta_{{\mathscr B}{\cal U}}$ that squares to zero with the deformed product, giving the resolution of ${\mathscr B}{\cal U}$. Reducing this differential modulo ${\cal I}$ gives the differential $k^*\delta_{{\mathscr B}{\cal U}}$, in the resolution of ${\mathscr B}E_{\cal U}$ in \eqref{keyb}. 


\subsubsection{}
Knowing the complex in \eqref{keya} suffices to compute link homology. Under \eqref{firsthf}, ${\cal U}$ corresponds to a simple ${\mathscr A}$-module: geometrically, the ${\cal U}=k_*I_{\cal U}$ brane intersects precisely one ${\cal T}_{i={\cal U}}$-brane, at one point. Applying $hom(-,{\cal U})$ to \eqref{keya} therefore gives a complex of vector spaces with contributions only from the ${\cal T}_{\cal U}$-summands. By adjunction, this is the same calculation as applying $hom(-,I_{\cal U})$ to the complex of $T_i=k^*{\cal T}_i$-branes. 

\subsubsection{}
 Working in terms of ${\cal T}$ branes makes the homological mirror symmetry manifest, and puts $\MDX$ and $\MDY$ at the same footing.
 
 The same resolution \eqref{keya} computes the invariant in $\MDX$. The morphisms of $\MDX$ are derived homs. Per definition, those are computed by finding the resolution of the brane in the first entry, ${\mathscr B}{\cal U}$ brane for us, of the $Hom$ in terms of projectives. This is nothing but \eqref{keya}. Constructing a projective resolution directly in the coherent-sheaf description is generally difficult, as is computing the Floer differential by counting disks. The geometric algorithm of \cite{ALR} supplies the resolution needed for both calculations.

 So, for the specific categories at hand, ordinarily unsolvable problems - explicitly finding resolutions and computing $Hom$'s of the derived category, become solvable explicitly.

\subsection{Chern-Simons link invariants}
To evaluate the Euler characteristic of the homology in \eqref{MP}, one simply counts the intersection points of the Lagrangians, keeping track of gradings. 
For links contained in a three-ball in $M_{3,{\rm CS}}$, the equivariant grading in \eqref{MP} reduces to a single ${\mathbb Z}$-grading.
The Euler characteristic becomes:
\beq\label{EG}
\chi({\mathscr B} E_{\mathcal U},I_{\mathcal U}) = \sum_{{\mathcal P} \in {\mathscr B} E_{\mathcal U} \cap I_{\mathcal U}}
(-1)^{M({\mathcal P})}{\fq}^{J({\mathcal P})},
\eeq
where $M({\mathcal P})$ and $J({\mathcal P})$ are the Maslov and $\fq$-grading of the point ${\mathcal P}$.
\subsubsection{}

The Euler characteristic depends only on the K-theory class, so it can be computed from the graded terms of the downstairs resolution \eqref{keyb}. Their relative gradings are already fixed by the initial differential $\delta^o_{{\mathscr B}{\cal U}}$, which is read from the geometry. Other than supplying the relative gradings, the differential is irrelevant. Since $I_{\cal U}$ intersects only $T_{\cal U}$, only the grading shifts of the $T_{\cal U}$-summands in $k^*{\mathscr B}{\cal U}({\cal T})$ contribute.
\subsubsection{}
This gives a new way to obtain ${U_{\fq}}({\fg})$ invariants of the link in ${\mathbb R}^3$ for any ${\fg}$. While it is guaranteed to coincide with Chern-Simons invariants, it does not rely on knowing them - unlike the approaches to categorification by \cite{Kh, KhR, Elias}. As we will see, it generalizes to any $\fg$, not just the simply laced ones, and to arbitrary $\Sigma$. The only restriction is that the links need to be colored by minuscule representations of $\fg$ .

\subsubsection{}
For ${\mathfrak g}={\mathfrak{su}}_2$ and links colored by the defining two-dimensional representation $V_w$, the preceding constructions recover the Jones polynomial and its categorification. On ${\cal X}$, the cup brane ${\cal U}$ is the structure sheaf of a product of $d$ non-intersecting ${\mathbb P}^1$’s. Its mirror on ${\cal Y}$ is the $\mathfrak q$-equivariant adaptation of a product of $d$ symplectic $S^2$’s. A detailed description of the theory, the algorithm for computing its homology using the resolution above, and many examples can be found in \cite{ALR}. The resulting homology was proven to agree with Khovanov homology in \cite{LeS}. The agreement is far from manifest, and the proof is involved.

 \section{Non-simply laced $U_{\fq}({\fg})$ link homology}\label{s-nsl}
String theory naturally extends the construction of the previous sections from simply laced to non-simply laced Lie algebras. A non-simply laced Lie algebra ${\fg'}$ can be obtained by folding the Dynkin diagram of a simply laced Lie algebra ${\fg_0'}$ by an admissible automorphism $h$. This folding has a direct realization in string theory: one starts with the six-dimensional $(2,0)$ theory of type ${\fg_0'}$ and compactifies it on a circle with an $h$-twist, so that going around the circle identifies the fields up to the action of $h$.
The resulting link homology will categorify $U_{\fq}(\fg)$ invariants where ${\fg}$ is the Lie algebra Langlands dual of ${\fg'}$.

The twist acts differently in the two mirror descriptions of the self-dual string theory.
One side is the orbifold sigma model in the $h$-twisted sector. In the other, only the $h$-fixed fields survive and one obtains a Landau-Ginzburg model of the fixed-point locus. These are two mirror descriptions the self-dual string in the twisted $(2,0)$ theory.

The corresponding B-side category remembers the folding, while the A-side admits a direct description in terms of the moduli space of monopoles the folded Lie algebra ${\fg'}={\fg}_0/H$. Their equivalence, Theorem \ref{T2}, is the non-simply-laced form of the homological mirror symmetry. 

\subsection{Non-simply laced ${\fg}$ from $(2,0)$ theory}
Let ${\sf R}'_0$ be the ADE surface associated to the simply laced Lie algebra $\fg'_0$. 
Compactify IIB string on a fibration of ${\sf R}'_0$ over a circle:
\beq\label{R0}{{\sf R}_0' \times S^1({ R}_\tau)\over H}.
\eeq
$H$ acts on ${\sf R}_0'$ by a hyper-K\"ahler automorphism, and on $S^1(R_\tau)$ by an order $m=|H|$ translation. Forgetting ${\sf R}_0'$, quotient by $H$ makes the circle radius $R_{\tau}$ $m$ times smaller.
\subsubsection{}
The hyper-Kahler automorphism of ${\sf R}_0'$ corresponds to an automorphism of the Dynkin diagram of ${\fg}_0'$ for which no two vertices belonging to the same orbit are adjacent \cite{WF}. The groups $H$ that arise are cyclic, generated by one element $h$, with $h^m=1$ and $m=2$ or $3$.
The automorphism permutes the compact and noncompact two-cycles of ${\sf R}'_0$, according to its action on the Dynkin diagram. 

\subsubsection{}\label{osa}
D3 branes wrapping two-cycles of ${\sf R}_0'$ lead to self-dual strings in six dimensions, as before. Consider a D3 brane wrapping ${\sf C} \times S^1(R_{\tau})$. For the D3 brane to be compatible with the twist, the two cycle ${\sf C}$ must be $h$-invariant. ${\sf C}$ is therefore the sum of all two-cycles in a single $h$-orbit.
$H$-orbits have either one element or $m=|H|$ elements. A two-cycle in an orbit of length one is already invariant. If
\beq
{\sf C}_0,h({\sf C}_0),\ldots,h^{m-1}({\sf C}_0)
\eeq
is an orbit of length $m$, the corresponding invariant cycle is
\beq\label{gf}
{\sf C}=\sum_{i=0}^{m-1}h^i({\sf C}_0).
\eeq
For compact cycles the D3-branes give 't Hooft-Polyakov monopoles, for noncompact cycles they give Dirac monopoles.  The compact and non-compact 2-cycles in ${\sf R}'_0$ carry the coroot and coweight charges of $\fg'_0$. Their $H$-invariant combinations give the coroot and coweight lattices of the folded Lie algebra
\beq
\fg'=\fg'_0/H.
\eeq
For example, corresponding to \eqref{gf} is a coroot of ${\fg}'$, written as
\beq
e^\vee_a=\sum_{i=0}^{m-1}h^i(e^\vee_{a,0}).
\eeq
 Orbits have length either one or $m$; correspondingly the invariant coroots have two possible lengths, whose squared-length ratio is $m$, the lacing number. The roots $e_a$ of ${\fg}'$ follow immediately from the definition of coroot,
$
e_a={2e_a^{\vee}}/{\langle e_a^{\vee},e_a^{\vee}\rangle}.
$
Correspondingly, the roots of ${\fg}'$ are sums of roots of ${\fg}_0'$ in the same orbit of $H$, divided by the length of the orbit. The Cartan matrix of ${\fg}'$, given by $C_{ab} = \langle e_a^{\vee}, e_b\rangle$, follows from the Cartan matrix of ${\fg}_0'$ and folding.

\subsubsection{}

The twisted fibration thus provides the geometric origin of the folding, which yields a non-simply laced Lie algebra ${\fg}'$ from a simply laced Lie algebra ${\fg}_0'$. The folding relates $({\fg}_0',H)$ and ${\fg}'$ as follows:
\begin{align}\label{folding}
(A_{2n-1}, {\mathbb Z}_2)&\rightarrow C_n\notag\\
(D_{n+1},{\mathbb Z}_2)& \rightarrow B_n \notag\\
(D_{4},{\mathbb Z}_3)& \rightarrow G_2\notag\\
(E_{6},{\mathbb Z}_2)& \rightarrow F_4
\end{align}
The Langlands dual Lie algebras ${\fg}$ are $B_n$, $C_n$, $G_2$ and $F_4$, respectively.

\subsection{KZ equation for non-simply laced ${\fg}$ from self-dual strings}\label{s-nslKZ}

We now place the $(2,0)$ theory of type ${\fg}_0'$ on
\beq
M_6=\Sigma\times{\sf S}\times{\sf D},
\eeq
as in \eqref{6mani}, with ${\sf D}$ viewed as a long cigar with monodromy $h$ around its tip. The circle at infinity of ${\sf D}$ is the twisted circle of \eqref{R0}, which we continue to denote by $S^1(R_\tau)$. Equivalently, if $\widetilde{\sf D}$ is the $m$-fold cover of ${\sf D}$ branched at the tip, on which $H$ acts by rotations about the tip, then
\beq
\widetilde{\sf D}/H={\sf D}.
\eeq
The theory originates from IIB string theory on the fibration of the ADE surface ${\sf R}'_0$ over $M_6$ with this monodromy. As before, the theory on $M_6$ is placed in the $\Omega_B$ background.
\subsubsection{}

Include D3-branes wrapping $H$-invariant combinations of two-cycles in ${\sf R}'_0$ and supported on ${\sf D}$. The worldsheet theory has two T-dual descriptions: one is the sigma model on $({\cal X}_0,{\rm T}_0)$, obtained by T-dualizing $S^1(R)\subset\Sigma$; the other is the sigma model on $({\cal Y}_0,{\cal W}_0)$, obtained by T-dualizing $S^1(R^{\vee})\subset{\sf S}$. Here $({\cal X}_0,{\rm T}_0)$ and $({\cal Y}_0,{\cal W}_0)$ are the mirror pair of section \ref{ss-limit}, for the simply laced Lie algebra ${\fg}_0'$. Since both ${\cal X}_0$ and ${\cal Y}_0$ are monopole moduli spaces, both provide magnetic descriptions of the theory.

The essential new ingredient is the $H$-twist on ${\sf D}$. The fields of the worldsheet theory of the self-dual string, transported once around the tip of ${\sf D}$, come back to themselves only up to the action of $h$ on the target, induced by its action on ${\sf R}_0'$. The sigma model on ${\sf D}$ is therefore in the $h$-twisted sector.
\subsubsection{}
The partition function of the self-dual string theory on ${\sf D}$, in the $h$-twisted sector, should compute conformal blocks of the affine Lie algebra $\fgh_k$ on $\Sigma$, where ${\fg}$ is related to ${\fg}'$ by Langlands duality
\beq\label{TL}
\fg={}^{L}(\fg').
\eeq
This comes about as follows.
\subsubsection{}

We have the $(2,0)$ theory of type ${\fg}_0'$ on the twisted circle $S^1(R_\tau)$, times an untwisted circle $S^1(\widetilde{R})$, which is either $S^1(R)\subset\Sigma$ or $S^1(R^{\vee})\subset{\sf S}$. When the torus $S^1(R_\tau)\times S^1(\widetilde{R})$ is small, the resulting four-dimensional theory has two descriptions, depending on which circle is shrunk first \cite{VafaM, WF}.

Shrinking the twisted circle $S^1(R_{\tau})\subset{\sf D}$ first gives a five-dimensional gauge theory with Lie algebra ${\fg}$, rather than ${\fg}'$. The self-dual strings wrapping $S^1(R_\tau)$ become electrically charged particles. Their charges are the $H$-invariant classes, as in \eqref{gf}, which are coroots of ${\fg}'$, and hence roots of ${\fg}$. This is the description relevant to the KZ equation of $\fgh_k$.

To derive the worldsheet theory on the self-dual string, one instead uses T-duality to shrink $S^1(\widetilde{R})$. This does not require $S^1(R_{\tau})$ to be small; its radius is arbitrary, and can eventually be taken to infinity. The two choices of $S^1(\widetilde{R})$ give the two geometric descriptions above, related by mirror symmetry. In both, the same strings are ${\fg}'$ monopoles, with magnetic charges in the coroot lattice of ${\fg}'$.

The fact that the same objects are electrically charged in one description and magnetically charged in the other is why $\fg$ and $\fg'$ are related by the Langlands duality \eqref{TL}.

\subsubsection{}

In the description based on ${\cal X}_0$, the theory on ${\sf D}$ is the sigma model to the orbifold ${\cal X}_0/H$ in the $h$-twisted sector, with an A-twist in the interior of ${\sf D}$ and a B-type boundary condition at infinity. It is equivariant with respect to the subtorus ${\rm T}\subset{\rm T}_0$ that commutes with $H$. Its equivariant parameters are $\lambda_0$, and one $\lambda_a$ for each $H$-orbit of nodes of the Dynkin diagram of ${\fg}_0'$. For the non-simply laced Lie algebra ${\fg}$, the KZ connection is identified with the orbifold quantum connection of the $h$-twisted sector of ${\cal X}_0/H$.

\subsubsection{}

Recall that in orbifold Gromov-Witten theory of ${\cal X}_0/H$, each marked point on the worldsheet carries an element $h_i\in H$, together with a class on the fixed locus ${\cal X}_0^{h_i}$, and in genus zero $\prod_i h_i=1$.

The quantum connection in \eqref{qdifg} comes about as follows. Differentiating the cigar partition function with respect to a K\"ahler modulus brings down an insertion of the corresponding divisor class near the tip of ${\sf D}$. The two insertions near the tip of the cigar, the pre-existing one and the one corresponding to the divisor class, bubble off onto a sphere attached to the tip of the cigar via a node. The quantum connection is the three-point function on this sphere. 

Its three points, the divisor, the insertion from the tip and the node, necessarily carry twists $(1,h,h^{-1})$. The divisor is an $H$-invariant class of ${\cal X}_0$; the other two carry classes in $H^*_{\rm T}({\cal X}_0^h)=H^*_{\rm T}({\cal X})$, where
$$
{\cal X}={\cal X}_0^h={\cal X}_0^{h^{-1}}.
$$

\subsubsection{}\label{s-Kinv}

The fixed locus ${\cal X}={\cal X}_0^h$ is the moduli space of ${\fg}'={\fg}'_0/H$ monopoles on ${\mathbb R}^3$. Start with ${\cal X}_0$, the moduli space of ${\fg}'_0$ monopoles on the same three-manifold. $H$-invariance requires ${\fg}_0'$ monopoles whose charges are exchanged by $H$ to sit at the same point in ${\mathbb R}^3$, where they form a single ${\fg}'$ monopole. An $h$-invariant monopole configuration therefore takes values in the $h$-invariant subalgebra of ${\fg}_0'$, which is ${\fg}'$. Its charges are $H$-invariant combinations of co-roots of ${\fg}_0'$, which are the co-roots of ${\fg}'$. Conversely, every configuration of ${\fg}'$ monopoles is $H$-invariant (see also \cite{WF}, section 4.3). 

Associated to the untwisted point is, a priori, the $H$-invariant cohomology of ${\cal X}_0$. We need only its divisor classes. The divisor classes of ${\cal X}_0$ correspond to directions in its complexified K\"ahler moduli space, as in section \ref{s-qde}. The K\"ahler moduli of ${\cal X}_0$ are the relative positions of the singular ${\fg}_0'$ monopoles on ${\mathbb R}\subset{\mathbb R}^3$, and their complexification is the configuration space of the marked points on $\Sigma$. The $H$-invariant part of this configuration space is the configuration space of the singular ${\fg}'$ monopoles. This is the complexified K\"ahler moduli space of ${\cal X}$, since ${\cal X}$ is the moduli space of ${\fg}'$ monopoles. The K\"ahler form of ${\cal X}$ is the restriction of that of ${\cal X}_0$, so the $H$-invariant divisor classes of ${\cal X}_0$ restrict to the divisor classes of ${\cal X}$, and all divisor classes of ${\cal X}$ arise in this way.

\subsubsection{}

The orbifold quantum connection of ${\cal X}_0/H$ in the $h$-twisted sector is thus a connection on a vector bundle with fibers $H^*_{\rm T}({\cal X})$, over the complexified K\"ahler moduli of ${\cal X}$, and it is identified with the KZ connection of $\fgh_k$. 

The KZ connection of $\fgh_k$ is not simply the ordinary quantum connection of ${\cal X}$, as one can easily verify by naively extending the work of \cite{Danilenko} from a simply laced to a non-simply laced ${\fg}$. (There is an analogous failure at the level of categories, already for the unknot. See section \ref{s-unknot}.)

Indeed, orbifold Gromov-Witten theory also sums over twisted maps to ${\cal X}_0$, and these give additional contributions to the quantum connection. They are required to reproduce the KZ connection for the non-simply laced Lie algebra ${\fg}$.

\subsubsection{}

The orbifold three-point function counts $H\times{\rm T}$-equivariant maps to ${\cal X}_0$ from the $m$-fold cover of the bubbled-off sphere, branched at its two twisted points; we denote this cover, which is again a ${\mathbb P}^1$, by $\widetilde{\mathbb P}^1$. There are two types of such maps. 

The first are the $h$-invariant maps from $\widetilde{\mathbb P}^1$ to the fixed-point locus ${\cal X}$, the moduli space of ${\fg}'$ monopoles. Counting only these gives the ordinary quantum connection of ${\cal X}$, which, as explained above, is not the Knizhnik-Zamolodchikov connection of the non-simply laced Lie algebra ${\fg}$.

The second are genuinely $h$-twisted maps, for which the action of $h$ on $\widetilde{\mathbb P}^1$ is accompanied by a nontrivial action on ${\cal X}_0$. These give new instanton contributions absent from the ordinary Gromov-Witten theory of ${\cal X}$. One can prove \cite{Danilenkounpublished} that, after summing over both types of maps, the orbifold quantum connection of ${\cal X}_0/H$ coincides with the Knizhnik-Zamolodchikov connection of the non-simply laced Lie algebra ${\fg}$.

\subsection{The mirror description}
In the description based on $({\cal Y}_0,{\cal W}_0)$, the theory reduces to the fixed-point Landau-Ginzburg model
\beq\label{hydef}
{\cal Y}=({\cal Y}_0)^h,\qquad
{\cal W}={\cal W}_0|_{\cal Y}.
\eeq
The theory on $\widetilde {\sf D}$ has a B-twist in the interior, for which only the constant modes contribute to amplitudes. Twisting by $h$ around the tip projects out all fields that are not $h$-invariant. The resulting theory is therefore the Landau-Ginzburg model on ${\sf D}=\widetilde {\sf D}/H$ with target $({\cal Y},{\cal W})$. 
Whereas ${\cal Y}_0$ is the moduli space of ${\fg}'_0$ monopoles on $\Sigma\times{\mathbb R}$, the fixed-point locus ${\cal Y}$ is the moduli space of ${\fg}'={\fg}'_0/H$ monopoles on the same three-manifold, for the same reason as in section \ref{s-Kinv}. 

Since the bulk of ${\sf D}$ maps to ${\cal Y}$, the boundary condition at $\partial{\sf D}$ is an A-brane supported on a Lagrangian ${\cal L}\subset{\cal Y}$.

\subsubsection{}\label{s-foldgeom}
The fixed locus ${\cal Y}=({\cal Y}_0)^h$ fibers over $Y=\prod_a {\rm Sym}^{d_a}(\Sigma)$, with points colored by simple roots $e_a$ of $\fg$, and with fibers $({\mathbb C}^{\times})^d$ over generic points of $Y$. The fiber coordinate $u_{a,\alpha}$ is the common value of the fiber coordinates $u'$ of ${\cal Y}_0$ in the corresponding $H$-orbit, and similarly for $y_{a,\alpha}$. The way the fibers degenerate differs from section \ref{ccoord} when roots of different lengths are involved. Let
$$
m_a={\langle e_a,e_a\rangle/ 2},
$$
so that $m_a=1$ for a short root and $m_a=m$ for a long root, with the inner product normalized as in section \ref{normalize}. When a pair of points colored by the same root $e_a$ collide, the fiber coordinates behave as in \eqref{cc1}. When points colored by $e_a$ and $e_b$, with $\langle e_a,e_b\rangle\neq 0$, collide,
\beq\label{cc2f}
u_a^{m_a}u_b^{m_b}\sim x_{ab}\,(y_a-y_b)^{-\langle e_a,e_b\rangle},\qquad x_{ab}\neq 0,
\eeq
and when a point colored by $e_a$ approaches a marked point $a_i$ colored by $\mu_i$,
\beq\label{cc3f}
u_a^{m_a}\sim x_{a}\,(y_a-a_i)^{\langle e_a,\mu_i\rangle},\qquad x_{a}\neq 0.
\eeq
For roots of the same length, \eqref{cc2f} is equivalent to \eqref{cc2}.

The relation \eqref{cc2f} for a long root $e_a$ and a short root $e_b$ follows from folding. A point colored by $e_b$ meets $m$ points of ${\cal Y}_0$ at once, one for each root in the $H$-orbit corresponding to $e_a$, which are forced by $h$-invariance to coincide. At such a collision, the functions $x_{ab}$ in \eqref{cc2} are not separately non-vanishing; only the product over the orbit is. Restricted to the fixed locus, where the $m$ fiber coordinates of the orbit agree, this gives $u_a^{m}u_b\sim x_{ab}(y_a-y_b)^{m}$. This is the local geometry of the $G'$ open Zastava space \cite{ALSZ}.

\subsubsection{}
To compute ${\cal W}={\cal W}_0|_{\cal Y}$, recall that the superpotential ${\cal W}_0$ from \eqref{supUPb} is the sum of the fiber and base superpotentials:
\beq\label{W0split}
{\cal W}_0 = W_{F,0} +W_0.
\eeq
The fiberwise superpotential \eqref{WF} is
$$W_{F,0}=\sum_{a } \sum_{\alpha} \bigl( \lambda_0 \log (u'_{a, \alpha})^{-1} + u'_{a, \alpha} \bigr),
$$
and the base superpotential \eqref{Wbase} is
$$W_0 = \sum_{a} \sum_{\alpha}\bigl(\lambda_a  \log y'_{a, \alpha} + \lambda_0\log  f_{a, \alpha}(y') \bigr),
$$
where
\beq\label{dressedag}
f_{a, \alpha} (y')=  {\displaystyle\prod_{ i}\;\;\;(1- a_i/y'_{a,\alpha})^{\langle e_a, \mu_i \rangle}\over \displaystyle \prod_{(b,\beta) \neq (a, \alpha)} \!\!\!\!(1- y'_{b,\beta}/y'_{a,\alpha})^{\langle e_a, e_b\rangle/2}},
\eeq
and where $e_a$ are the simple roots of ${\fg}_0'$. Written in this form, $W_0$ agrees with \eqref{Wbase} up to a shift of the $\lambda_a$ and an additive constant.

Next, we identify $(y'_{a, \alpha}, u'_{a, \alpha})$ which are in long orbits of the $H$ action. The long orbits have length $m$, where $m$ is the lacing number of ${\fg}$; the short orbits have length $1.$ Since $h$ acts by permutations, restricting to the $h$-invariant sector means identifying the fields in a single orbit of the $h$-action. Contributions to ${\cal W}$ associated with long (short) roots of ${\fg}$ come from contributions to ${\cal W}_0$ which are in long (short) orbits of $h$.

It follows by inspection that, written in terms of the local $(u,y)$ coordinates, the superpotential becomes
\beq\label{conjpot}
{\cal W} =W_F+W,
\eeq
where
\beq\label{conjpotF}\begin{aligned}
W_F = &\sum_{a , short} \sum_{\alpha} \bigl(\lambda_0 \log u_{a, \alpha}^{-1}
+ u_{a,\alpha} \bigr) \\
+m &\sum_{a , long} \sum_{\alpha} \bigl(\lambda_0 \log u_{a, \alpha}^{-1}
+ u_{a,\alpha}\bigr).
\end{aligned}
\eeq
In \eqref{conjpotF}, ``long" and ``short" refer to the lengths of orbits of $H$.
This is the same as the length of roots of ${\fg}$: any term in which the long root appears contributes $m$ times what it would have, were it only to involve short roots.

The downstairs potential $W$ is given by the same formula as $W_0$,
\beq\label{Wfold}
W = \sum_{a} \sum_{\alpha}\bigl(\lambda_a  \log y_{a, \alpha} + \lambda_0\log  f_{a, \alpha}(y) \bigr),
\eeq
with $f_{a,\alpha}(y)$ as in \eqref{dressedag},
$$
f_{a,\alpha} (y)= {\displaystyle\prod_{ i}\;\;\;(1- a_i/y_{a,\alpha})^{\langle e_a, \mu_i \rangle}\over \displaystyle \prod_{(b,\beta) \neq (a, \alpha)} \!\!\!\!(1- y_{b,\beta}/y_{a,\alpha})^{\langle e_a, e_b\rangle/2}},
$$
except $e_a$ and $\mu_i$ are now understood as simple roots and highest weights of ${\fg}$, or equivalently the co-roots and co-weights of its Langlands dual Lie algebra ${\fg}'$, and $\lambda_a$ is the sum of the $\lambda$'s of the nodes in the $a$-th orbit. While the formulas look the same, the matrix $\langle e_a, e_b\rangle$ that enters them is different -- it now coincides with the symmetrized Cartan matrix of the non-simply laced Lie algebra ${\fg}$.

In computing $f_{a,\alpha}(y)$ in ${\cal W}$, we used the fact that the simple roots and the fundamental weights of ${\fg}$ coincide with the simple co-roots and the fundamental co-weights of its Langlands dual ${\fg}'={^L{\fg}}$, respectively. The latter are, in turn, simply the sums of the simple co-roots and the fundamental co-weights of ${\fg}_0'$ lying in a single orbit of $H$. The co-weight lattice is the lattice dual to the root lattice; the co-root lattice the dual of the weight lattice.

With \eqref{cc2f} and \eqref{cc3f}, the multivalued terms in ${\cal W}$ are logarithms of functions which are regular and non-vanishing on ${\cal Y}$, as in the simply laced case.
\subsubsection{}\label{normalize}
As it comes out of the computation, $\langle e_a, e_a\rangle$ ends up normalized so that short roots of ${\fg}$ have squared length $2$, whereas long roots have squared length $2m$, with $m$ the lacing number of ${\fg}$. The overall length of the roots is a convention, which depends on the choice of normalization of the inner product, but their relative lengths cannot be changed. For example, from ${\fg}_0'=\mathfrak{su}_{2n}$, which has $2n-1$ simple roots forming $n-1$ orbits of length $2$ and one orbit of length $1$, we end up with ${\fg}=\mathfrak{so}_{2n+1}$, which has $n-1$ long simple roots, of squared length $4$, and one short simple root, of squared length $2$.

Recall that, by definition, $\langle w_a, e_b^{\vee}\rangle = \delta_{ab}$, where $w_a$ is the fundamental weight, and $e_b^{\vee}=2e_b/\langle e_b,e_b\rangle$ is the co-root corresponding to $e_b$. In the current normalization, a short root coincides with its co-root, whereas a long root is $m$ times its co-root. It follows that $\langle e_a, w_a\rangle=1$ if $e_a$ is a short root, and $\langle e_a, w_a\rangle=m$ if it is a long root.

In this normalization, ${\fq}$ is the quantum group parameter associated with short roots: ${\fq}_a={\fq}^{\langle e_a,e_a\rangle/2}$ equals ${\fq}$ for short roots and ${\fq}^m$ for long roots, as in the standard normalization of $U_{\fq}({\fg})$. Correspondingly, $k+h^\vee$ in \eqref{l0def} is replaced by $m(k+h^\vee)$, where $k$ is the level of $\fgh_k$ in the normalization in which long roots have squared length $2$.
\subsubsection{}
As before, the equivariant central charge function of a brane ${\cal L}$ in $({\cal Y},{\cal W})$ is
\beq
{\cal Z}[{\cal L}]=\int_{\cal L}\Omega\, e^{-{\cal W}},
\eeq
where $\Omega=\bigwedge_{a,\alpha}{du_{a,\alpha}/u_{a,\alpha}}\wedge{dy_{a,\alpha}/y_{a,\alpha}}$ is the top holomorphic form on ${\cal Y}$. The vector-valued blocks are obtained as in \eqref{intV}, by inserting at the tip of ${\sf D}$ operators $\Phi_\alpha$ mirror to the states of the $h$-twisted sector, the classes in $H^*_{\rm T}({\cal X}_0^h)=H^*_{\rm T}({\cal X})$.

As in section \ref{downsuperpotential}, the fiber superpotential $W_F$ in \eqref{conjpotF} has a unique nondegenerate critical point, at $u_{a,\alpha}=\lambda_0$, for long and short orbits alike; the factor $m$ only rescales the long-orbit terms. Integrating out the fiber directions gives integrals of the form \eqref{SXa} over the base, with $W$ in \eqref{Wfold}. Since $W$ depends on ${\fg}$ only through its symmetrized Cartan matrix and the highest weights $\mu_i$, these are the integral representations of conformal blocks of $\fgh_k$ of \cite{SV}, which hold for non-simply laced ${\fg}$ as well, with the level as in section \ref{normalize}.
\subsection{Folding and homological mirror symmetry}
The A-side description is based on the Fukaya-Seidel category
\beq
D{\rm Fuk}({\cal Y},{\cal W})
\eeq
of the fixed-point Landau-Ginzburg model $({\cal Y}, {\cal W})$.
The target ${\cal Y}$ is the moduli space of ${\fg}'={}^L\fg$ monopoles, and the superpotential ${\cal W}$ in \eqref{conjpot} depends on $\fg$ through its Weyl-invariant inner product, normalized as in section \ref{normalize}, so that the short roots have squared length $2$ and the long roots squared length $2m$.

\subsubsection{}
In direct generalization of the simply laced case, $D{\rm Fuk}({\cal Y},{\cal W})$ contains branes ${\cal T}_{\cal C}$, each of which is a product of real-line Lagrangians on $\Sigma$, colored by simple roots of $\fg$, paired with real-line Lagrangians in the fibers. Let
\beq
{\cal T}=\bigoplus_{\cal C}{\cal T}_{\cal C}
\eeq
be the sum of all inequivalent such branes, where equivalence is isotopy in the theory with superpotential ${\cal W}$. We define
\beq
\MDY\subset D{\rm Fuk}({\cal Y},{\cal W})
\eeq
to be the subcategory split-generated by the ${\cal T}$-branes. As in the simply-laced case, one expects $\MDY$ to be all of $D{\rm Fuk}({\cal Y},{\cal W})$. Let
\beq
{\mathscr A}={\rm Hom}^{*}_{\MDY}({\cal T},{\cal T})
\eeq
be the endomorphism algebra of ${\cal T}$.

The algebra ${\mathscr A}$ can be computed explicitly. In \cite{ALSZ}, we show that it coincides with the cylindrical KLRW algebra of $\fg$, the cylindrical analogue of the planar KLRW algebra of \cite{KhL, webster}. This gives:

\begin{theorem}[A.-LePage-Shende-Zhou]\label{T1}
$$
{\mathscr D}_{\mathscr A}\simeq \MDY,
$$
where ${\mathscr A}$ is the cylindrical KLRW algebra of $\fg$.
\end{theorem}
\subsubsection{}\label{s-hB}

On the B-side, the category associated to the non-simply laced ${\fg}$ is obtained by folding the category of B-branes on ${\cal X}_0$. We denote it by
\beq\label{hB}
{\mathscr D}_{{\cal X}_0}^{[h]}.
\eeq
The brackets are to indicate that it is not the subcategory of $h$-invariant objects and morphisms.
\subsubsection{}
Recall from section \ref{HMS} that ${\mathscr D}_{{\cal X}_0}=DCoh_{{\rm T}_0}({\cal X}_0)$ is generated by a tilting bundle ${\cal T}_0=\bigoplus_i{\cal T}_{0,i}$, whose endomorphism algebra ${\mathscr A}_0$ is the cylindrical KLRW algebra of ${\fg}'_0$ \cite{W1,W2}.  Mirror symmetry identifies 
$ {\mathscr D}_{{\cal X}_0} \cong {\mathscr D}_{{\mathscr A}_0} \cong {\mathscr D}_{{\cal Y}_0}$ and maps the components of the tilting bundle to ${\cal T}_0$-branes on ${\cal Y}_0$.

The summands ${\cal T}_{0,i}$ of ${\cal T}_0$ are labeled by sequences of simple roots of ${\fg}'_0$, interleaved with the weights of the heavy strings. The automorphism $h$ relabels the summands of ${\cal T}_0$ and permutes the basis of ${\mathscr A}_0$. Exchanging two adjacent orthogonal roots in a label gives an isomorphic summand of ${\cal T}_0$, because the corresponding crossing is invertible.

\subsubsection{}

The generators of the folded category ${\mathscr D}_{{\mathscr X}_0}^{[h]}$ are the $h$-invariant summands of ${\cal T}_0$. A summand is mapped to an isomorphic one exactly when the roots in its label come in complete $H$-orbits, with the roots of each orbit consecutive.

The morphisms of ${\mathscr A}_0^{[h]}$ are spanned by the basis diagrams of ${\mathscr A}_0$ fixed by $h$ individually. This is smaller than the $h$-invariant subspace, as the later contains sums over orbits of diagrams, which get projected out of  ${\mathscr A}_0^{[h]}$.

The composition of morphisms in ${\mathscr D}_{{\cal X}_0}^{[h]}$ is obtained from the composition in ${\mathscr A}_0$, but not naively. The product of two $h$-fixed diagrams in ${\mathscr A}_0$ contains diagrams that are not fixed. Reduce the product of two fixed diagrams to the basis, one relation at a time, applying each relation of ${\mathscr A}_0$ to all strands of an $H$-orbit at once. Of the terms this generates, keep only those in which the same term is chosen on every strand of the orbit. This restricts back to $h$-fixed diagrams. Moreover, give each kept term the sign it carries on a single strand. This gives an associative algebra ${\mathscr A}_0^{[h]}$ \cite{ALSZ2}. Its underlying vector space, the span of the fixed diagrams, is a subspace of ${\mathscr A}_0$, but its product is not the restriction of the product of ${\mathscr A}_0$, so ${\mathscr A}_0^{[h]}$ is not a subalgebra and $({\mathscr D}_{{\cal X}_0})^{[h]}$ is not a subcategory of ${\mathscr D}_{{\cal X}_0}$.\footnote{The folding of KLR algebras in \cite{McNamara} is a different construction, which does not produce the KLR algebra of ${\fg}$.}
\subsubsection{}\label{s-T2}
The two descriptions of the category coincide:
\begin{theorem}[A.-LePage-Shende-Zhou]\label{T2}
$$
{\mathscr D}_{{\cal X}_0}^{[h]}
\simeq{\mathscr D}_{{\mathscr A}_0^{[h]}}
\simeq{\mathscr D}_{\mathscr A}
\simeq\MDY .
$$
\end{theorem}
\noindent The first equivalence holds by construction, since ${\mathscr D}_{{\cal X}_0}^{[h]}$ is generated by the $h$-invariant summands of ${\cal T}_0$ with endomorphism algebra ${\mathscr A}_0^{[h]}$, and the last is Theorem \ref{T1}. The content of the theorem is the isomorphism of algebras
\beq\label{foldA}
{\mathscr A}_0^{[h]}\simeq{\mathscr A},
\eeq
between the folded KLRW algebra of ${\fg}_0'$ and the KLRW algebra of ${\fg}$. Under it, the strands colored by the roots of an $H$-orbit become a single strand colored by the corresponding simple root of ${\fg}$, a dot placed on all of them at once becomes a dot of degree $\langle e_a,e_a\rangle$, and the relations of ${\mathscr A}_0$, reduced by the rule of the previous subsection, become the relations of ${\mathscr A}$ with the symmetrized Cartan matrix of ${\fg}$ \cite{ALSZ2}.

The same folding can be carried out on the A-side, starting from the ${\cal T}_0$-branes on ${\cal Y}_0$, mirror to the summands of the tilting bundle. By construction, the resulting category ${\mathscr D}_{{\cal Y}_0}^{[h]}$ is again ${\mathscr D}_{{\mathscr A}_0^{[h]}}$, so Theorem \ref{T2} identifies it with $\MDY$ as well.  ${\mathscr D}_{{\cal Y}_0}^{[h]}$  should be given a direct geometric interpretation.
\subsubsection{}
In the two geometric descriptions, ${\mathscr D}_{{\cal X}_0}^{[h]}$ and $\MDY$, it is the magnetic Lie algebra ${\fg}'$ that is manifest; in particular, ${\cal Y}$ is the moduli space of ${\fg}'$ monopoles. The algebraic description ${\mathscr D}_{\mathscr A}$ has the electric Lie algebra ${\fg}$ manifest instead, since ${\mathscr A}$ is the cylindrical KLRW algebra of ${\fg}$.

\subsection{Folding and $U_{\fq}(\fg)$ link homology}
The categories $\MDY$ and ${\mathscr D}_{{\cal X}_0}^{[h]}$ make sense for arbitrary simple Lie algebra ${\fg}$, and get related by mirror symmetry, per Theorem \ref{T2}.

To obtain $U_{\fq}(\fg)$ link homology, we need to restrict to Lie algebras ${\fg}$ which admit minuscule representations. For non-simply laced ${\fg}$, these are $B_n$ with the spinor representation and $C_n$ with the defining representation. 

The spinor representation of $B_n$ is obtained by folding the $n$-th antisymmetric representation of ${\fg}_0'=A_{2n-1}$ and the defining representation of $C_n$ by folding the defining representation of ${\fg}_0'=D_{n+1}$, with ${\fg}_0'$ as in \eqref{folding}; see section \ref{s-unknot}.

\subsubsection{}
In $\MDY$, the construction of link homology in section \ref{lsl} carries over to non-simply laced ${\fg}$ without change. Take a link $K\subset M_{3,CS}$ colored by minuscule representations of ${\fg}$. The cup and cap branes ${\cal U}$ and ${\mathscr B}{\cal U}$ in $\MDY$ are obtained from ${\fg}'$ monopole bubbling, exactly as in section \ref{ccs}; the braid group acts by braiding the marked points on $\Sigma$. The link homology is
\beq\label{knot-fixedA}
{\cal H}^{*,*}(K)=Hom^{*,*}_{\MDY}({\mathscr B}{\cal U},{\cal U}).
\eeq
The graded Euler characteristic of ${\cal H}^{*,*}(K)$ is the $U_{\fq}({\fg})$ invariant of $K$, since by section \ref{s-nslKZ} the central charges of branes in $\MDY$ are conformal blocks of $\fgh_k$.
\subsubsection{}

The link homology ${\cal H}^{*,*}(K)$ has the B-side realization, as the subspace of morphisms between $h$-invariant braided cup ${\mathscr B}_0{\cal U}_0$ and cap ${\cal U}_0$ branes of ${\mathscr D}_{{\cal X}_0}$, which are individually fixed by the $h$-action. This defines the space of morphisms between them, viewed as objects of ${\mathscr D}_{{\cal X}_0}^{[h]}$: 
\beq\label{knot-fixedB}
{\cal H}^{*,*}(K)=Hom^{*,*}_{{\mathscr D}_{{\cal X}_0}^{[h]}}({\mathscr B}_0{\cal U}_0,{\cal U}_0).
\eeq
We will give explicit examples in the next section. To compute the space of morphisms, one would need a projective resolution of ${\mathscr B}_0{\cal U}_0$ in terms of $h$-invariant ${\cal T}_0$ branes on ${\cal X}_0$. This is equivalent to working in the 
${\mathscr D}_{{{\mathscr A}_0}^{[h]}}$ category; and equivalent to working in terms of ${\mathscr D}_{{{\cal Y}_0}^{[h]}}$. The latter perspective makes it manifest that, the braid group action taking ${\mathscr U}_0$-brane to ${\mathscr B}_0{\cal U}_0$, preserves $h$-invariance.
\subsubsection{}
Isotopy invariance of ${\cal H}^{*,*}(K)$ defined by \eqref{knot-fixedA} or \eqref{knot-fixedB} can be shown by methods from \cite{ALR}, which would be very worthwhile to do. The task is to prove that the cup branes obey certain slide moves. 

At a conceptual level, isotopy invariance follows as in section \ref{lsl}. Cutting the circle open, the $h$-twist does not act on $M_{4,CS}=M_{3,CS}\times {\mathbb R}_{\tau}$ which remains topological along $M_{4,CS}$; rather $h$ just acts as a projection to a subset of states. Thus, the heavy strings supported on $K\times{\mathbb R}_\tau$ preserve that invariance exactly as before.
\subsubsection{}
The resulting $U_{\fq}(\fg)$ link homology should, as a consequence of Theorem \eqref{T2}, coincide with that defined by Webster in \cite{webster}, in terms of the category of modules of the KLRW algebra (for links embedded in ${\mathbb R}^3$, which is what \cite{webster} corresponds to).

For ${\fg}=B_n$ and $V$ a spinor representation, a different link homology was recently defined in \cite{Elias}. While the approach of \cite{Elias} also uses folding, starting with $\fg_0'=A_{2n-1}$, the resulting homology is bigger.

\section{$U_{\fq}(\fg)$ unknot homology}\label{s-unknot}

As a simple illustration, we compute the homology of the unknot colored by a minuscule representation $V$ of an arbitrary simple Lie algebra $\fg$. For simply laced $\fg$, it was computed in \cite{A1}. The non-simply laced cases, for $\fg=B_n$ with $V$ the spinor representation and $\fg=C_n$, with $V$ the vector representation, are new. 

For the unknot, the link homology can be computed directly, without resolving the branes in terms of ${\cal T}$-branes as in \cite{ALR}. We begin with the A-side computation, which treats all Lie algebras at once and determines both the homological and the equivariant gradings. We then identify the same states on the B-side, starting from the simply laced computation of section 6.3 of \cite{A1}. In the non-simply laced case, the action of $h$ on these states makes Langlands duality explicit.

\subsection{A-side}\label{s-slA}

The unknot homology is the space of endomorphisms of the cup brane ${\cal U}$ in $\MDY$, the category of A-branes on $({\cal Y},{\cal W})$ from sections \ref{sec3} and \ref{s-nsl}. Here ${\cal Y}$ is the moduli space of $G'$ monopoles on $\Sigma\times{\mathbb R}$, where the Lie algebra $\fg'$ of $G'$ is the Langlands dual of $\fg$. As explained in section \ref{EHMS}, this space can be computed as a space of morphisms in $\MDy$,
\beq\label{unknot-mirror}
\operatorname{Hom}^{*,*}_{\MDY}({\cal U},{\cal U})
=\operatorname{Hom}^{*,*}_{\MDy}(E_{\cal U},I_{\cal U}),
\eeq
with the cap represented by the figure-eight brane $E_{\cal U}$ and the cup by the interval brane $I_{\cal U}$. This is \eqref{MPud} with the trivial braid. We will show that all intersection points of the two branes have Maslov gradings of the same parity, so the Floer differential vanishes on degree grounds. Computing the homology then amounts to labeling the intersection points and computing their gradings.

\subsubsection{}
The intersection points 
$${\cal P}_\nu\in E_{\cal U}\cap I_{\cal U}$$
are labeled by pairs of weights $(\nu,\nu^*)$ of $V$ and $V^*$ that add up to zero, $\nu^*=-\nu$. (We work in terms of the roots and weights of $\fg$, whether or not it is simply laced.) The labeling comes about as follows.

The brane $E_{\cal U}$ can be viewed as a complex of branes associated to $V\otimes V^*$, which we will call the stable envelope branes. The stable envelope brane associated to a pair of weights $(\nu,\nu^*)$ is represented in the base by a collection of $U$-shaped branes, colored by simple roots. Those surrounding the marked point labeled by $\mu$, the highest weight of $V$, are ordered according to a sequence of lowering operators producing $\nu$ from $\mu$, and similarly for $V^*$. Since $V$, and hence $V^*$, is minuscule, different sequences are related by exchanges of orthogonal roots, which do not change the brane. This corresponds to the fact the weights of $V$ and $V^*$ have multiplicity one. Stretching $E_{\cal U}$ gives a collection of these branes connected at infinity, as in figure \ref{f_minuscule}; the interpretation in terms of complexes and connected sums is explained in \cite{ALR}.

Each intersection point ${\cal P}_\nu\in E_{\cal U}\cap I_{\cal U}$ selects one such stable envelope brane. Its two weights add up to zero because the charge of the light strings is $\mu+\mu^*=\sum_a d_a e_a$, as in \eqref{chargew}, which is what allows the cup to fuse $V\otimes V^*$ to the trivial representation. Since the weights of $V$ have multiplicity one, there is one intersection point for each weight $\nu$ of $V$.

\subsubsection{}
The relative grading is topological and easy to compute. It is determined by the change of phase of $\Omega e^{-{\cal W}}$ around the boundary of an interpolating disk: $\Omega$ determines the homological grading and $e^{-{\cal W}}$ the equivariant grading \cite{A2}. For this calculation, we may take the disks to be constant in the fiber directions, where the two fiber Lagrangians intersect at one point. It therefore suffices to examine the disk in the base $Y$ and work in terms of $\Omega_Y e^{-{W}}$. When only one colored point moves, we work in $\Sigma$ itself.
\subsubsection{}

\begin{figure}[h]
\begin{center}
\includegraphics[scale=0.45]{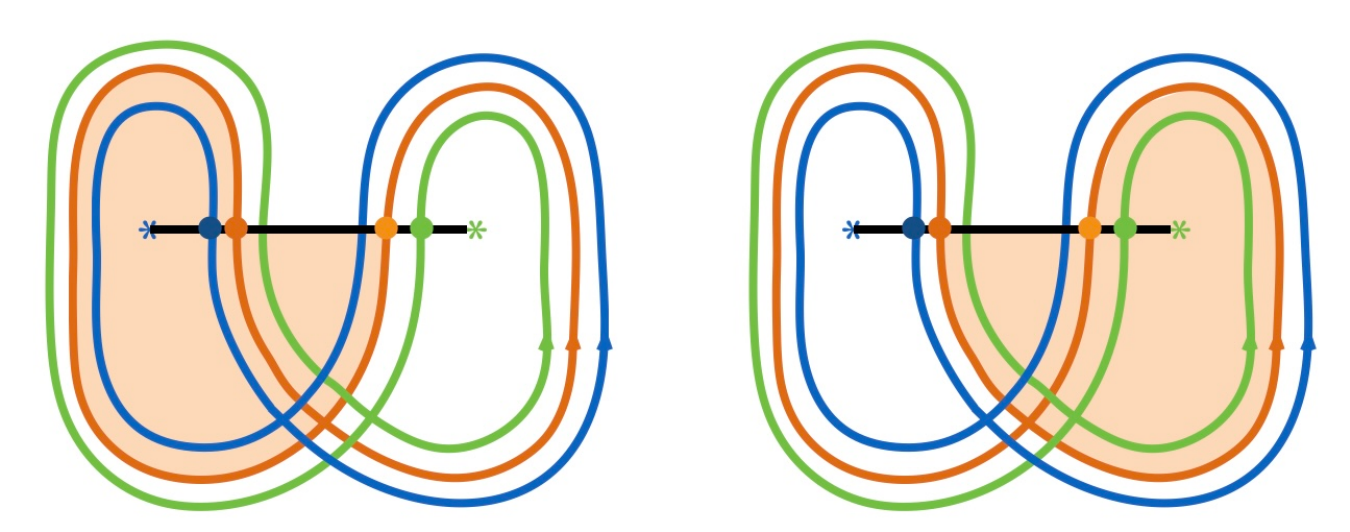}
\caption{Two intersection points that differ by a single simple root, and a pair of disks interpolating between them.}\label{f_minuscule}
\end{center}
\end{figure}

Take a pair of weights of $V$ differing by one simple root,
\beq\label{take}
\nu=\nu'-e_*.
\eeq
The two disks in Fig.\ref{f_minuscule} interpolate between ${\cal P}_{\nu'}$ and ${\cal P}_{\nu}$ by moving the point colored by $e_*$.
Going around the boundary of either disk, the tangent direction of the moving point makes one full turn. The Maslov gradings therefore differ by
\beq\label{Msim}
M({\cal P}_\nu)-M({\cal P}_{\nu'})=2.
\eeq
This does not depend on the length of $e_*$.
\subsubsection{}
The equivariant grading does depend on the root length. The marked point labeled by $\mu$ and the colored points enclosed by the first disk contribute their total charge $\nu'=\nu+e_*$. The monodromy of the superpotential gives
\beq\label{take1}
J({\cal P}_\nu)-J({\cal P}_{\nu'})
=-\langle\nu+e_*,e_*\rangle.
\eeq
The second disk instead encloses the marked point labeled by $\mu^*$ and the colored points around it, of total charge $\nu^*$. It gives
\beq\label{take2}
J({\cal P}_\nu)-J({\cal P}_{\nu'})
=-\langle\nu^*,e_*\rangle,
\qquad \nu^*=-\nu.
\eeq
The minuscule condition ensures that these two answers agree. Since $\nu$ and $\nu+e_*$ are both weights of $V$, 
\beq
\langle\nu+e_*,e_*^\vee\rangle
=\langle-\nu,e_*^\vee\rangle=1,
\qquad
e_*^\vee={2e_*\over\langle e_*,e_*\rangle}.
\eeq
This is because, for every weight $\lambda$ of a minuscule representation,  $\langle\lambda,e_*^\vee\rangle\in\{-1,0,1\}$.
Thus both disks give
\beq\label{jfinal}
J({\cal P}_\nu)-J({\cal P}_{\nu'})
=-{1\over2}\langle e_*,e_*\rangle.
\eeq
With short roots of squared length $2$ and long roots of squared length $2m$, the equivariant degree decreases by $1$ for a short-root lowering operator and by $m$ for a long-root lowering operator. The homological degree increases by $2$ in either case.

\subsubsection{}
The homological and equivariant gradings are
\beq\label{nsM}
M({\cal P}_\nu)=2\langle\rho^\vee,\mu-\nu\rangle,
\eeq
\beq\label{nsJ}
J({\cal P}_\nu)=\langle\rho,\nu\rangle,
\eeq
where $\rho$ and $\rho^\vee$ are the Weyl vector and Weyl covector of $\fg$:
\beq
\rho={1\over2}\sum_{\alpha>0}\alpha,
\qquad
\rho^\vee={1\over2}\sum_{\alpha>0}\alpha^\vee.
\eeq
The gradings $M({\cal P}_\nu)$ and $J({\cal P}_\nu)$ satisfy \eqref{Msim} and \eqref{jfinal} since the Weyl vector and Weyl covector are characterized by $\langle\rho,e_a^\vee\rangle=1$ and $\langle\rho^\vee,e_a\rangle=1$ for every simple root $e_a$. In particular, $M({\cal P}_\nu)$ is twice the number of lowering operators needed to produce $\nu$ from $\mu$.
\subsubsection{}
The gradings of $\operatorname{Hom}^{*,*}_{\MDY}({\cal U},{\cal U})$ do not depend on the choice of grading of ${\cal U}$. The intersection point ${\cal P}_\mu$ is the identity morphism of the brane, and as such its natural Maslov and equivariant degrees are $0$. In \eqref{nsM}, ${\cal P}_\mu$ has Maslov degree zero.

As in \cite{A1}, the shift is chosen so that link homology is invariant under Serre duality, which maps a link $K$ to its mirror image $K^*$. Serre duality identifies  $\operatorname{Hom}_{\MDY}({\mathscr B}{\cal U},{\cal U}[n]\{k\})$ with $\operatorname{Hom}_{\MDY}({\cal U},{\mathscr B}{\cal U}[2D-n]\{-D^\vee-k\})$, where 
$$
D=2\langle\rho^{\vee},\mu\rangle, \qquad D^\vee=2\langle\rho,\mu\rangle.
$$
In \eqref{nsJ}, we gave the equivariant grading an overall shift by $D^\vee/2$ relative to the one in which ${\cal P}_\mu$ has degree zero. The unknot is its own mirror image, so its homology should be invariant under Serre duality. This fixes the shift. Since $w_0$ permutes the weights of $V$ and takes $\rho$ and $\rho^\vee$ to their negatives, the degrees in \eqref{nsM} and \eqref{nsJ} are symmetric under $(M,J)\to(2D-M,-J)$. With the absolute grading of the unknot homology fixed, the absolute gradings of homologies of all other links may be determined by methods described in \cite{ALR}.

\subsubsection{}

Since all Maslov gradings in \eqref{nsM} are even, the Floer differential vanishes on degree grounds. Consequently, the intersection points ${\cal P}_\nu$ labeled by weights $\nu$ of representation $\fg$ are the $U_{\fq}(\fg)$ homology groups
of the unknot $K=\bigcirc$:
\beq\label{ubasic}
{\cal H}^{*,*}(\bigcirc,V)
=\bigoplus_{\nu\in V}
\bigl({\mathbb C}\,{\cal P}_\nu\bigr) 
[ M({\cal P}_\nu)]\{J({\cal P}_\nu)\},
\eeq
with the Maslov and equivariant gradings in \eqref{nsM} and \eqref{nsJ}.
\subsubsection{}
The Euler characteristic of $H^{*,*}(\bigcirc,V)$ is
\beq\label{unknot-euler}
\begin{aligned}
\chi(\bigcirc,V)
&=\sum_{\nu\in V}(-1)^{ M({\cal P}_\nu)}
\fq^{J({\cal P}_\nu)}\\
&=\sum_{\nu\in V}\fq^{\langle\rho,\nu\rangle}
=\operatorname{Tr}_V\fq^\rho
=J_{\bigcirc,V}(\fq).
\end{aligned}
\eeq
This is the $U_{\fq}(\fg)$ invariant of the unknot in blackboard framing, given as $\rm{Tr}_V\fq^\rho$, the quantum dimension of representation $V$ of $\fg$. The computation illustrates the new way of obtaining $U_{\fq}(\fg)$ link invariants, for any $\fg$ and links colored by its minuscule representations: as signed, $\fq$-graded counts of intersection points of branes.
\subsection{B-side, simply laced}\label{s-slB}

Suppose first that $\fg$ is simply laced; the non-simply laced theory is obtained from it by folding. Near the cup brane, the target of the worldsheet sigma model is
\beq
{\cal X}=T^*U,\qquad U=G'/P,
\eeq
where $U$ is the minuscule Grassmannian of $G'$, and $P\subset G'$ is the maximal parabolic subgroup associated, under Langlands duality, to the highest weight $\mu$ of $V$. The cup brane
\beq
{\cal U}={\cal O}_U
\eeq
is the structure sheaf of $U$, supported on the zero section.

\subsubsection{}
The homology of the unknot $K=\bigcirc$ is the space of endomorphisms of ${\cal U}$
$$
{\cal H}^{*,*}(\bigcirc,V)=\operatorname{Hom}^{*,*}_{\MDX}({\cal U},{\cal U})=\bigoplus_{k,j}\operatorname{Hom}_{\MDX}({\cal U},{\cal U}[k]\{j\}),
$$
in $\MDX$, the category of ${\rm T}$-equivariant B-branes on ${\cal X}$. The morphisms can be computed in $T^*U$, since $U$ is embedded in ${\cal X}$ as the zero section of its local $T^*U$ neighborhood. By remark 5.11 of \cite{CK2}, the nonzero morphism groups are
\beq\label{key1}
\operatorname{Hom}_{\MDX}({\cal U},{\cal U}[2j]\{-j\})=H^{j,j}(U),
\eeq
for $0\leq j\leq\dim_{\mathbb C}U$, with the identity morphism in degree zero.\footnote{Equation (6.3) of \cite{A1} has the opposite sign of the equivariant degree; \eqref{key1} is what follows from remark 5.11 of \cite{CK1}, with the fibres of $T^*U$ scaling with $\fq$.}

By \eqref{key1}, the class labeled by $\nu$ has homological degree $2j(\nu)$, equal to the Maslov degree $M({\cal P}_\nu)$ in \eqref{nsM}, and equivariant degree $-j(\nu)$.

\subsubsection{}
Thus the unknot homology is the cohomology of the minuscule Grassmannian $U$.
In section 6.3.2 of \cite{A1} $H^*(U)$ is computed by Morse theory. 

A generic one-parameter subgroup of ${\rm T}$ defines a Morse function whose critical points are the torus fixed points. (The associated attracting cells are known as the Schubert cells. The Poincar\'e duals of their closures form the Schubert basis of $H^*(U)$.) 
The torus fixed points are labeled by the weights $\nu$ of $V$, just as the intersection points ${\cal P}_\nu$ on the A-side. We choose the one-parameter subgroup so that the fixed point labeled by the highest weight $\mu$ is the minimum of the Morse function, to match the A-side conventions. Then, the class labeled by $\nu$ lies in $H^{j,j}(U)$ with
\beq\label{jform}
j(\nu)={1\over2}D-\langle\rho^\vee, \nu\rangle = \langle\rho^\vee, \mu-\nu\rangle ,
\eeq
where $\rho^{\vee}$ is the Weyl covector, and $D=\dim_{\mathbb C}U=2\langle\rho^\vee, \mu\rangle $. Since we took $\fg$ to be a simply laced Lie algebra, the Weyl covector and vector agree
\beq\label{re}
\rho^\vee=\rho.
\eeq
Hence, the unknot homology, given by \eqref{key1} and \eqref{jform}, agrees with what we had found on the A-side, with the homological and equivariant degrees written in \eqref{nsM} and \eqref{nsJ}, up to an overall degree shift, which is a convention.
\subsubsection{}
The equivariant Euler characteristic of $\MDX$ is given in equation 6.4 of \cite{A1}; with the sign convention in \eqref{key1}, the shift of the equivariant degree there becomes $+D/2$. Applied to the present setting this gives
\beq\label{unknot prediction}
\begin{aligned}
\chi(\bigcirc,V)
&=\sum_{n,k}(-1)^n\,\fq^{k+D/2}\,\dim\operatorname{Hom}_{\MDX}({\cal U},{\cal U}[n]\{k\})\\
&=\operatorname{tr}_V\fq^{\rho^\vee}
=\operatorname{tr}_V\fq^\rho
=J_{\bigcirc,V}(\fq).
\end{aligned}
\eeq
This agrees with the A-side result \eqref{unknot-euler}.

\subsection{B-side, non-simply laced}\label{s-nslB}
Let $V$ be a minuscule representation of $\fg$, obtained by folding an $H$-invariant minuscule representation $V_0$ of ${\fg}'_0$. The highest weight $\mu_0$ of $V_0$ is $H$-invariant and corresponds under folding to the highest weight $\mu$ of $V$, and the $h$-invariant weights of $V_0$ are the weights of $V$. Recall the weights of $\fg$ are the coweights of $\fg'$. Folding identifies these with the $H$-invariant coweights of $\fg'_0$. These are weights of $\fg'_0$, since $\fg'_0$ is simply laced.

\subsubsection{}
The B-side is the folded category ${\mathscr D}_{{\cal X}_0}^{[h]}$ of \eqref{hB}. Before folding, near the cup brane, the target of the worldsheet sigma model and the cup brane are
\beq
{\cal X}_0=T^*U_0,\qquad {\cal U}_0={\cal O}_{U_0},
\eeq
where $U_0=G'_0/P_0$ is the minuscule Grassmannian of $G'_0$, and $P_0\subset G'_0$ is the maximal parabolic subgroup associated to $\mu_0$. Since $\mu_0$ is $h$-invariant, $h$ preserves $P_0$. It therefore acts on $U_0$ and preserves the brane ${\cal U}_0$.

\subsubsection{}
As in section \ref{s-hB}, the morphisms of the folded category ${\mathscr D}_{{\cal X}_0}^{[h]}$ are spanned by the basis elements fixed individually by $h$; here the basis is the Schubert basis, which $h$ permutes. Hence
\beq\label{unknot-fixed}
\operatorname{Hom}^{*,*}_{{\mathscr D}_{{\cal X}_0}^{[h]}}({\cal U}_0,{\cal U}_0)\cong H^*(U_0)^{[h]},
\eeq
where $H^*(U_0)^{[h]}$ denotes the span of the Schubert classes fixed by $h$. It is smaller than the $h$-invariant subspace of $H^*(U_0)$, which also contains sums over nontrivial orbits. 

As a vector space, $H^*(U_0)^{[h]}$ in \eqref{unknot-fixed} is isomorphic to the cohomology of the fixed locus $U=U_0^h$, which is the minuscule Grassmannian of $G'$ associated to $\mu$: $H^*(U_0)^{[h]}\cong H^*(U)$. Indeed, the torus fixed points of $U$ are the $h$-fixed torus fixed points of $U_0$, so both spaces have bases labeled by the weights $\nu$ of $V$. This parallels section \ref{s-nslKZ}, where the states of the $h$-twisted sector are the classes on the fixed locus ${\cal X}={\cal X}_0^h$.

\subsubsection{}
As graded vector spaces, however, $H^*(U_0)^{[h]}$ and $H^*(U)$ differ. This parallels the fact that the KZ connection of $\fg$ differs from the ordinary quantum connection of ${\cal X}$.

At the fixed point labeled by $\nu$, the repelling directions of $U_0$ come in $H$-orbits. On $U_0$, by \eqref{key1}, a repelling direction adds $2$ to the cohomological degree and $-1$ to the equivariant degree. An orbit of length $m$ of repelling directions on $U_0$ therefore adds $(2m,-m)$. On $U_0^h$, the length $m$ orbit is a single repelling direction, and adds $(2,-1)$. In the folded category, as we now show, the orbit adds $(2,-m)$: $2$ to the homological degree, as on $U_0^h$, and $-m$ to the equivariant degree, as on $U_0$.

The equivariant degree is inherited from $U_0$, since the morphisms of the folded category are spanned by morphisms of ${\mathscr D}_{{\cal X}_0}$, with the same equivariant degrees, see section \ref{s-hB}. By \eqref{jform}, applied to $U_0$, the class labeled by $\nu$ has equivariant degree $-\langle\rho_0,\mu_0-\nu\rangle$, minus the number of repelling directions, where $\rho_0$ is the Weyl vector of $\fg'_0$. This agrees with the equivariant degree on the A-side, $-\langle\rho,\mu-\nu\rangle$ from \eqref{nsJ} before the shift. Under folding, $\mu_0$ and the $h$-fixed weights $\nu$ of $V_0$ are identified with the highest weight $\mu$ and the weights $\nu$ of $V$; with the normalization of section \ref{normalize}, the identification preserves the inner product. Moreover, the Weyl vector $\rho_0$ of $\fg'_0$ coincides with the Weyl vector of $\fg$:
\beq\label{unknot-rho}
\rho_0=\rho.
\eeq
Indeed, a Weyl vector is the sum of the fundamental weights of the corresponding Lie algebra. The Weyl vector $\rho_0$ of $\fg'_0$ is $h$-invariant, and the sums of the fundamental weights of $\fg'_0$ over $H$-orbits are the fundamental coweights of $\fg'$, see section \ref{osa}. These are the fundamental weights of $\fg$, so $\rho_0$ is their sum, $\rho$. Hence $\langle\rho_0,\mu_0-\nu\rangle=\langle\rho,\mu-\nu\rangle$.

The homological degree is not inherited from $U_0$. By Theorem \ref{T2}, the folded category is equivalent to $\MDY$, with ${\cal U}_0$ corresponding to the cup brane ${\cal U}$ of section \ref{s-unknot}. The class labeled by $\nu$ therefore has the homological degree \eqref{nsM}, $2\langle\rho^\vee,\mu-\nu\rangle$. By \eqref{jform}, applied to $U_0^h$, this is twice the codimension of the Schubert cell of $\nu$ in $U_0^h$, that is, twice the number of $H$-orbits of repelling directions.

The folded B-side therefore recovers the states and gradings found on the A-side, up to the overall shift of the equivariant degree.

\subsubsection{}

The equivariant Euler characteristic of $\operatorname{Hom}^{*,*}_{{\mathscr D}_{{\cal X}_0}^{[h]}}({\cal U}_0,{\cal U}_0)$ is the twisted trace
\beq\label{twistedt}
\chi({\cal U}_0,{\cal U}_0)^{[h]}=\operatorname{tr}_{V_0}h\,\fq^{\rho_0}.
\eeq
Only the individually fixed classes contribute to the Euler characteristic, since $h$ acts by permuting the Schubert basis of $H^*(U_0)$ as permutes the weights of $V_0$.
Because of the identification in \eqref{unknot-rho},
\beq\label{unknot-twisted}
\chi({\cal U}_0,{\cal U}_0)^{[h]}=\operatorname{Tr}_{V_0}h\,\fq^{\rho_0}=\operatorname{Tr}_V\fq^\rho=J_{\bigcirc,V}(\fq).
\eeq 
In \eqref{unknot-twisted} we shifted the equivariant degree is shifted as on $U_0$, by ${1\over 2}\dim_{\mathbb C}U_0=\langle\rho_0,\mu_0\rangle = \langle\rho,\mu\rangle$. This is distinct from ${1\over2}\dim_{\mathbb C}U$.

The equality of the twisted and ordinary characters is the twining-character identity \cite{orbit}. It holds for any $H$-invariant highest weight; here we use the minuscule case, for which the states are realized by the cup brane above.

\subsubsection{}
A different categorification of the spinor-colored $\mathfrak{so}_{2n+1}$ invariant is due to Bodish, Elias and Rose \cite{Elias}. It starts from the same unfolded data, the ${\fg}'_0=\mathfrak{su}_{2n}$ homology of a link colored by the $n$-th antisymmetric representation $V'_0$, and equips it with an involution whose eigenspaces are link invariants. The $\mathfrak{so}_{2n+1}$ invariant is the Euler characteristic of the $+1$ eigenspace minus that of the $-1$ eigenspace. This is proven for $n\leq3$, and relies on conjectures for $n\geq4$.

For $n\geq2$, the homology of \cite{Elias} differs from ours already for the unknot. Ours is $H^*(U_0)^{[h]}$, of dimension $2^n$. Their unknot homology is the entire $V'_0$-colored $\mathfrak{su}_{2n}$ homology of the unknot, $H^*({\rm Gr}(n,2n))=H^*(U_0)$, of dimension $\binom{2n}{n}$, split into eigenspaces of dimensions $\frac12\big(\binom{2n}{n}\pm2^n\big)$. 

More precisely, \cite{Elias} start with $\mathfrak{su}_{2n}$ Khovanov-Rozansky homology \cite{KhR} colored by any antisymmetric representation \cite{Wu, Yonezawa}. The homologies from \cite{KhR, Wu, Yonezawa} are expected to be equivalent to our link homology in type ${\fg}=A_{2n-1}$, as they have been shown to be equivalent to Webster's categorification \cite{MackaayWebster}.

 \section{General $\Sigma$}\label{GS}

The construction generalizes to categorified $U_{\fq}(\fg)$ invariants of links in
\beq\label{gen3m}
M_{3,CS} = \Sigma\times{\mathbb R}_{CS},
\eeq
where $\Sigma$ is a non-compact, but otherwise arbitrary, Riemann surface, and $\fg$ is simply laced or obtained by folding as in section \ref{s-nsl}. We start with the $(2,0)$ conformal theory on $M_6$, as in \eqref{6mani}, and replace $\Sigma={\mathbb R}\times S^1(R)$ there by the more general surface.

\subsection{The setup}\label{2bt}

To preserve supersymmetry we apply the twist from section \ref{tt}, specialized to $M_{4, CS}=\Sigma\times {\sf D}$. The spin connection of $\Sigma$ is canceled by an $SO(2)_R$ twist, which rotates two of the five scalar fields in \eqref{mod}. Given the choices we have already made, the only $SO(2)$ R-symmetry available for twisting is the one under which
\beq\label{ts}
\varphi_a(y)=\int_{{\sf C}_a(y)}\left(m_s^4\omega_I/g_B+im_s^2 B_{\rm RR}\right)
\eeq
has charge $1$. Here ${\sf C}_a(y)$ is the cycle ${\sf C}_a$ over the point $y\in\Sigma$.

\subsubsection{}

Namely, we have abelianized the $(2,0)$ theory by working in the background where the scalar
$$\int_{{\sf C}_a}m_s^2B_{\rm NS}/g_B > 0$$
is turned on.
This gives the self-dual strings non-zero tension $M_a^2$, from \eqref{tension}. In the IIA frame, it becomes the monopole field $\Psi_a$ of section \ref{s-monopoles}.

We have equipped the theory with the ${\fq}$-grading, by identifying $SO(2)$ acting by rotations of ${\sf S}={\mathbb C}$ with $SO(2)_R$ symmetry acting on the complex scalars 
$$\int_{{\sf C}_a} \omega_{\sf R}^{2,0},
$$
as in section \ref{firsttwist},
where $\omega_{\sf R}^{2,0}=\omega_J+i\omega_K$ is the holomorphic $(2,0)$ form of the surface ${\sf R}$.

Having made these choices, the only available pair of scalars is the complex combination in \eqref{ts}. The twist exists only in the $(2,0)$ conformal theory. In IIB string theory and in the $(2,0)$ little string theory, the periods of $B_{\rm RR}$ are periodic scalars, so no continuous symmetry rotates them into the scalar coming from $\omega_I$. In the conformal limit, $m_s^2\int_{{\sf C}_a}B_{\rm RR}$ decompactifies, and this rotation becomes part of $SO(5)_R$.
\subsubsection{}
The twisted compactification of the $(2,0)$ theory on $\Sigma$ turns $\varphi_a(y)$ from \eqref{ts} into a one-form
$$\varphi_a=\varphi_a(y)\,dy,$$
where $y$ is a local coordinate on $\Sigma$. More invariantly, $\varphi_a$ is a section of the canonical bundle $K_\Sigma$.
Since the bulk $(2,0)$ theory is abelianized, $\varphi_a$ is not dynamical in six dimensions, but part of specifying the background.

 At the outset, we set $\varphi_a$ to zero, and describe the moduli space of self-dual strings in this background in section \ref{s-Ysigma}. In section \ref{s-Wsigma} we turn it back on. As we will see, it adds to the effective superpotential on self-dual strings.

\subsubsection{}\label{pi}

In what follows, $\Sigma$ is a non-compact surface of genus $g$, obtained from a compact Riemann surface $\overline{\Sigma}$ by deleting a nonempty set of points,
\beq\label{ns}
\Sigma = \overline{\Sigma}\backslash \{p_I\}.
\eeq
The deleted points will be associated to Verma module representations, as $0$ and $\infty$ were in $\Sigma = {\mathbb C}^{\times}$. We also pick a basis of $H_1(\overline \Sigma, {\mathbb Z})$, a polarization, and a collection of $g$ $A$-cycles on $\Sigma$, $\#(A_k, A_\ell) =0$. It is convenient to pick out one of the deleted points and call it $p_*$, the point at infinity, and to call the remaining ones, which we continue to denote by $p_I$, the finite points.

\subsubsection{}\label{s-phibackground}

The background value of $\varphi_a$ is a meromorphic one-form on $\overline{\Sigma}$, holomorphic on $\Sigma$, with at most simple poles at the deleted points. To specify it uniquely (see \cite{GH} Ch 2.2), we pick its residues and fix its periods around the $A$-cycles. Thus we choose ${\rm rk}\,{\fg}$ complex numbers $\lambda_{a,I}$ for each finite point and $\lambda^{A}_{a,k}$ for each $A_k$ cycle, and set
\beq\label{data}
{1\over 2\pi i}\oint_{\gamma_{p_I}} \varphi_a = \lambda_{a,I}, \qquad {1\over 2\pi i}\oint_{A_k} \varphi_a = \lambda^{A}_{a,k},
\eeq
where $\gamma_{p_I}$ is a small loop encircling the point $p_I$. Since the sum of the residues of a meromorphic one-form on a compact Riemann surface vanishes, the residue at $p_*$ is determined by the residues at the finite points:
\beq\label{pinfty}
{1\over 2\pi i}\oint_{\gamma_{p_*}} \varphi_a = \lambda_{a,*}=-\sum_I \lambda_{a,I},
\eeq
where $\gamma_{p_*}$ is a small loop around $p_*$.

\subsubsection{}
The first set, the $\lambda_{a, I}$ for each $p_I$, will be the highest weight of a Verma module inserted there. For $\Sigma={\mathbb C}^{\times}$, the background value was $\varphi_a=\lambda_a\,dy/y$, where $\lambda_a$ is the highest weight of the Verma module at $y=0$, as in \eqref{Wbase}. 

The second set, the $\lambda^{A}_{a,k}$, will correspond to the holonomies of the Chern-Simons gauge field around the cycles $A_k$. In Chern-Simons theory, the holonomies around the other half of the cycles are their conjugate variables.

\subsection{Self-dual strings as monopoles on ${\Sigma}\times{\mathbb R}$}\label{s-Ysigma}

We follow the strategy of sections 4 and 5, to describe the $(2,0)$ theory on $M_6$ in terms of theories on its self-dual strings, probing $M_4=\Sigma\times {\sf S} \subset M_6$. 

\subsubsection{}
To derive the description of the sigma model, we choose the second of the two T-dualities of section \ref{ss-limit}, which exists for any $\Sigma$. As we did there, replace ${\sf S}={\mathbb C}$ by ${\mathbb C}^{\times}={\mathbb R}_t\times S^1(R^{\vee})$. T-dualizing $S^1(R^{\vee})$ turns the theory on the self-dual strings into a non-linear sigma model on
\beq
{\cal Y}={\cal Y}(\Sigma).
\eeq
${\cal Y}(\Sigma)$ appears in the geometric Langlands program, as the open  $G'$ Zastava space of $\Sigma$ \cite{Gaitsgory08, Raskin}. As before, $G'$ is the Lie group with Lie algebra $\fg'$ Langlands dual of $\fg$. $G'$ is of adjoint type; its Langlands dual is the Chern-Simons gauge group $G$. One should regard ${\cal Y}(\Sigma)$ as
the moduli space of $G'$ monopoles on
\beq\label{monman}
M_3=\Sigma \times {\mathbb R}_t,
\eeq
where the Lie algebra ${\fg}'$ of $G'$ is the Langlands dual of ${\fg}$. We will describe ${\cal Y}(\Sigma)$ only as a complex manifold. 

 A mirror description for general $\Sigma$ should exist, as in mirror symmetry for higher genus curves \cite{Seidelg}. We will not pursue it here.

\subsubsection{}
The fact that ${\cal Y}(\Sigma)$ should be the moduli space of monopoles follows as in section \ref{s-monopoles}. At energies below $1/R^{\vee}$, the $(2,0)$ theory on $S^1(R^{\vee})$ is described by a five-dimensional gauge theory with gauge group $G'$ and coupling $1/g_{5d}^2=1/R^{\vee}$, on a five manifold obtained by forgetting $S^1(R^{\vee})$.

The self-dual strings probing $M_4 = \Sigma\times {\sf S}$ carry magnetic charges on $M_3 =\Sigma \times {\mathbb R}_t$ and are mutually supersymmetric. The latter is the case because the supersymmetry the strings preserve is determined by the background in which only the monopole field $\Psi_a>0$ from \eqref{monopolefield} is turned on. As a result of supersymmetry, the corresponding bulk fields solve the Bogomolny equation \eqref{bogomolny}, now on $M_3$. (The three-manifold $M_3$ should be distinguished from $M_{3,CS}$ in \eqref{gen3m}, since their origin in $M_6$ and their roles are different.) 

We first describe ${\cal Y}(\Sigma)$ as a complex manifold, and then explain, starting from the Bogomolny equation, why it has the factorization property used in the description. For details, see \cite{ALSZ}.

\subsubsection{}
An important property of the self-dual string theory is locality. 
The interaction of self-dual strings in an open patch $D\subset \Sigma$ may be derived 
from interactions of wrapped D3 branes of IIB string on ${\sf R}$ and taking the little string and conformal limits. The latter come from integrating out massive open strings. Since the winding string modes decouple in the $m_s\rightarrow \infty$ limit, there are no new effects from compact cycles of $\Sigma$.
It follows that for disjoint open sets 
$D, D'$ of $\Sigma$, where $D\cap D'=\emptyset$, the moduli space should factorize:
\beq\label{factoriz}
{\cal Y}(D\cup D')\simeq{\cal Y}(D)\times{\cal Y}(D').
\eeq
The factorization property implies the global geometry of ${\cal Y}(\Sigma)$ is obtained by assembling these local pieces associated to open sets $D\subset \Sigma$ subject to factorization. This fixes the geometry of ${\cal Y}(\Sigma)$ as follows.

\subsubsection{}
${\cal Y}(\Sigma)$ has a projection
\beq
\pi: \;{\cal Y}(\Sigma)\longrightarrow Y(\Sigma)=\prod_{a=1}^{{\rm rk}(\fg)}{\rm Sym}^{d_a}(\Sigma),
\eeq
which records the positions of strings as divisors on $\Sigma$, see section \ref{ybf}. The divisor is a collection of $d=\sum_a d_a$ points on $\Sigma$, of which $d_a$ are colored by the simple coroot $e_a$ of ${\fg}'$, equivalently the simple root of $\fg$, the charge of the self-dual string.

Over a divisor with distinct points, the fiber of ${\cal Y}(\Sigma)$ is
$$
Y_F=(\mathbb C^\times)^d.
$$
The radial direction in each $\mathbb C^\times$ factor comes from the position of the corresponding self-dual string along ${\mathbb R}_t\subset{\sf S}^\times$. Its circle and the circle in ${\sf S}^{\times}$ are T-dual.

\subsubsection{}
In the complement of $\Delta\subset Y(\Sigma)$, where $\Delta$ is the divisor where additional light open strings stretching between D3 branes become massless, the fibration has the simple description
\beq
{\cal Y}^{o}(\Sigma)=\prod_a{\rm Sym}^{d_a}({\mathbb C}^{\times}\times\Sigma)\setminus \pi^{-1}(\Delta).
\eeq
To obtain the full space ${\cal Y}(\Sigma)$, we fill $\pi^{-1}(\Delta)$ back in, allowing strings to collide on $\Sigma$ (while staying away from the deleted points in $\Sigma = \overline{\Sigma}\backslash \{p_I\}$).

By the factorization property \eqref{factoriz} of ${\cal Y}(\Sigma)$, the fibers over a component of $\Delta\subset Y(\Sigma)$ corresponding to a pair of colored points coming together on $\Sigma$ 
are fixed by the local behavior of the fibers over an open coordinate patch $D\subset\Sigma$ containing them, which we may model by ${\mathbb C}$. The local geometry is therefore the same as in section \ref{ccoord}. The coordinates $(u,y)$ used below give a local trivialization of the fibration; in particular, the $u$'s are not globally defined coordinates on ${\cal Y}(\Sigma)$. The description does not depend on the choice of the local coordinate $y$ on the patch: under a change of coordinate, differences such as $y_a-y_b$ change by a non-vanishing factor, which is absorbed by the non-vanishing functions $x$ below, with the $u$'s unchanged. Thus the descriptions on overlapping patches agree. There are three cases to consider.
\subsubsection{}
Let a pair of points colored by the same root $e_a$ come together, with coordinates $y_a$ and $y'_a$. The fiber coordinates $u_a$ and $u'_a$, which are ${\mathbb C}^{\times}$-valued away from the collision, may go to infinity as
\beq\label{c1}
u_a={x_a\over y_a-y'_a},\qquad u'_a={x'_a\over y'_a-y_a},
\eeq
for $x_a,x'_a\in{\mathbb C}^{\times}$, while
\beq
u_a+u'_a={x_a-x'_a\over y_a-y'_a}
\eeq
remains finite, which requires $x_a-x'_a$ to vanish at $y_a=y'_a$.

\subsubsection{}
Next, suppose that a pair of points colored by different roots $e_a$ and $e_b$ come together at $y_a=y_b$. Then
\beq\label{c2}
u_a^{m_a}u_b^{m_b}\sim x_ax_b(y_a-y_b)^{-\langle e_a,e_b\rangle},
\eeq
where $x_ax_b\neq0$ and
$
m_a={\langle e_a,e_a\rangle\over 2}.$ Here $m_a=1$ for a short root and $m_a=m$ for a long root, where $m$ is the lacing number.

If the roots are orthogonal, $\langle e_a,e_b\rangle=0$, fibers do not degenerate over the collision point. If the roots are not orthogonal and have the same length, the degeneration over $y_a=y_b$ is as in the simply laced case, since then $m_a=m_b =-\langle e_a,e_b\rangle$. If they have different lengths and are not orthogonal, taking $e_a$ long and $e_b$ short gives $m_a=m$, $m_b=1$, and $\langle e_a,e_b\rangle=-m$.

\subsubsection{}
Finally, when a point colored by $e_a$, at $y_a$, approaches a marked point $q_i$ colored by a representation of highest weight $w_i$, the local coordinates satisfy
\beq\label{c3}
u_a^{m_a}=x_a(y_a-q_i)^{\langle e_a,w_i\rangle},
\qquad x_a\neq0.
\eeq
The appearance of the marked points in this formula does not mean that ${\cal Y}(\Sigma)$ is the moduli space of smooth monopoles in the presence of singular ones. All the singular monopoles are located at $t\rightarrow -\infty$, the end of ${\mathbb R}_t$ corresponding to the origin of ${\sf S}$, for the same reason as in section \ref{s-heavyinf}: they are T-dual to defects at the origin of ${\sf S}$. Their positions on $\Sigma$ only enter the local $u$-coordinates and the superpotential below.

\subsubsection{}
We pause to explain the origin of factorization in \eqref{factoriz}
from the perspective of the Bogomolny equation in \eqref{bogomolny}. 

Recall from section \ref{s-monopoles} that we can describe the monopole moduli space, as a complex manifold, by holomorphic scattering data on \(\Sigma\). The data come from solving $\mathcal D_tV=0$, with $\mathcal D_t=D_t-i\Psi$, along the line $\{y\}\times{\mathbb R}_t$, for each $y\in\Sigma$, where $V$ is a section of the bundle associated to a representation of ${\fg}'$. The equation transports $V$ along \(t\), and the complex equation in \eqref{twoe} ensures that this transport preserves the holomorphic structure defined by $D_{\bar y}$.

For \(G'=SU(2)\), with $V$ in the two-dimensional representation, the solutions of $\mathcal D_tV=0$ form a holomorphic rank-two bundle \(E\) over $\Sigma$. The solutions decaying at \(t\to\pm\infty\) determine a pair of line sub-bundles \(L_\pm\subset E\), the scattering lines. The monopoles are located at the points of $\Sigma$ where the two lines fail to be transverse. Asymptotic framing at $t\rightarrow -\infty$ fixes the normalization of the solutions decaying there, and so identifies \(L_-\) with a fixed line bundle, the same for all monopole configurations. Since \(\det E\simeq{\cal O}\), the projection of $L_+$ to the quotient $E/L_-$ is a map
\[
b:L_+\longrightarrow E/L_-\simeq L_-^{-1},
\]
which vanishes exactly where $L_+$ coincides with $L_-$. Its zero divisor \(P\) is the divisor of positions of the monopoles on $\Sigma$.

Away from monopoles, $L_\pm$ are transverse, so that $E=L_-\oplus L_+$. Then the projection $b$ is invertible: every vector in \(L_-^{-1}\) has a unique lift lying in \(L_+\). Thus, on $\Sigma\setminus P$, $E$ with its two scattering lines is canonically isomorphic to $L_-\oplus L_-^{-1}$ with its two summands. We will refer to $L_-\oplus L_-^{-1}$ as the reference configuration \cite{ALSZ}. The existence of this reference configuration is what allows the data at separated monopole positions to be glued coherently. 

Concretely, given $V_+ = a V_- + b W_-$ in the notation of section \ref{s-monopoles}, where $V_\pm$ are sections of $L_\pm$ and the class $[W_-]$ of $W_-$ is a section of $E/L_-\simeq L_-^{-1}$, the projection maps $V_+$ to $[bW_-]$, so the lift of an arbitrary vector $[cW_-]$ is ${c\over b}V_+$. The frame
\beq\label{locsplit}
(V_-,\; V_+/b)
\eeq
of $E$ is adapted to both lines and involves no choices: $V_-$ is fixed by the framing, and $V_+/b$ does not change when $V_+$ is rescaled. At the points of $P$ the frame is singular:
$$
V_+/b=W_-+S\,V_-,\qquad S=a/b,
$$
where $S$ has poles on $P$. The remaining holomorphic information is therefore how the reference configuration is modified at the points of \(P\). It is encoded in the polar parts of $S$, since changing $W_-$ by a holomorphic multiple of $V_-$ changes $S$ by a holomorphic function. Thus, $S$, a meromorphic section of $L_-^{\otimes 2}$, generalizes the rational map, and $b$ its denominator polynomial, in the description in section \ref{s-monopoles}. 

For general \(G'\), the scattering flags replace the two lines. Where they are transverse, they determine a locally adapted frame up to torus rescaling; the framing removes that freedom. They therefore identify the bundle, with its two flags, with a fixed reference configuration away from the colored divisor \(P\). The divisor \(P\) is the divisor of positions of light self-dual strings/'t Hooft-Polyakov monopoles, of degree $d=\sum_a d_a$. 

No component of the divisor $P$ is allowed to coincide with the deleted points in $\Sigma=\overline{\Sigma}\backslash\{p_I\}$, so both scattering flags extend regularly over the $p_I$'s and remain transverse.
\subsubsection{}
Let $P=P_1+P_2$, with $P_1$ and $P_2$ supported in disjoint disks $D,D'\subset\Sigma$. Choose the scattering data on $D$ and on $D'$ independently: on each disk, a bundle with its two scattering lines, with monopoles at $P_1$ and at $P_2$, respectively. The bundles on $D\setminus P_1$ and $D'\setminus P_2$, with their scattering lines, have a canonical identification, as we explained above, with the reference configuration. These identifications glue the bundles on both disks to the reference configuration on $\Sigma\setminus P$, with no further choices. This constructs the global bundle $E$, together with its scattering lines. Independent choices of scattering data on disjoint disks give the factorization property \eqref{factoriz}.

\subsubsection{}
The fiber coordinates $u$ in \eqref{c1}--\eqref{c3} are related to the scattering data as follows. For $G'=SU(2)$, $S$ is a meromorphic section of $L_-^{\otimes 2}$, with a simple pole at each point $y_i$ of $P$ when these points are distinct. For the coefficients of its poles to be numbers, which do not depend on the choice of a local coordinate on $\Sigma$, $S$ has to be a one-form. We therefore take the fixed line bundle of the framing to be such that
\beq\label{frameK}
L_-^{\otimes 2}\simeq K_\Sigma\otimes{\cal O}\Bigl(-\sum_j\langle e,w_j\rangle\,q_j\Bigr),
\eeq
the bundle of one-forms on $\Sigma$ which vanish to order $\langle e,w_j\rangle$ at the marked points $q_j$, where $e$ is the simple root. Then
\beq\label{ures}
u_i={\rm Res}_{y=y_i}\,S.
\eeq
The zeros of $S$ at the marked points are how the positions of the singular monopoles at $t\to-\infty$ enter the $u$-coordinates; they give \eqref{c3} \cite{ALSZ}. In a local coordinate $y$, away from the marked points,
$$
S=\sum_i {u_i\over y-y_i} dy+\ldots,
$$
where the omitted terms are holomorphic, as in section \ref{s-monopoles}. For general $G'$, there is one such one-form $S_a$ for each simple root $e_a$, with zeros of order $\langle e_a,w_j\rangle/m_a$ at the $q_j$, as in \eqref{c3}.

\subsubsection{}
The string construction predicts that \(\mathcal Y(\Sigma)\), equipped with the appropriate metric, is the moduli space of $G'$ monopoles on $M_3=\Sigma\times{\mathbb R}_t$. It implies that the real equation in \eqref{twoe} has a solution for every point of ${\cal Y}(\Sigma)$, with suitable boundary conditions inherited from the string construction. The analysis of the real equation, and the construction of the Kahler metric on ${\cal Y}(\Sigma)$ that its solutions would give, are left for another day, as they are not needed for our paper.

\subsection{Superpotential on ${\cal Y}(\Sigma)$}\label{s-Wsigma}

The sigma model on ${\cal Y}(\Sigma)$ has a superpotential
\beq
{\cal W}={\cal W}_{\rm top}+{\cal W}_{\rm equiv}.
\eeq
The two terms have different origins.

\subsubsection{}
The topological superpotential ${\cal W}_{\rm top}$ is generated by filling in the origin of ${\sf S}$, as in section \ref{upsuperpotential}. Recall that the fibers $Y_F=({\mathbb C}^{\times})^d$ over $Y(\Sigma)$ are mirror to the positions of the $d$ self-dual strings on ${\sf S}^{\times} = {\sf S}\backslash 0$. The value of ${\cal W}_{\rm top}$ on the fiber over a generic point $P\in Y(\Sigma)$ is
\beq\label{wtg}
{\cal W}_{\rm top}|_P = \sum_i u_i.
\eeq
This generalizes the potential in \eqref{Wtop}. Over less generic points, this extends to a well defined function 
$$ {\cal W}_{\rm top}: {\cal Y}(\Sigma) \rightarrow {\mathbb C},$$
as in section \ref{ccoord}, using the collision formulas: by \eqref{c1}, the sum $u_a+u'_a$ remains finite when two points of the same color collide, and by \eqref{c2} and \eqref{c3}, the $u$'s themselves remain finite in the other two cases. The potential ${\cal W}_{\rm top}$ is known as the canonical map of \cite{Raskin}.

\subsubsection{}
The equivariant superpotential ${\cal W}_{\rm equiv}$ contains the logarithmic, multivalued terms in ${\cal W}$. As in section \ref{upsuperpotentiald}, it is the sum of a term that depends on the fiber coordinates and the base superpotential $W$,
\beq\label{WS}
{\cal W}_{\rm equiv}=\lambda_0\sum_{a,i}m_a\log u_{a,i}^{-1}+W.
\eeq
The first term mirrors the equivariant action on ${\sf S}$, with parameter ${\fq}=e^{2\pi i\lambda_0}$, and is the same for any $\Sigma$. 

The base superpotential $W$ is associated to deforming the theory by turning on $\varphi_a\neq 0$ from section \ref{s-phibackground}. For $\lambda_0\neq 0$, it has additional quantum contributions from interactions between the self-dual strings. They are needed to make ${\cal W}_{\rm equiv}$ regular on ${\cal Y}(\Sigma)$; the first term alone is not, since the $u$'s have zeros and poles over $\Delta \subset {\cal Y}(\Sigma)$.

\subsubsection{}
Take a single self-dual string, arising from a D3 brane wrapping ${\sf C}_a\subset{\sf R}$, placed at a point $y\in\Sigma$. The charge of such a string is the simple root $e_a$ of $\fg$. The effective value of $\varphi_a$ seen by the string is related to ${\cal W}$ via $\varphi_a = d_{y_a} {\cal W}$. When $\Sigma$ is flat and $\varphi_a$ a constant, its contribution to ${\cal W}$ is just the contribution to the (effective twisted) superpotential induced by turning on a complexified real Fayet-Iliopoulos parameter, as in \eqref{firsta}.

For several self-dual strings placed at points $y_{a,i}\in\Sigma$, the net superpotential satisfies
\beq\label{multi}
\varphi_a(y_{a,i})= d_{y_{a,i}} {\cal W},
\eeq
where $d_{y_{a,i}}=dy_{a,i}\,\partial_{y_{a,i}}$ is the differential with respect to the position of one string, with the other strings and the fiber coordinates held fixed. For $\lambda_0\neq 0$, the effective value of $\varphi_a$ seen by a string is not, however, just its background value. It receives additional, quantum contributions, coming from interactions with the other strings. They are determined by locality, and the requirement that $e^{-{\cal W}}$ be defined on ${\cal Y}(\Sigma)$. Since ${\cal W}_{\rm top}$ is already regular, the constraint is on ${\cal W}_{\rm equiv}$.

\subsubsection{}
The effective one-form $\varphi_a$, seen by a string of charge $e_a$, has a first-order pole at the position $y_{b,j}\in\Sigma$ of any other string, of charge $e_b$, with residue
\beq\label{corpot}
{1\over 2\pi i}\oint_{\gamma_{y_{b,j}}}\varphi_a=-\lambda_0\langle e_a,e_b\rangle.
\eeq
It has a pole at the position $q_i$ of every non-compact string of charge $w_i$, with residue
\beq\label{corpot2}
{1\over 2\pi i}\oint_{\gamma_{q_i}}\varphi_a=\lambda_0\langle e_a,w_i\rangle.
\eeq
The residue at infinity, $p_*\in\overline{\Sigma}$, is determined by these and by the topology of $\Sigma$, as we explain in section \ref{s-charge}.

For an open set $D\subset \Sigma$, we derived these contributions in the IIB description by integrating out massive open strings stretching between the corresponding D3 branes and then taking the conformal limit. They extend to $\Sigma$, by locality.

Another way to derive them is by asking for ${\cal W}_{\rm equiv}$ in \eqref{WS} to be regular over $\Delta$, where the $u$'s behave as in \eqref{c1}--\eqref{c3}.

\subsubsection{}\label{s-Gamma}
The superpotential ${\cal W}$ is multivalued, and so is $e^{-{\cal W}}$: when the strings are taken around a closed loop in ${\cal Y}(\Sigma)$, $e^{-{\cal W}}$ comes back multiplied by a factor, its holonomy around the loop. Let
\beq\label{Yhat}
\widehat{\cal Y}(\Sigma)\ \longrightarrow\ {\cal Y}(\Sigma)
\eeq
be the smallest covering space on which $e^{-{\cal W}}$ is single valued, and $\Gamma$ its group of deck transformations. As we will explain below, $\Gamma$ is the group in which the equivariant gradings of ${\mathscr D}_{{\cal Y}(\Sigma)}$ take values.

\subsubsection{}\label{gamma}
The group $\Gamma$ is generated by
\beq\label{gens}
\fq,\qquad \mathfrak{r}_{a,I},\qquad \mathfrak{a}_{a,k},\qquad \mathfrak{b}_{a,k}.
\eeq
The first one, $\fq$, multiplies $e^{-{\cal W}}$ by $\fq=e^{2\pi i\lambda_0}$, and we denote it by the same letter. By \eqref{corpot}, \eqref{corpot2} and \eqref{WS}, $e^{-{\cal W}}$ is multiplied by
\beq\label{srcmult}
\fq^{\langle e_a,e_b\rangle},\qquad \fq^{-\langle e_a,w_j\rangle}\qquad\textup{and}\qquad\fq^{m_a},
\eeq
when a string of charge $e_a$ goes once around a string of charge $e_b$, when it goes once around a marked point $q_j$ of weight $w_j$, and when its fiber coordinate goes once around zero, counterclockwise in each case.

The generators $\mathfrak{r}_{a,I}$, $\mathfrak{a}_{a,k}$ and $\mathfrak{b}_{a,k}$ correspond to loops of a string of charge $e_a$. Take $\mathfrak{r}_{a,I}$ to correspond to a small loop around $p_I$, which multiplies $e^{-{\cal W}}$ by $e^{-2\pi i\lambda_{a,I}}$, by \eqref{data}. Take $\mathfrak{a}_{a,k}$ to correspond to the loop around the $A_k$ cycle which multiplies $e^{-{\cal W}}$ by $e^{-2\pi i\lambda^{A}_{a,k}}$. Fix $\mathfrak{b}_{a,k}$ by choosing a starting point and a specific path in the $B_k$ homology class; another choice changes it by a power of $\fq$. Here $B_k$, for $k=1,\ldots,g$, are the cycles dual to the $A_k$, with intersection numbers $\#(A_k, B_\ell)=\delta_{k\ell}$ and $\#(B_k, B_\ell)=0$.

\subsubsection{}
The generators $\fq$ and $\mathfrak{r}_{a,I}$ commute with everything, the $\mathfrak{a}_{a,k}$'s commute with each other, and so do the $\mathfrak{b}_{a,k}$'s. The holonomies around the $A_k$-cycles and around the $B_{\ell}$-cycles, however, do not commute:
\beq\label{Heis}
\mathfrak{a}_{a,k}\,\mathfrak{b}_{b,\ell}=\fq^{-\langle e_a,e_b\rangle\delta_{k\ell}}\;\mathfrak{b}_{b,\ell}\,\mathfrak{a}_{a,k},
\eeq
where the loops in a product are taken from left to right. Thus, $\Gamma$ is a Heisenberg-type group. It is the central extension, by the powers of $\fq$, of the lattice $H_1(\Sigma,{\mathbb Z})\otimes\Lambda_{{\rm rt},+}$, where $\Lambda_{{\rm rt},+}$ is spanned by the simple roots $e_a$ with $d_a>0$. The extension is defined by the pairing on this lattice which is the intersection pairing on $H_1(\Sigma,{\mathbb Z})$ times the inner product on $\Lambda_{{\rm rt},+}$.

For $\Sigma$ of genus zero, there are no $\mathfrak{a}$'s and $\mathfrak{b}$'s, and $\Gamma$ is commutative. Its generators $\fq$ and $\mathfrak{r}_{a,I}$ correspond to the $\fq$-grading and to the gradings associated to the highest weights of Verma modules in \cite{A2}.

\subsubsection{}\label{s-handles}
The relation \eqref{Heis} comes about as follows. Take a string of charge $e_a$ once around $A_k$, then a string of charge $e_b$ once around $B_k$, and then each of them back. Since $A_k$ and $B_k$ have a point in common, this loop is the first string going once, clockwise, around the second. By \eqref{srcmult}, the corresponding element of $\Gamma$ multiplies $e^{-{\cal W}}$ by $\fq^{-\langle e_a,e_b\rangle}$. For any other pair of cycles, the paths of the two strings do not meet, and the loop is trivial. This leads to \eqref{Heis} for two strings, of charges ${e_a}$ and ${e_b}$.

For a pair of strings of equal charge, this leads to the following subtle, but crucial effect. In ${\cal Y}(\Sigma)$, strings of the same charge are not labeled. Up to powers of $\fq$, we cannot distinguish a loop in which one of the strings goes around the $A_k$ cycle, the other around a $B_\ell$ cycle, from a loop in which one string stays put and the other one traverses both the $A_k$ and $B_{\ell}$ cycles. It follows that \eqref{Heis} holds for one and the same string as well. In a theory with only one string of charge $e_a$, it holds by locality -- the requirement of compatibility with factorization in \eqref{factoriz}.

\subsubsection{}

Because of \eqref{Heis}, $\mathfrak{b}_{b,k}$ cannot multiply $e^{-{\cal W}}$ by a constant. When a string of charge $e_b$ goes once around $B_k$, the pole in \eqref{corpot} crosses $A_k$, so the effective one-form seen by a string of charge $e_a$ changes by
\beq\label{Bshift}
\varphi_a\ \longrightarrow\ \varphi_a-2\pi i\lambda_0\langle e_a,e_b\rangle\,\omega_k,
\eeq
where $\omega_k$ is one of the $g$ holomorphic one-forms on $\overline{\Sigma}$ normalized by 
$$\oint_{A_\ell}\omega_k=\delta_{k\ell}.
$$
This replaces 
$\lambda^{A}_{a,k}$ by $\lambda^{A}_{a,k}-\lambda_0\langle e_a,e_b\rangle.$ The one-form $\varphi_b$, seen by the string that went around $B_k$, changes in the same way, by $-2\pi i\lambda_0\langle e_b,e_b\rangle\,\omega_k$. This is a consequence of \eqref{Heis} for $e_a=e_b$. 

Thus, for $g>0$, $e^{-{\cal W}}$ is a section of a line bundle, rather than a function with constant multipliers.
\subsubsection{}\label{s-charge}
The relation \eqref{Heis} determines the holonomy around $p_*$, and the highest weight $w_*$ of the Verma module placed there. The loop
$$
\mathfrak{l}_{a,k}= \mathfrak{a}_{a,k}\mathfrak{b}_{a,k}\mathfrak{a}_{a,k}^{-1}\mathfrak{b}_{a,k}^{-1}
$$
takes a string of charge $e_a$ around the boundary of the empty $k$-th handle of $\Sigma$. By \eqref{Heis}, the product of these loops over the $g$ handles multiplies $e^{-{\cal W}}$ by $\fq^{-g\langle e_a,e_a\rangle}$.

The product of loops $\prod_{k=1}^g \mathfrak{l}_{a,k}$ around the handles is homotopic to a loop that takes the string once, counterclockwise, around all the other points of $\Sigma$: the deleted points, $p_*$ included, the marked points and the positions of other strings. It follows that the residues of $\varphi_a(y_{a,i})$ at all these points add up to $g\langle e_a,e_a\rangle\lambda_0$. 

Since the residues are determined by the charges, the equality of residues, for every $e_a$, is a consequence of charge conservation. We are learning that for $g>0$, the topology of the surface $\Sigma$ itself is a source of ``background" charge.

\subsubsection{}

Let $w_x$ be the charge at a marked or deleted point $x$, $p_*$ included. If $x$ is a marked point $q_j$, $w_x$ is the highest weight $w_j$ of the finite dimensional representation coloring it, by \eqref{corpot2}; at a finite deleted point $p_I$, it is the highest weight of the Verma module there, with $\lambda_{a,I}=\lambda_0\langle e_a,w_x\rangle$. The residues at these points add up to $\lambda_0\langle e_a,\sum_xw_x\rangle$. The other strings have total charge $\sum_bd_be_b-e_a$, so the residues at them add up to $-\lambda_0\langle e_a,\sum_bd_be_b-e_a\rangle$. The homotopy of the two loops in section \ref{s-charge} implies the residues must add up to $g\langle e_a,e_a\rangle\lambda_0$ for every $e_a$.

It follows from this that the charge conservation for the self-dual strings requires
\beq\label{chargecons}
\sum_x w_x=(2g-2)\rho+\sum_b d_b e_b,
\eeq
where we used $\langle e_a,e_a\rangle=2m_a=2\langle e_a,\rho\rangle$, with $\rho$ the Weyl vector of $\fg$.

\subsubsection{}
In \eqref{chargecons}, $p_*$ is on the same footing as all the other points; we let the weights at the other points and the charges of the strings determine the weight $w_*$ at $p_*$. If $p_*$ is the only deleted point, so that there are no other Verma modules, $w_*=(2g-2)\rho+\sum_ad_ae_a-\sum_jw_j$ is an integral weight.

For a link $K$ in $\Sigma\times{\mathbb R}$, the weights of the marked points, at the ends of the cups and the caps, add up to the charges of the strings, as in \eqref{dranks}, so $w_*=(2g-2)\rho$. Thus, for $g>1$, obtaining link invariants requires an additional string insertion on $\Sigma$, at $p_*$. This is analogous to what happens in Heegaard Floer theory \cite{Osz, Rasmussen}.

For $g>0$ and $\fq\neq 1$, the residues of $\varphi_a(y_{a,i})$ do not add up to zero. It is not a meromorphic one-form on ${\Sigma}$: by \eqref{Bshift}, it is not globally defined, since $e^{-{\cal W}}$ is a section of a line bundle, rather than a function with constant multipliers.

\subsubsection{}\label{s-eW}
The residues and the holonomies above determine $e^{-{\cal W}}$. Let $E(p,q)$ be the prime form \cite{Fay}, the $(-1/2,-1/2)$ differential on $\overline{\Sigma}\times\overline{\Sigma}$ with a simple zero along the diagonal, normalized with respect to the $A$-cycles, and let $\sigma$ be the $g/2$-differential on $\overline{\Sigma}$ without zeros or poles, given in \cite{Fay} by
\beq\label{sigdef}
\sigma(y)=\exp\Bigl(-\sum_{k=1}^g\oint_{A_k}\omega_k(y')\,\log E(y',y)\Bigr).
\eeq
Then
\beq\label{eW}
\begin{aligned}
e^{-{\cal W}}=e^{-{\cal W}_{\rm top}}&\prod_{(a,i)}\biggl[\,e^{-2\pi i\sum_k\lambda^{A}_{a,k}\int_{p_*}^{y_{a,i}}\omega_k}\;\prod_x E(y_{a,i},x)^{-\lambda_0\langle e_a,w_x\rangle}\;\Bigl({u_{a,i}\over\sigma(y_{a,i})^2}\Bigr)^{\lambda_0m_a}\biggr]\\
\times&\prod_{\{(a,i),(b,i')\}}E(y_{a,i},y_{b,i'})^{\lambda_0\langle e_a,e_b\rangle},
\end{aligned}
\eeq
where the first product is over the strings, the product over $x$ is over all the marked points and all the deleted points, with the weights in \eqref{chargecons}, and the last product is over the pairs of distinct strings.

\subsubsection{}
The expression for $e^{-{\cal W}}$ in \eqref{eW} breaks down as follows:
\begin{itemize}
\item[--]
The first factor in the square bracket gives the periods in \eqref{data}, the second the residues at the marked and the deleted points, and the last product the residues \eqref{corpot} at the other strings; the factors of $\sigma$ have no zeros or poles. 
\item[--]
Since $E(y,x)$ is a $-1/2$ differential in $y$, and $\sigma(y)^2$ is a $g$-differential, \eqref{chargecons} is the condition for the right hand side of \eqref{eW} to be a function of the positions of the strings, rather than a differential. 
\item[--]
By \cite{Fay}, when $y$ goes once around $A_k$, $E(y,x)$ and $\sigma(y)$ come back to themselves up to a sign. Since, by \eqref{chargecons}, they enter \eqref{eW} with total weight zero, the signs cancel. It follows that a string of charge $e_a$ going once around $A_k$ multiplies $e^{-{\cal W}}$ by
$$\mathfrak{a}_{a,k}\, e^{-{\cal W}} = 
e^{-2\pi i\lambda^{A}_{a,k}} \, e^{-{\cal W}}.
$$
This depends on the chosen representative $A_k$ cycle which defines $\lambda^{A}_{a,k}$. A different choice gives a power of $\fq$ which depends on the path. 
\item[--]
When $y$ goes once around $B_k$, $E(y,x)^2$ is multiplied by $\exp\bigl(-2\pi i\,\tau_{kk}-4\pi i\int_x^{y}\omega_k\bigr)$, where $\tau_{k\ell}=\oint_{B_k}\omega_\ell$ is the period matrix of $\overline{\Sigma}$, and $\sigma(y)^2$ by a constant times $\exp\bigl(4\pi i(g-1)\int_{p_*}^{y}\omega_k\bigr)$. This reproduces the fact, from \eqref{Bshift}, that, a properly normalized $\mathfrak{b}_{a,k}$, multiplies $e^{-{\cal W}}$ by
\beq\label{Bmult}
\mathfrak{b}_{a,k}\, e^{-{\cal W}}=
\exp\Bigl(2\pi i\lambda_0\sum_{(b,i')}\langle e_a,e_b\rangle\int_{p_*}^{y_{b,i'}}\omega_k\Bigr) \, e^{-{\cal W}},
\eeq
where the sum runs over all the strings, the one that moved included. 
This gives \eqref{Bshift} and, since $\int_{p_*}^{y}\omega_k$ increases by one when $y$ goes once around $A_k$, \eqref{Heis}.
 
\item[--]
Over $\Delta$, the $u$'s in \eqref{eW} are to be replaced by the non-vanishing functions $x$, using \eqref{c1}--\eqref{c3}. The zeros and poles of the $u$'s then cancel those of the prime forms, so ${\cal W}$ is finite there.
\item[--]The point $p_*$ enters as a puncture associated with weight $w_*$ determined from charge conservation \eqref{chargecons}. If $w_*$ ends up dominant, it corresponds to the highest weight of a finite dimensional representation $V_*$ of $\fg$. The behavior of the holomorphic form $\Omega$ distinguishes it from a marked point. 
\end{itemize}
\subsubsection{}
The superpotential ${\cal W}$ given by equation \eqref{eW} gives the unique $e^{-{\cal W}}$ with the residues and the holonomies described above, consistent with locality, and up to a constant factor and up to shifts of the $\lambda^{A}_{a,k}$ by integers and by integer multiples of $\lambda_0$, which only relabel the equivariant gradings. This is because the ratio of two functions $e^{-{\cal W}_{1}}$ and $e^{-{\cal W}_{2}}$ with such properties is, as a function of the position of any one string, a function on $\overline{\Sigma}$ without zeros or poles, which is multiplied by constants when the string goes around the $A_k$ and the $B_k$. It follows that $e^{{\cal W}_{1}-{\cal W}_{2}}$ is the exponential of the integral of a holomorphic one-form.

\subsubsection{}\label{s-su2}
It is useful to spell this out for $\fg=\mathfrak{su}_2$. There is one simple root $e$, with $\langle e,e\rangle=2$ and $\rho=e/2$. Take $d$ strings of charge $e$ at points $y_i\in\Sigma$, with fiber coordinates $u_i$. The charge conservation \eqref{chargecons} becomes
\beq\label{chargesu2}
\sum_x\mu_x=2g-2+2d,
\eeq
where $x$'s are the positions of marked points and holes, associated with charges $w_x$, with $\mu_x=\langle e,w_x\rangle$. If $x=p_I$ is a finite deleted point $p_I$, the residue $\lambda_I$ in \eqref{data} equals $\lambda_0\mu_x$. If $p_*$ is the only deleted point, all the exponents in the square bracket are integers, and the one at $p_*$ is
$
\mu_*=2g-2+2d-\sum_j\mu_j.
$
With this, \eqref{eW} becomes
\beq\label{eWsu2}
e^{-{\cal W}}=e^{-{\cal W}_{\rm top}}\,\prod_{i}e^{-2\pi i\sum_k\lambda^{A}_{k}\int_{p_*}^{y_i}\omega_k}\,\biggl[\prod_{i,x}E(y_i,x)^{\mu_x}\,\prod_{i<i'}E(y_i,y_{i'})^{-2}\,\prod_i{\sigma(y_i)^2\over u_i}\biggr]^{-\lambda_0}.
\eeq

Then, without the $u$'s, as a function of the position of one string, the expression in the square bracket has a zero of order $\mu_j$ at the marked point $q_j$, a zero of order $\mu_*$ at $p_*$, or a pole if $\mu_*$ is negative, a double pole at each of the other strings, and no other zeros or poles. For a link, $\mu_*=2g-2$. This corresponds to inserting $V_*$, which is a spin $g-1$ representation.
\subsubsection{}
The group $\Gamma$ in this $\mathfrak{su}_2$ example is generated by $\fq$, $\mathfrak{r}_I$, $\mathfrak{a}_k$ and $\mathfrak{b}_k$, with
\beq\label{Heissu2}
\mathfrak{a}_{k}\,\mathfrak{b}_{\ell}=\fq^{-2\delta_{k\ell}}\;\mathfrak{b}_{\ell}\,\mathfrak{a}_{k},
\eeq
and all the other pairs commuting. By \eqref{Bmult}, $\mathfrak{b}_k$ multiplies $e^{-{\cal W}}$ by a constant times $\fq^{2s_k}$, where $s_k=\sum_i\int_{p_*}^{y_i}\omega_k$. For a class $\gamma=\sum_k(m_kA_k+n_kB_k)+\sum_Il_I\gamma_{p_I}$ in $H_1(\Sigma,{\mathbb Z})$, let
\beq\label{egamma}
e_\gamma=\fq^{\sum_km_kn_k}\;\prod_k\mathfrak{a}_k^{m_k}\;\prod_k\mathfrak{b}_k^{n_k}\;\prod_I\mathfrak{r}_I^{l_I}.
\eeq
Up to a power of $\fq$, $e_\gamma$ is the holonomy of $e^{-{\cal W}}$ when a string goes once around a loop of class $\gamma$. Every element of $\Gamma$ is $\fq^{j}e_\gamma$ for exactly one integer $j$ and one class $\gamma$, and
\beq\label{eprod}
e_\gamma\,e_{\gamma'}=\fq^{-\#(\gamma,\gamma')}\,e_{\gamma+\gamma'}.
\eeq

\subsection{Top holomorphic form on ${\cal Y}(\Sigma)$}\label{s-Omega}

The target ${\cal Y}(\Sigma)$ can be equipped with a nowhere vanishing top holomorphic form $\Omega$. It is needed to equip ${\mathscr D}_{{\cal Y}(\Sigma)}$ with a ${\mathbb Z}$-valued homological grading. Such a form exists thanks to the non-compactness of $\Sigma$.

\subsubsection{}
Since $\Sigma$ is an open surface, $H^2(\Sigma,\mathbb Z)=0$, so its canonical bundle $K_{\Sigma}$ has no topological obstruction to being trivial. Its trivialization can be chosen holomorphically, giving a nowhere vanishing holomorphic section
\beq\label{triv}
\omega_\Sigma \in H^0(\Sigma, K_{\Sigma}).
\eeq
For $\Sigma={\mathbb C}^{\times}$, we can take $\omega_\Sigma=dy/y$, as in \eqref{tf}.

\subsubsection{}
A smooth monopole contributes two complex coordinates to ${\cal Y}(\Sigma)$: its position $y\in\Sigma$ and a fiber coordinate $u$, which is ${\mathbb C}^{\times}$-valued away from $\Delta$. The ${\mathbb C}^{\times}$ factor has the canonical holomorphic one-form ${du/u}$. Together with $\omega_\Sigma$ in \eqref{triv}, it equips a smooth monopole at $(u_{a,i},y_{a,i})$ with a holomorphic two-form ${du_{a,i}\over u_{a,i}}\wedge\omega_\Sigma(y_{a,i})$. On ${\cal Y}^{o}(\Sigma)$, where the monopoles are separated, the top holomorphic form is
\beq\label{Os}
\Omega=\bigwedge_{a,i}\left({du_{a,i}\over u_{a,i}}\wedge\omega_\Sigma(y_{a,i})\right).
\eeq
Although the $u_{a,i}$ have zeros and poles over $\Delta$, by \eqref{c1}--\eqref{c3} this expression extends over $\Delta$ and defines a nowhere-vanishing top holomorphic form on ${\cal Y}(\Sigma)$.

\subsubsection{}
As we will see below, the category ${\mathscr D}_{{\cal Y}(\Sigma)}$ of A-branes on ${\cal Y}(\Sigma)$ with superpotential ${\cal W}$ does not depend on the full choice of $\omega_\Sigma$, but only on the homotopy class of the line field that it determines. This is the same line field that enters the definition of Fukaya categories of surfaces in \cite{ALSZ}. 

A line field on $\Sigma$ is a continuous choice, at every point $y\in\Sigma$, of an unoriented real line
\beq
\ell_y\subset T_y\Sigma.
\eeq
To get a line field from $\omega_{\Sigma}$, locally write
\beq
\omega_\Sigma=f(y)\,dy.
\eeq
This determines a distinguished direction in $T_y\Sigma$, the direction along which $\omega_\Sigma$ is real. The line field is unoriented, and the same line field is determined by both $f(y)dy$ and $-f(y)dy$. Thus the line field is naturally determined by the phase of $\omega_\Sigma^2$. In this sense, $\omega_\Sigma$ is a convenient holomorphic representative of an underlying topological choice, the homotopy class of the line field.

In the expression for $\Omega$ in \eqref{Os}, the same nowhere vanishing one-form $\omega_\Sigma$ appears for every monopole, so its phase induces the same line field on $\Sigma$ for every symmetric product factor of $Y(\Sigma)$, precisely as in \cite{ALSZ}.

\subsubsection{}
The choice of the line field on $\Sigma$ has a direct interpretation from the perspective of the self-dual strings, as a global choice of $R$-symmetry.

A self-dual string located at $y\in\Sigma$ has a complex scalar $y$ describing its position in the two directions of $\Sigma$ transverse to the string. Locally, rotations
\beq
y\longrightarrow e^{i\theta}y
\eeq
are an $R$-symmetry of the effective two-dimensional theory on the string. Choosing a reference for these rotations amounts to choosing a real direction in the $y$-plane. 

On a general curved $\Sigma$, there is no preferred coordinate $y$ that makes this choice globally. Choosing a line field supplies a reference direction in every tangent plane $T_y\Sigma$, and so a way to compare the local $R$-symmetries as the self-dual string moves over $\Sigma$.

\subsection{Conformal blocks of $\fgh_k$ on $\Sigma$}\label{s-ff}
The theory with superpotential ${\cal W}$ in \eqref{eW} and the top holomorphic form $\Omega$ in \eqref{Os} gives rise to conformal blocks of $\fgh_k$ on $\Sigma$, as it did for $\Sigma={\mathbb C}^{\times}$ in section \ref{ss-limit}. The conformal blocks are the vector vertex functions ${\cal V}_{\alpha}[{\cal L}]$ of the theory on self-dual strings supported on ${\sf D}$ in $M_6$, thought of as a long cigar. They are given, as in \eqref{intV} by,
\beq\label{vva}
{\cal V}_{\alpha}[{\cal L}] = \int_{\cal L} \Phi_{\alpha} \Omega e^{-{\cal W}},
\eeq
with suitable single valued insertions $\Phi_{\alpha}$.

\subsubsection{}
Integration over the brane in the fibers $Y_F$ has the same effect as evaluating ${\cal W}$ at its unique fiberwise critical point, as in section \ref{downsuperpotential}. This  strips the $u$-dependent factors from  \eqref{eW}, so up to a constant,
$$e^{-\cal W}|_{Y_{F, crit}} = e^{-W}.$$ 
What remains is an integral of the form \eqref{SXa}, over a Lagrangian in $Y(\Sigma)$, with the base superpotential $W$ from \eqref{WS} and $\Omega_Y=\prod_{a,i}\omega_{\Sigma}(y_{a,i})$.

 The resulting integral reproduces the free field representation of $\fgh_k$ conformal blocks on a surface of genus $g$ from \cite{Gerasimov}. The conformal blocks correspond to having a vertex operator of weight $w_x$ at each marked and deleted point, and a fixed momentum through each cycle $A_k$, determined by the $\lambda^{A}_{a,k}$. Charge conservation \eqref{chargecons} is the condition for the free field correlator of screening charges not to vanish. The insertions $\Phi_{\alpha}$ correspond to the correlators of the remaining free fields in \cite{Gerasimov}. 

\subsubsection{}\label{s-basis}
The free field formalism describes conformal blocks with Verma modules running through the surface. Recall how a basis of the space of conformal blocks on a surface of genus $g$ with $n$ insertions is labeled. 

Choose a decomposition of the surface into pairs of pants or, equivalently, a trivalent graph with $g$ loops and $n$ external legs, one ending at each insertion. One should think of the graph as living inside a handle body obtained by filling the $A_k$-cycles of the surface in. Label every edge of the graph by a representation, with the external legs labeled by the representations at the insertions, in such a way that the three representations meeting at each vertex are compatible with fusion. For $\fg$ of rank greater than one, one also chooses an intertwiner at each vertex.

Since representations running inside the surface are Verma modules, the handle body with the graph removed is topologically $\Sigma \times {\mathbb R}_+$, with the conformal block the state obtained at the boundary. The conformal blocks with all such labelings form a basis. A different decomposition into pairs of pants gives a different basis of the same space.

\subsubsection{}\label{s-handle}
We have picked in section \ref{pi} the $A_k$ cycles that get filled in. In the absence of self-dual strings, we only get Verma modules running along the dual $B_k$ cycles of highest weight determined by $\lambda^{A}_{a,k}$ in \eqref{data} and the Verma modules associated to the holes in \eqref{ns}. The rest of the graph labeling is fixed by charge conservation, which is compatible with any pants decomposition.

The vector vertex function \eqref{vva} describes conformal blocks obtained by adding self-dual strings. Adding self-dual strings we get ${\cal Y}(\Sigma)$ as their moduli space. To get the partition function of the self-dual strings supported on ${\sf D}$ we need to specify the brane at the boundary ${\partial {\sf D}}$ at infinity, corresponding to choosing a Lagrangian ${\cal L}$ in ${\cal Y}(\Sigma)$. The choice of this Lagrangian brane determines the integration cycle in \eqref{vva}, and the conformal block. Only the homology class of its downstairs counterpart $L$ in $Y(\Sigma)$ affects which conformal block we get, up to normalization, for the same reason as in section \ref{ss-limit}.
\subsubsection{}
\begin{figure}[h!]
\begin{center}
\includegraphics[width=\textwidth]{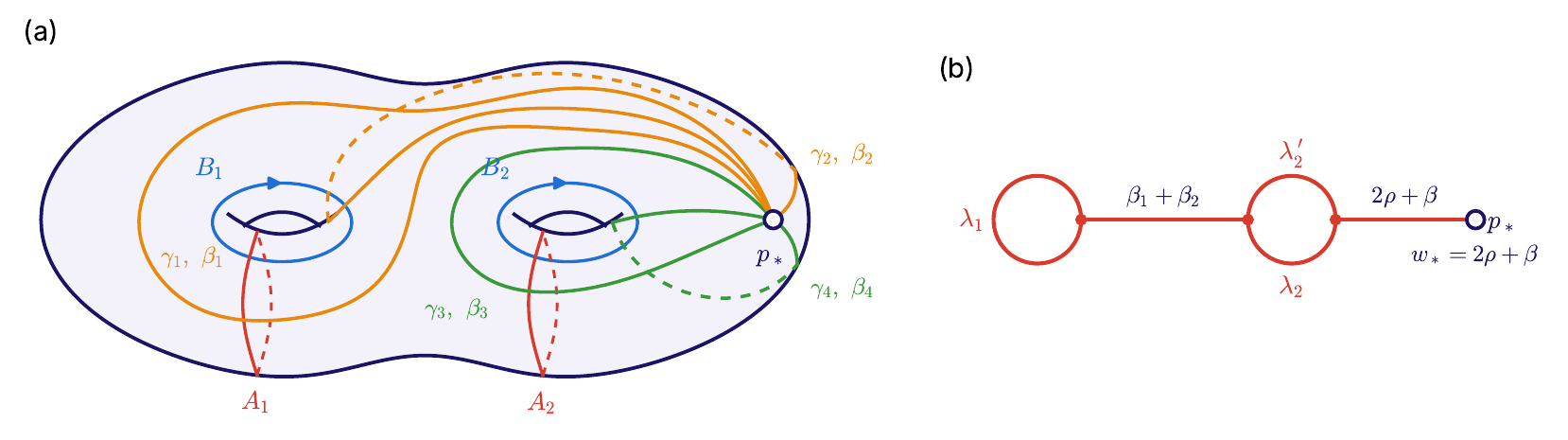}
\caption{(a) The arcs supporting the ${\cal T}_{\cal C}$ branes on a surface $\Sigma$ of genus two, with $p_*$ the only deleted point. The arc ${\gamma}_i$ carries self-dual strings of total charge $\beta_i=\beta(\gamma_i)$. (b) The corresponding conformal block, with every edge labeled by the highest weight of the Verma module running through it. Here $\lambda^{A}_{a,k}=\lambda_0\langle e_a,w^{A}_k\rangle$, $\beta=\sum_{i}\beta_i$, and $w^{A\prime}_2=w^{A}_2+2\rho+\beta_1+\beta_2$.}\label{f_Tblocks}
\end{center}
\end{figure}

The collection of ${\cal T}_{\cal C}$ branes, supported on arcs $\gamma_i$ described in section \ref{s-exact}, should span the space of conformal blocks with fixed numbers of screening charges given by \eqref{chargecons}. Presence of branes changes the Verma module weights. The ${\cal T}_{\cal C}$ branes are non-compact, so a basis can be provided by branes supported on arcs $\gamma_i$ which begin and end at $p_*$ and cross one of the $A_k$ or $B_k$ cycles once, as in figure \ref{f_Tblocks}. 

\subsection{Gradings in a Fukaya category}\label{GF}
Let us briefly review, following \cite{A2} and references therein, the gradings in the category $\MDY$ of A-branes on some ${\cal Y}$, with superpotential ${\cal W}$ and top holomorphic form $\Omega$.

Objects of $\MDY$ are Lagrangians ${\cal L}$ in ${\cal Y}$, equipped with a lift of the phase of
\beq\label{oW}
\Omega e^{-{\cal W}}|_{\cal L}
\eeq
to a real-valued function on ${\cal L}$. The phase of $\Omega$, restricted to a Lagrangian ${\cal L}$, gives the familiar cohomological, or Maslov, grading of the corresponding A-brane. If the superpotential ${\cal W}$ is multivalued, its monodromies give additional equivariant gradings, coming from the need to define the phase of $e^{-{\cal W}}$ restricted to the brane. Thus the cohomological and equivariant gradings have different origins, in $\Omega$ and ${\cal W}$, but enter the A-brane theory only through the combination in \eqref{oW}.

\subsubsection{}
There are two distinct choices involved in the cohomological grading. The first is the global choice of the homotopy class of the line field on $\Sigma$. Once one line field $\ell_0$ is chosen, any other differs from it by a pointwise rotation, giving a continuous map
\beq
\Sigma\longrightarrow S^1.
\eeq
The homotopy classes of continuous maps from $\Sigma$ to $S^1$ are
\beq
H^1(\Sigma,\mathbb Z).
\eeq
The set of homotopy classes of line fields cannot be canonically identified with this group, as there is no distinguished zero line field. Choosing an $\ell_0$ supplies an origin. In other words, the homotopy classes of line fields form a torsor for $H^1(\Sigma,\mathbb Z)$. Geometrically, classes in $H^1(\Sigma,\mathbb Z)$ measure the additional winding of the line field around the nontrivial $1$-cycles of $\Sigma$. For example, for $\Sigma=\mathbb C$ there is only one homotopy class of line fields. For $\Sigma={\mathbb C}^{\times}$, there is a ${\mathbb Z}$-ambiguity corresponding to trading $\omega_\Sigma = {dy\over y}$ for $y^n {dy\over y}$, for $n\in {\mathbb Z}$.

\subsubsection{}
The second choice is the grading of an individual A-brane. After the line field, and hence the grading structure on ${\cal Y}$, has been fixed, there remains a choice of grading for any Lagrangian ${\cal L}$ in ${\cal Y}$. An A-brane is not just a Lagrangian ${\cal L}\subset{\cal Y}$, but a Lagrangian together with a choice of grading. This grading is a lift of the phase of $\Omega$ along ${\cal L}$ to a real-valued function. Different lifts differ by integer multiples of $\pi$, giving the usual shift
\beq
{\cal L}\longrightarrow {\cal L}[n].
\eeq
Unlike the choice of line field, this choice is always present, even for $\Sigma=\mathbb C$.

For an explanation of how fixing the cohomological gradings of branes fixes the degrees of all their intersection points, see \cite{A2}, which follows \cite{Auroux}.

\subsubsection{}
The non-single-valued superpotential ${\cal W}$ introduces equivariant gradings in the category of A-branes \cite{A2}. When the superpotential ${\cal W}$ is not single valued, to define a brane ${\cal L}$ as an object of $\MDY$, we need to define a lift of ${\cal W}$ as a single-valued function on ${\cal L}$. Only Lagrangians for which such a lift can be given are objects of the category. This lets us compare the lift of ${\cal W}$ between any two points ${\cal P},{\cal Q}\in {\cal L}$:
\beq
{\cal W}({\cal P})-{\cal W}({\cal Q})=\int_{\cal Q}^{\cal P}d{\cal W},
\eeq
where the integral is along a path in ${\cal L}$. For a Lagrangian admitting a lift of ${\cal W}$, this integral is path independent. Thus, once the lift is fixed at one point of ${\cal L}$, it is fixed everywhere on ${\cal L}$. The choice of lift is the choice of equivariant grading of the brane ${\cal L}$.

\subsubsection{}
Assume that $d{\cal W}/2\pi i$ is a closed one-form on ${\cal Y}$ (we will relax this assumption in the next subsection, as it does not hold for ${\cal Y}(\Sigma)$ with $\Sigma$ of genus $g>0$). In a basis of integral cohomology,
\beq\label{dw}
{d{\cal W}\over 2\pi i}\simeq\sum_i\lambda_i c_i,
\eeq
where $c_i\in H^1({\cal Y},\mathbb Z)$ have integer periods, $\lambda_i\in\mathbb C$, and the equality is up to exact terms coming from the single-valued part of ${\cal W}$. Equivalently, choose dual one-cycles $\gamma_i\in H_1({\cal Y},\mathbb Z)$, with
\beq
\int_{\gamma_i}c_j=\delta_{ij}.
\eeq
Then
\beq\label{lf}
\lambda_i={1\over 2\pi i}\oint_{\gamma_i}d{\cal W}.
\eeq
\subsubsection{}
The category of A-branes has one $\mathbb Z$-valued equivariant grading for each $c_i$ in \eqref{dw}, with the corresponding ${\lambda_i}$ not zero.

Let ${\cal L}\{n_i\} \in \MDY$ be obtained from the A-brane ${\cal L}\in \MDY$ by shifting its equivariant gradings by $\{n_i\}$, with $n_i\in\mathbb Z$. By definition,
\beq
{\cal W}|_{{\cal L}\{n_i\}}-{\cal W}|_{\cal L}=2\pi i\sum_i\lambda_i n_i.
\eeq
Having picked a grading for all objects, gradings of all elements of Floer complexes are automatically fixed. In particular, the grading of an intersection point 
$$
{\cal P} \in {\cal L}_0\cap {\cal L}_1
$$
is determined by the difference of values of ${\cal W}$ at ${\cal P}$:
$$
2\pi i \,\sum_i \lambda_i  n_i({\cal P})  = {\cal W}({\cal P})|_{{\cal L}_1} - {\cal W}({\cal P})|_{{\cal L}_0}.
$$
This enters the Euler characteristic, which is a Laurent polynomial in one variable for every equivariant grading:
\beq
\chi({\cal L}_0,{\cal L}_1)=\sum_{{\cal P}\in {\cal L}_0\cap {\cal L}_1}(-1)^{M({\cal P})}\prod_i\fq_i^{n_i({\cal P})},
\eeq
where values of the equivariant variables ${\fq_i}$ are determined by the actual superpotential by \eqref{lf} and
\beq
\fq_i=e^{2\pi i\lambda_i}.
\eeq

\subsection{Gradings in ${\mathscr D}_{{\cal Y}(\Sigma)}$}\label{s-GS}

Let us now specialize the general discussion of section \ref{GF} to our ${\cal Y}(\Sigma)$.

\subsubsection{}
The equivariant gradings of ${\mathscr D}_{{\cal Y}(\Sigma)}$ take values in $\Gamma$, the group of monodromies of $e^{-{\cal W}}$ defined in section \ref{s-Gamma}, with generators \eqref{gens}.

For $\Sigma$ of genus zero, $\Gamma$ is commutative, generated by $\fq$ and the $\mathfrak{r}_{a,I}$, and we are in the setting of section \ref{GF} and of \cite{A2}. The generator $\fq$ gives the quantum $\fq$-grading, which comes from both the base $Y(\Sigma)$ and the fibers $Y_F$, by \eqref{srcmult}. The generators $\mathfrak{r}_{a,I}$ give the gradings associated to highest weights of Verma modules, one for every simple root $e_a$ and every finite deleted point $p_I$, which come from the base alone. Since the highest weights are generic, these generators are independent, and an element of $\Gamma$ is a collection of integers, one for each of them.

\subsubsection{}
For $\Sigma$ of genus $g>0$, $\Gamma$ is non-commutative. Its generators $\mathfrak{a}_{a,k}$ and $\mathfrak{b}_{a,k}$, which correspond to a string of charge $e_a$ going once around the cycle $A_k$, and once around the dual cycle $B_k$, commute only up to a power of $\fq$, by \eqref{Heis}. 

Correspondingly, the assumption made in section \ref{GF}, that $d{\cal W}/2\pi i$ is a closed one-form, does not hold on ${\cal Y}(\Sigma)$: by \eqref{Bshift}, $d{\cal W}$ does not come back to itself when a string goes once around a $B_k$ cycle.

\subsubsection{}\label{s-GG}
As always, ${\cal W}$ is well defined on a brane. Objects of ${\mathscr D}_{{\cal Y}(\Sigma)}$ are those Lagrangians ${\cal L}$ which admit a lift of ${\cal W}$ to a single-valued function on ${\cal L}$. The choice of lift is the choice of equivariant grading of ${\cal L}$.

A lift of ${\cal W}$ as a single-valued function on ${\cal L}$ is the same as a lift of ${\cal L}$ to the covering space $\widehat{\cal Y}(\Sigma)$ in \eqref{Yhat}. By the definition of $\Gamma$ as the group of deck transformations of the covering space, any two choices differ by an element of $\Gamma$. Unlike in section \ref{GF}, the two lifts of ${\cal W}$ need not differ by a constant: if the lifts of ${\cal L}$ differ by $\mathfrak{b}_{a,k}$, then, by \eqref{Bmult}, the lifts of ${\cal W}$ differ by a function on ${\cal L}$.

Having picked the lifts of ${\cal L}_0$ and ${\cal L}_1$, the equivariant grading of an intersection point ${\cal P}\in{\cal L}_0\cap{\cal L}_1$ is the element $\mathfrak{n}({\cal P})$ of $\Gamma$ which takes the point over ${\cal P}$ on the lift of ${\cal L}_0$ to the point over ${\cal P}$ on the lift of ${\cal L}_1$. Since a holomorphic disk in ${\cal Y}(\Sigma)$ lifts to $\widehat{\cal Y}(\Sigma)$, the Floer differential preserves the equivariant grading, and the composition of morphisms multiplies their equivariant gradings, in the order in which the morphisms are composed.

\subsubsection{}
Thus, for $g>0$ the equivariant grading is not a collection of integers, but an element of the group $\Gamma$. Only its image in the lattice 
$$H_1(\Sigma,{\mathbb Z})\otimes\Lambda_{{\rm rt},+},$$ 
of which, as explained in section \ref{s-Gamma}, $\Gamma$ is the central extension by the powers of $\fq$, is a collection of integers. The power of $\fq$ is not a grading on its own, since by \eqref{Heis} powers of $\fq$ are commutators in $\Gamma$.

\subsubsection{}
The Euler characteristic
\beq\label{chiG}
\chi({\cal L}_0,{\cal L}_1)=\sum_{{\cal P}\in {\cal L}_0\cap {\cal L}_1}(-1)^{M({\cal P})}\,\mathfrak{n}({\cal P})
\eeq
is an element of the group ring of $\Gamma$. 

The fact that $\Gamma$ is a non-commutative group for $g>0$ decategorifies to the fact that, in Chern-Simons theory on $M_{3, CS}=\Sigma\times{\mathbb R}$, the holonomies around intersecting 1-cycles do not commute; see section \ref{s-example}. In particular, the $U_{\fq}(\fg)$ link invariants in $\Sigma\times{\mathbb R}$ are themselves not numbers; they are valued in the Skein algebra of the surface $\Sigma$.

For $\Sigma$ of genus zero, every element of $\Gamma$ multiplies $e^{-{\cal W}}$ by a constant. Replacing $\mathfrak{n}({\cal P})$ by this constant, which is $\prod_i\fq_i^{n_i({\cal P})}$, we recover the Euler characteristic of section \ref{GF}.

\subsection{Exact solution}\label{s-exact}
It is shown in \cite{ALSZ} that the category of A-branes on ${\cal Y}(\Sigma)$ is exactly solvable, as in the special case $\Sigma={\mathbb C}^{\times}$. It has a generating brane, which we will again denote by ${\cal T}$, whose endomorphism algebra is computable explicitly. The algebra is the KLRW algebra assigned to the surface $\Sigma$.

\subsubsection{}
The ${\cal T}={\cal T}(\gamma)$-brane depends on a choice of arcs on $\Sigma$. Let
\beq
\gamma=(\gamma_1,\ldots,\gamma_N)
\eeq
be a collection of disjoint arcs on $\Sigma$, each running between a pair of deleted points $p_I$ and $p_J$ in \eqref{ns}, possibly the same. The arcs cut $\Sigma$ into disks, such that each disk contains at most one marked point. For $\Sigma={\mathbb C}^{\times}$, we can take the $\gamma_i$ to be straight real lines on the cylinder, one between each pair of consecutive marked points.

Take a collection of parallel copies of the arcs, colored by simple roots of $\fg$, with $d_a$ copies colored by $e_a$. Their product is a Lagrangian in $Y(\Sigma)$. Pairing each arc with a real-line Lagrangian in the ${\mathbb C}^{\times}$ fiber over it, we get an A-brane ${\cal T}_{\cal C}(\gamma)$ in ${\cal Y}(\Sigma)$, where ${\cal C}$ records the sequence of colors along each $\gamma_i$. As in section \ref{TA}, let
\beq
{\cal T}(\gamma)=\bigoplus_{\cal C}{\cal T}_{\cal C}(\gamma)
\eeq
be the sum of all inequivalent such branes.

\subsubsection{}
The endomorphism algebra of ${\cal T}(\gamma)$,
\beq
{\mathscr A}(\Sigma,\gamma)={\rm Hom}^{*,*}_{{\mathscr D}_{{\cal Y}(\Sigma)}}({\cal T}(\gamma),{\cal T}(\gamma)),
\eeq
is computed in \cite{ALSZ}, extending the computation of \cite{ADZ} for $\Sigma={\mathbb C}^{\times}$. We will call ${\mathscr A}(\Sigma,\gamma)$ the KLRW algebra of the surface $\Sigma$. Like the cylindrical KLRW algebra, it is defined in terms of strand diagrams. The strand diagrams represent open strings stretching between the ${\cal T}_{\cal C}(\gamma)$-branes on $\Sigma$. The intersections along the fibers over $\Sigma$ are represented by dots decorating the strands. The stretched strings can be traded for intersection points of branes, using the fact that ${\mathscr D}_{{\cal Y}(\Sigma)}$ is the wrapped Fukaya category, with maximal wrapping at each hole on $\Sigma$, corresponding to a deleted point $p_I$ in \eqref{ns}. These algebras are new for any $\Sigma$ other than ${\mathbb C}^{\times}$, for which ${\mathscr A}(\Sigma,\gamma)$ becomes the cylindrical KLRW algebra of \cite{W1,W2}, and the strip ${\mathbb R}\times [0,1]$, for which it recovers the original KLRW algebra of \cite{KhL, Rh, webster}. The strip case requires extending the theory to non-compact Riemann surfaces with non-compact boundaries, which \cite{ALSZ} provides.

\subsubsection{}
As in section \ref{HMS}, we identify ${\mathscr D}_{{\cal Y}(\Sigma)}$ with its subcategory split-generated by the ${\cal T}(\gamma)$-branes. This gives:
\begin{theorem}[A.-LePage-Shende-Zhou]\label{sige}
$$
{\mathscr D}_{{\cal Y}(\Sigma)}\simeq{\mathscr D}_{{\mathscr A}(\Sigma,\gamma)},
$$
where ${\mathscr D}_{{\mathscr A}(\Sigma,\gamma)} = D({\mathscr A}(\Sigma,\gamma){\rm -mod})$ is the derived category of modules of ${\mathscr A}(\Sigma,\gamma)$.
\end{theorem}
\noindent The category does not depend on the choice of the arcs $\gamma$, although the algebra does.

The correspondence assigns to an A-brane ${\cal L}\in {\mathscr D}_{{\cal Y}(\Sigma)}$ a module for ${\mathscr A}(\Sigma,\gamma)$; one should think of the two as equivalent objects, as in section \ref{why}.

\subsubsection{}

The construction of link homology in section \ref{lsl} carries over to general $\Sigma$. Take a link $K\subset M_{3,CS}=\Sigma\times{\mathbb R}$ in \eqref{gen3m}, colored by minuscule representations of $\fg$. Translate the link to a pair of matchings on $\Sigma$. Each matching now lives inside the filling of $\Sigma$ together with a labeled graph from section \ref{s-basis}; in particular the two can braid. This gives us ${\cal Y}(\Sigma)$, the category ${\mathscr D}_{{\cal Y}(\Sigma)}$, and a pair of branes ${\cal U}$ and ${\mathscr B}{\cal U}$.

Take the brane ${\cal U}$ to describe the configuration where the cups link neither each other nor the Verma module graph; this is the natural analogue of a plat closure from section 2. The domain wall corresponding to the cups is obtained from monopole bubbling, as in section \ref{ccs}. The ${\mathscr B}{\cal U}$ brane corresponds to the collection of cups braided with each other and with the Verma graph. The link homology is
\beq\label{SH}
{\cal H}^{*,*}(K)={\rm Hom}^{*,*}_{{\mathscr D}_{{\cal Y}(\Sigma)}}({\mathscr B}{\cal U},{\cal U}).
\eeq
Its equivariant grading takes values in the group $\Gamma$, as explained in section \ref{s-GG}. 

The invariant depends not only on the link, but also on the choices of the graphs inside the two fillings, and the representations coloring them. 
\subsubsection{}
As for $\Sigma={\mathbb C}^{\times}$, the ${\mathscr B}{\cal U}$ brane can be resolved in terms of ${\cal T}(\gamma)$-branes, as in \eqref{keya}. We have a choice of both the matchings and the arcs $\gamma$. A convenient choice is the one for which the ${\cal U}$ brane, viewed as an $I_{\cal U}$ brane with $U$-shaped branes in the fibers, intersects a single ${\cal T}_{\cal C}(\gamma)$-brane, at one point. This makes ${\cal U}$ a simple module of the ${\mathscr A}(\Sigma,\gamma)$ algebra. Then, ${\cal H}^{*,*}(K)$ is the cohomology of the complex of vector spaces obtained by applying $hom^{*,*}(-,{\cal U})$ to the resolution of ${\mathscr B}{\cal U}$. Topological invariance of the theory along $M_{4, CS}$ guarantees that ${\cal H}^{*,*}(K)$ is an invariant of the framed link $K\subset M_{3, CS}$.

\subsubsection{}\label{s-blocks}
The Euler characteristic of ${\cal H}^{*,*}(K)$, defined as in \eqref{chiG}, is a $U_{\fq}(\fg)$ invariant of the link $K$ in $M_{3, CS}=\Sigma\times{\mathbb R}$. As in \eqref{PKqa}, it is the matrix element
\beq\label{PKS}
\chi({\mathscr B}{\cal U},{\cal U})=\langle {\mathfrak B}{\mathfrak U}|{\mathfrak U}\rangle,
\eeq
where $|{\mathfrak U}\rangle$ is the state in the Chern-Simons Hilbert space on $\Sigma$ corresponding to the brane ${\cal U}$, and ${\mathfrak B}$ is the $U_{\fq}(\fg)$ braiding matrix that decategorifies ${\mathscr B}$. 

The novel aspects, for $g>0$, are that $\chi({\mathscr B}{\cal U},{\cal U})$ is valued in the group ring of $\Gamma$, and that the invariant depends not only on the link, but also on a pair of Verma graphs. Together, this should reproduce the fact that the Chern-Simons or $U_{\fq}(\fg)$ invariants of a link in $M_{3, CS}$ are valued in the Skein module of $\Sigma$.

\subsection{Examples}\label{s-example}

\begin{figure}[h!]
\begin{center}
\includegraphics[width=\textwidth]{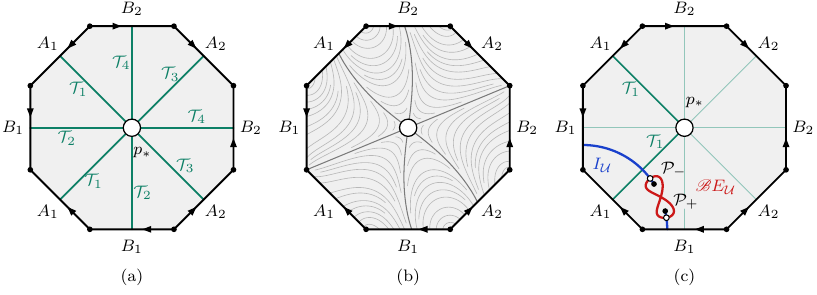}
\caption{(a) The surface $\Sigma$ of genus two, cut along the cycles $A_k$ and $B_k$ into an octagon. Sides with the same label are identified, and the deleted point $p_*$ is at the center. The four arcs, each running from $p_*$ across one of the cycles and back to $p_*$, cut $\Sigma$ into a disk, as in section \ref{s-exact}. They are labeled by the corresponding branes ${\cal T}_i$, for $Y(\Sigma)=\Sigma$. (b) The line field. It is transverse to the cycles $A_k$ and $B_k$, and six of its leaves end at $p_*$. (c) The knot $K_\gamma$ with $\gamma=A_1$ runs alongside a side $A_1$, from one side $B_1$ to the other. Its branes are the interval brane $I_{\cal U}$ along the blue segment, and the figure-eight brane ${\mathscr B}E_{\cal U}$ around the red segment, which lies between the two marked points. The brane $I_{\cal U}$ intersects a single ${\cal T}_{\cal C}(\gamma)$-brane, ${\cal T}_1$, at one point.}\label{f_g2knot}
\end{center}
\end{figure}

As an example, take $\fg=\mathfrak{su}_2$, with links colored by the two-dimensional representation $V$, and $\Sigma$ of genus two, with $p_*$ the only deleted point. We draw $\Sigma$ as in figure \ref{f_g2knot}, as a domain in the complex plane, with identifications. Without any marked points, the arcs $\gamma_i$ are the collection of four in figure \ref{f_g2knot}$(a)$. Addition of marked points requires introducing additional arcs.
A line field of section \ref{s-Omega} is shown in figure \ref{f_g2knot}$(b)$.

For simplicity, we will pick simple knots and links in $M_{3,CS}$, for which the homology can be computed directly, without resolving the branes in terms of ${\cal T}(\gamma)$-branes. The resolution itself works as in \cite{ALR}. For the same reason, we will consider only very special conformal blocks where the Verma graph is that determined by the background alone, so the ${\cal U}$ branes only correspond to the collection of cups.

Since we will not get the most general conformal blocks on $\Sigma$ in this way, we will not get the most general link invariants. The point of the example is to illustrate how equivariant gradings being valued in a non-commutative group $\Gamma$ accounts for the fact that $U_{\fq}(\fg)$ link invariants are valued in the Skein module of $\Sigma$.

\subsubsection{}
Let $K_\gamma$ be the knot which runs once around $\gamma$, one of the cycles $A_k$ or $B_k$ in figure \ref{f_g2knot}. Two marked points divide $\gamma$ into a red and a blue segment. As in \eqref{MPud}, ${\cal H}^{*,*}(K_\gamma)$ is the space of morphisms from the figure-eight brane ${\mathscr B}E_{\cal U}$ to the interval brane $I_{\cal U}$, on 
$$Y(\Sigma)=\Sigma.
$$
The branes intersect at two points ${\cal P}_{\pm}$, one for each weight of $V$, as for the unknot. Unlike there, no disk interpolates between them: the path from ${\cal P}_-$ to ${\cal P}_+$ along $I_{\cal U}$, and back along ${\mathscr B}E_{\cal U}$, is the cycle $\gamma$. The equivariant degrees of the two points differ by the holonomy of $e^{-{\cal W}}$ around $\gamma$, an element of $\Gamma$ in the class $\gamma$, which we may take to be $e_\gamma$. The Floer differential preserves the equivariant degree, so it vanishes. The Maslov degrees of the two points are the same, since $\gamma$ is transverse to the line field. Fixing the overall shifts so that the homological degree is zero and the equivariant degrees are symmetric, as in \eqref{nsJ},
\beq\label{exk}
{\cal H}^{*,*}(K_\gamma)=\bigl({\mathbb C}\,{\cal P}_+\bigr)\{x_\gamma\}\oplus\bigl({\mathbb C}\,{\cal P}_-\bigr)\{x_\gamma^{-1}\},
\eeq
where we defined $ x_\gamma$ by
$$
 x_\gamma^2=e_\gamma.
$$
Since $ x_\gamma$ is not an element of $\Gamma$, one should view this as the result of shifting the gradings of the overall link homology by a formal $ x_\gamma^{-1}$. 
\subsubsection{}
\begin{figure}[h!]
\begin{center}
\includegraphics[width=0.85\textwidth]{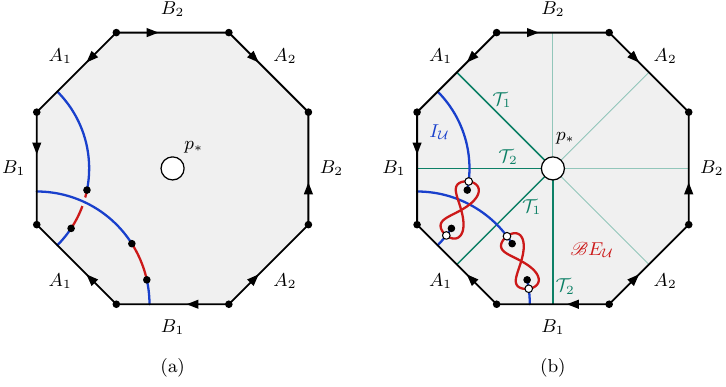}
\caption{(a) The link $K$ and its bi-coloring; the component around $A_1$ passes over the one around $B_1$. (b) The branes $I_{\cal U}$ and ${\mathscr B}E_{\cal U}$. Each of their four intersection points in ${\rm Sym}^2(\Sigma)$ is a pair of the points marked by open circles, one from each component. The brane $I_{\cal U}$ intersects a single ${\cal T}_{\cal C}(\gamma)$-brane, the product of ${\cal T}_1$ and ${\cal T}_2$, at one point.}\label{f_g2link}
\end{center}
\end{figure}

Next, let $K$ be the link in figure \ref{f_g2link}, whose two components run once around $A_1$ and $B_1$, the first passing over the second where they cross. Now 
$$Y(\Sigma)={\rm Sym}^2(\Sigma),
$$
and the two branes intersect at four points ${\cal P}_{\epsilon_1\epsilon_2}$, with $\epsilon_1,\epsilon_2=\pm1$: one of the two points above for each component. They have the same Maslov degree, and the differential vanishes as before, so
\beq\label{exl}
{\cal H}^{*,*}(K)=\bigoplus_{\epsilon_1,\epsilon_2=\pm1}\bigl({\mathbb C}\,{\cal P}_{\epsilon_1\epsilon_2}\bigr)\{x_{A_1}^{\epsilon_1}x_{B_1}^{\epsilon_2}\},
\eeq
The order of the factors matters, since as a consequence of \eqref{eprod}
\beq\label{ncom}
\qquad\qquad x_\gamma\,x_{\gamma'}=\fq^{-\#(\gamma,\gamma')/4}\,x_{\gamma+\gamma'}.
\eeq
As the string of the first component goes around $A_1$, it crosses the red segment of the second, between the two positions of its string. By \eqref{corpot}, the holonomy of $e^{-{\cal W}}$ around $A_1$ differs between the two by a factor of $\fq^{\langle e,e\rangle}$, which is the relation $e_{A_1}e_{B_1}=\fq^{-2}e_{B_1}e_{A_1}$ in $\Gamma$. If the second component passes over the first, the order is opposite; components around $A_1$ and $B_2$ do not meet, and the factors commute.

\subsubsection{}
These homology groups decategorify to $U_{\fq}(\fg)$ invariants of a link in $M_{3, CS}=\Sigma\times {\mathbb R}$. The Euler characteristic of ${\cal H}^{*,*}(K_\gamma)$ is 
\beq\label{us}
\chi(K_\gamma)=x_\gamma+x_\gamma^{-1}=\operatorname{Tr}_V x_\gamma,
\eeq
and that of ${\cal H}^{*,*}(K)$ is
\beq\label{exchi}
\chi(K)=\operatorname{Tr}_V x_{A_1}\cdot\operatorname{Tr}_V x_{B_1}=\fq^{-1/4}\,\operatorname{Tr}_V x_{A_1+B_1}+\fq^{1/4}\,\operatorname{Tr}_V x_{A_1-B_1},
\eeq
where we used \eqref{ncom}.

\subsubsection{}
The Euler characteristics in \eqref{us} and \eqref{exchi} are examples of \eqref{PKS}. Recall the $U_{\fq}(\mathfrak{su}_2)$ invariants in $M_{3, CS}$. They satisfy the same Skein relations as the Jones polynomial of a link in ${\mathbb R}^3$. Using the Skein relation, one gets to simplify the invariant of every link to the form $P_K(\fq) = \sum_{s} P_{K}(\fq, s) \prod_{c\in s}{\rm Tr}_V x_{c}$, where $s$ is a collection of disjoint non-self-intersecting loops on $\Sigma$, none of which bounds a disk. Above, ${\rm Tr}_V x_{c}=x_c+x_c^{-1}$ is the holonomy of the Chern-Simons gauge field along the loop $c$. Because in Chern-Simons theory holonomies around intersecting cycles do not commute, $x_{c}$'s satisfy the relation in \eqref{ncom}. The relation comes from canonical quantization of abelianized Chern-Simons theory, see e.g. \cite{Elitzur}. The result is an invariant valued in the Skein module of $\Sigma$.

In general, $x_c$ depends on the Verma graph, and its coloring, from section \ref{s-basis}. For the simple conformal blocks with trivial Verma graph considered here, there are relations between $x_c$'s which otherwise do not hold: for example, $x_{A_1}x_{A_2}$ equals $x_c$, up to a power of $\fq^{1/2}$, for a loop $c$ of class $A_1+A_2$. As a result, different collections $s$ can contribute with the same equivariant degree.

\subsubsection{}
For $\fg=\mathfrak{su}_2$, Asaeda, Przytycki and Sikora \cite{APS} generalized Khovanov homology to links in $M_{3, CS}$. Their homology groups $H_{ijs}(K)$ have, in addition to the two integer gradings $i$ and $j$ of \cite{Kh}, a grading $s$. The $s$-grading counts, with signs, the loops obtained from $K$ by the Skein relation which do not bound disks, separately for every isotopy class of loops on $\Sigma$. In Chern-Simons theory, this corresponds to completely generic Verma modules: the holonomies around all the loops are independent variables.

We expect that link homology ${\cal H}^{*,*}(K)$, corresponding to trivial Verma graphs as in this subsection, is the homology of the complex of \cite{APS} with an additional differential which merges two loops into one, or splits one loop into two, when the three loops bound a pair of pants on $\Sigma$. The differential of \cite{APS} vanishes unless one of the three loops bounds a disk. The two generators related by the additional contribution to the differential have different $s$, and the same equivariant degree.

When $K$ gives rise to no such triple of loops, the additional differential vanishes, and
$
{\cal H}^{*,*}(K)=\bigoplus_{i,j,s}H_{ijs}(K)\{x_s\},
$
up to an overall shift of the gradings. This is the case for the links in figures \ref{f_g2knot} and \ref{f_g2link}: \eqref{exk} and \eqref{exl} are the homology groups of \cite{APS}.

\subsection{Categorification of quantum Langlands correspondence}
The correspondence in Theorem \ref{sige} is a statement of the Langlands correspondence: On the left-hand side, only the $\fg'$ symmetry is manifest, as ${\cal Y}(\Sigma)$ is the $G'$ Zastava space. The right-hand side has the Langlands dual $\fg$ symmetry manifest, categorifying $U_{\fq}(\fg)$-modules on $\Sigma$.

It categorifies a key step of the quantum geometric Langlands correspondence.

\subsubsection{}
There are two formulations of the quantum Langlands correspondence, which are categorified by Theorem \ref{sige}.

Gaitsgory proved a local version of the quantum Langlands correspondence, conjectured by Lurie, for $\fq$ not a root of unity \cite{Gaitsgory08}. The correspondence is an equivalence between the category of $\fq$-twisted Whittaker sheaves of the group $G'$, associated to marked points $q_i$ on $\Sigma$ and labeled by dominant weights of $\fg$, and the category of modules of $U_{\fq}(\fg)$, one for each marked point. The irreducible twisted Whittaker sheaf with dominant weights $w_i$ of $\fg$ at the marked points $q_i$ corresponds to the irreducible modules of $U_{\fq}(\fg)$ with highest weights $w_i$.

The quantum Langlands correspondence of \cite{AFO} identifies conformal blocks of two chiral algebras on $\Sigma$: the affine Lie algebra $\fgh_k$ and the W-algebra ${\cal W}_{\beta}(\fg')$. Here $\beta$ is determined by $k$ as in \cite{AFO}, where our $\fg'$ and $\fg$ are called $\fg$ and $^L\fg$. The identification, constructed in \cite{AFO} for $\Sigma={\mathbb C}^{\times}$, works in families, is compatible with the action of braiding, and should hold for general, at least non-compact, $\Sigma$.
\subsubsection{}
The two formulations of the quantum Langlands correspondence are related by an equivalence of braided tensor categories: the category of twisted Whittaker sheaves of $G'$ is equivalent to the category of modules of ${\cal W}_{\beta}(\fg')$ generated by its irreducible modules labeled by dominant weights of $\fg$ \cite{FG, AF}. 

With it and the Kazhdan-Lusztig correspondence, Gaitsgory's theorem becomes an equivalence between categories of modules of the two chiral algebras whose conformal blocks are identified in \cite{AFO}. For general ${\fg}'$, this is a conjecture of \cite{AFO}. For simply laced $\fg$, it is a theorem of \cite{CDN}.
\subsubsection{}
Gaitsgory's proof \cite{Gaitsgory08} makes crucial use of the Zastava space ${\cal Y}(\Sigma)$.  Twisted Whittaker sheaves are sheaves on the space of $G'$ bundles on $\Sigma$ with a single framed flag. ${\cal Y}(\Sigma)$ maps to this space by forgetting the scattering flag at $t\to+\infty$. The irreducible twisted Whittaker sheaf with weights $w_i$ at the marked points $q_i$ pulls back to the sheaf on ${\cal Y}(\Sigma)$ generated by $e^{-{\cal W}}$. The Whittaker condition makes the sheaf depend on the bundle through $e^{-{\cal W}_{\rm top}}$. The twist of the sheaf is the multivalued factor $e^{-{\cal W}_{\rm equiv}}$.

Integrating the sheaf over the fibers of $\pi:{\cal Y}(\Sigma)\to Y(\Sigma)$, as in section \ref{downsuperpotential}, leaves $e^{-W}$ on $Y(\Sigma)$. Away from $\Delta$, $e^{-W}$ defines a local system, whose monodromies around strings and marked points are given in \eqref{srcmult}. Together with a sign under the exchange of two points of the same color, which comes from the top holomorphic form on $Y(\Sigma)$, these are the monodromies of the local system of Bezrukavnikov, Finkelberg and Schechtman \cite{BFS}, with their $\zeta$ given by $\fq=\zeta^{-2}$. How the local system is extended across $\Delta$, where strings collide with each other or with marked points, determines the quantum group module at each $q_j$. By \cite{BFS}, the minimal extension corresponds to the irreducible modules with highest weights $w_j$. Their formulas, continued naively to generic $k$, give the integrands of the free field formalism for $\fgh_k$ conformal blocks.

For $\Sigma$ of genus $g>0$, the monodromies around the $A_k$ and the $B_k$ cycles do not commute, by \eqref{Heis}. A local system with such monodromies must have rank greater than one; in \cite{BFS}, where the curve is compact and $\fq$ is a root of unity, it is the Heisenberg local system. Charge conservation \eqref{chargecons} is the counterpart of the condition in \cite{BFS} on the degree of the configuration of colored points on a curve of genus $g$. Our current setting extends this to surfaces with holes, labeled by Verma module representations.

The same sheaf $e^{-W},$ accompanied with $\Omega_Y = \prod_{i, \alpha} \omega_{\Sigma}(y_{i,\alpha})$, is the integrand in the free field formalism of the corresponding ${\cal W}_{\beta}(\fg')$ algebra conformal blocks, generalizing \cite{AFO}.

\subsubsection{}
Objects of ${\mathscr D}_{{\mathscr A}(\Sigma,\gamma)}$ categorify $U_{\fq}(\fg)$-modules associated to $\Sigma$ with collections of marked points, labeled by the finite dimensional representations of $\fg$, and holes, associated to Verma modules. By the Drinfeld-Kohno theorem \cite{Drinfeld, Koh} and the Kazhdan-Lusztig correspondence \cite{KL}, these correspond to conformal blocks of the affine Lie algebra $\fgh_k$, at the level $k$ determined by $\fq$.

Objects of ${\mathscr D}_{{\cal Y}(\Sigma)}$ are A-branes. Decategorification remembers only the homology class $[{\cal L}]$ of ${\cal L}\in {\mathscr D}_{{\cal Y}(\Sigma)}$. The homology class determines the equivariant central charge of the brane, ${\cal Z}[{\cal L}]=\int_{\cal L}\Omega e^{-{\cal W}}$. Upon integrating over the fibers $Y_F$ to get $\int_L\Omega_Ye^{-W}$, it is easily shown to give a conformal block of the ${\cal W}_{\beta}(\fg')$ algebra, generalizing \cite{AFO}.
The same brane also gives rise to the corresponding conformal block of $\fgh_k$ from section \ref{s-ff}, as 
the vector-valued ${\cal V}_{\alpha}[{\cal L}]=\int_{\cal L} \Phi_{\alpha}\, \Omega\, e^{-{\cal W}}$, for single valued $\Phi_{\alpha}$.

\subsubsection{}
The categories ${\mathscr D}_{{\cal Y}(\Sigma)}$ and  ${\mathscr D}_{{\mathscr A}(\Sigma,\gamma)}$ are topological. In the terminology of \cite{BZN}, Theorem \ref{sige} thus belongs to the Betti version of the quantum Langlands correspondence. The identification of conformal blocks in \cite{AFO}, which depends on the complex structure of $\Sigma$, belongs to the de Rham version.

The categories and the conformal blocks depend on $\Sigma$ differently, because they come from different twists. The conformal blocks ${\cal Z}[{\cal L}]$ and ${\cal V}_{\alpha}[{\cal L}]$ are associated to the pair of supercharges preserved near the tip of the underlying cigar ${\sf D}$ in $M_6$, as in section \ref{2dsusy}. For that pair, the twist on $\Sigma$ is holomorphic.
The categories ${\mathscr D}_{{\cal Y}(\Sigma)}$ and ${\mathscr D}_{{\mathscr A}(\Sigma,\gamma)}$ are associated to the two supercharges of the Donaldson twist, preserved near the infinity of the cigar. As in section \ref{tt}, this is a topological twist on $\Sigma$.

\subsubsection{}

A structure important for Langlands, but missing in the current discussion, is that it should come with factorization \cite{Gaitsgory08}. In particular, the correspondence between conformal blocks of a pair of chiral algebras is a consequence of equivalences of their module categories, viewed as braided tensor categories with a fusion product. Having the latter, the correspondence between conformal blocks would follow by taking invariants \cite{AFO}.

The upcoming work with LePage, Shende and Zhou \cite{ALSZ} extends the theory from the non-compact surfaces $\Sigma$ considered so far, without boundary, to non-compact surfaces with a non-compact boundary. Introducing boundaries on $\Sigma$ corresponds to modifying the superpotential ${\cal W}_{\rm top}$ to include terms with polynomial growth on $\Sigma$. This modifies the theory; in the Fukaya category, it corresponds to adding stops along the boundary components. (The difference between the superpotential in \cite{A2} and the similar one in \cite{GW} is of this kind; it results in a drastic modification of the category.)

The key theorem proven in \cite{ALSZ} is that ${\mathscr D}_{{\cal Y}(\Sigma)}$ and ${\mathscr D}_{{\mathscr A}(\Sigma, \gamma)}$ obey gluing over strips, so that Theorem \ref{sige} holds compatibly with the factorization structure corresponding to cutting $\Sigma$ into pieces and gluing it back from them. The cuts are along the arcs $\gamma_i$, and the gluing is over the strips $\gamma_i\times[0,1]$. It is also compatible with the braiding of marked points. The result is a 2d TQFT, valued in categories, which includes surfaces $\Sigma$ of arbitrary genus.

\subsubsection{}
Gluing over strips implies that the strip category, corresponding to the surface $\Sigma$ which is ${\mathbb R} \times [0,1]$, categorifies $U_{\fq}(\fg)_+$, the positive half of $U_{\fq}(\fg)$, as a Hopf category with its product and co-product.

The product structure comes from gluing strips. The strip category acts on any category associated to a surface with a boundary component. In particular, the standard finite dimensional and Verma representations of $U_{\fq}(\fg)$ and their tensor products come from a surface $\Sigma$ which is ${\mathbb R}\times {\mathbb R}_{\geq 0}$ with some interior marked points and holes.

\subsubsection{}
The co-product comes from gluing pairs of module categories via an open pair of pants \cite{ALSZ}. This provides a monoidal structure on the 2-category of categorified representations of $U_{\fq}(\fg)$, solving a long-standing open problem, posed by L. Crane and I. Frenkel in their 1994 paper \cite{CF} that started the categorification subject. The theory in \cite{ALSZ} solves the problem uniformly for all ${\fg}$. For previous attempts to solve the problem, see \cite{RouT, MR, McM, McM2, ER}.

In a Hopf algebra, the multiplication and co-multiplication are dual operations: the multiplication in $U_{\fq}(\fg)_+$ is the co-multiplication in its dual $U_{\fq}(\fg)_-$, and vice versa \cite{Resh}. In the solution of the problem in \cite{ALSZ}, they are not on the same footing. The co-multiplication is only a bimodule, not a map from $U=U_{\fq}(\fg)_+$ to $U\otimes U$; the two become dual only upon decategorification. As a result, the categorified theory has only a Hopfish \cite{TWZ}, rather than Hopf, structure. Presumably, this is the reason why the solution to the problem posed in \cite{CF} took so long to find.

\section*{Statement on AI use}
I wrote the manuscript and take full responsibility for its content.
Claude Fable 5.2 was a very helpful assistant for section \ref{GS}
(General $\Sigma$). It noticed a subtle error in an early draft of the section, worked out the correct formula for the superpotential $e^{-\cal W}$, and drew figures.
Here and elsewhere in the paper, it was used for editing my text.

\end{document}